\documentclass[12pt]{article} 

\usepackage{natbib}
\usepackage[ruled]{algorithm2e}
\usepackage{amsmath}
\usepackage{bm}
\usepackage{booktabs}
\usepackage{graphicx}
\usepackage[center]{subfigure}
\usepackage{caption}
\usepackage{url}
\usepackage{xspace}
\usepackage{multirow}
\usepackage{paralist}
\usepackage{enumitem}
\usepackage{authblk}
\newcommand{\revone}[1]{#1}
\setcitestyle{numbers,open={[},close={]}}

\newcommand{\lname}{Activity Decay Rate\xspace}

\newcommand{\mname}{Peer Influence Growth Rate\xspace}

\newcommand{\acname}{Critical Activity Threshold\xspace}
\newcommand{\qname}{Maximum Peer Activity Flow\xspace}

\newcommand{\ldesc}{\emph{\lname}\xspace}

\newcommand{\mdesc}{\emph{\mname}\xspace}

\newcommand{\acdesc}{\emph{\acname}\xspace}
\newcommand{\qdesc}{\emph{\qname}\xspace}

\newcommand{\adm}{\textit{Activity Dynamics Model}\xspace}

\newcommand{\kcaption}{Zachary's Karate Club Network\xspace}

\newcommand{\ratio}{$\lambda/\mu$\xspace}
\newcommand{\lam}{$\lambda$\xspace}
\newcommand{\mue}{$\mu$\xspace}
\newcommand{\kone}{$\kappa_1$\xspace}

\begin{document}


\title{Activity Dynamics in Collaboration Networks}

\author[1]{Simon Walk\thanks{simon.walk@tugraz.at}}
\author[2]{Denis Helic}
\author[2]{Florian Geigl}
\author[3,4]{Markus Strohmaier}
\affil[1]{IICM - Graz University of Technology}
\affil[2]{KTI - Graz University of Technology}
\affil[3]{GESIS - Leibniz Institute for the Social Sciences}
\affil[4]{University of Koblenz-Landau}
\renewcommand\Authands{ and }

\maketitle

\begin{abstract}
\textbf{Abstract} 
Many online collaboration networks struggle to gain user activity and become self-sustaining due to the ramp-up problem or dwindling activity within the system. Prominent examples include online encyclopedias such as (Semantic) MediaWikis, Question and Answering portals such as StackOverflow, and many others. Only a small fraction of these systems manage to reach self-sustaining activity, a level of activity that prevents the system from reverting to a non-active state. In this paper, we model and analyze activity dynamics in synthetic and empirical collaboration networks. Our approach is based on two opposing and well-studied principles:
\begin{inparaenum}[(i)]
\item without incentives, users tend to lose interest to contribute and thus, systems become inactive, and
\item people are susceptible to actions taken by their peers (social or peer influence).
\end{inparaenum}
With the activity dynamics model that we introduce in this paper we can represent typical situations of such collaboration networks. For example, activity in a collaborative network, without external impulses or investments, will vanish over time, eventually rendering the system inactive. However, by appropriately manipulating the activity dynamics and/or the underlying collaboration networks, we can jump-start a previously inactive system and advance it towards an active state. To be able to do so, we first describe our model and its underlying mechanisms. We then provide illustrative examples of empirical datasets and characterize the barrier that has to be breached by a system before it can become self-sustaining in terms of critical mass and activity dynamics. Additionally, we expand on this empirical illustration and introduce a new metric $p$---the \emph{Activity Momentum}---to assess the activity robustness of collaboration networks.	
\end{abstract}




\section{Introduction}

One of the major problems faced by both, new and existing online social and collaboration networks---such as Facebook or StackOverflow---revolves around efficiently identifying and motivating the appropriate users to contribute new content. In an optimal scenario, this newly contributed content provides enough incentive for other users to contribute, triggering further actions and contributions. Once such a self-reinforced state of increasing activity is reached, we can say that a system becomes self-sustaining, meaning that sufficiently high levels of activity are reached, which will keep the system active without further external impulses. For example, when looking at well-established collaborative websites, such as StackOverflow or Wikipedia, we already know that at some point in time, these systems have become self-sustaining (in terms of activity), evident in their steady growing number of supporters and overall activity.

However, these self-sustaining states are neither easy to reach nor guaranteed to last. For example, \citet{suh2009singularity} showed that the growth of Wikipedia is slowing down, indicating a loss in momentum and perhaps even first evidence of a collapse. Moreover, we typically lack the tools to properly analyze these trends in activity dynamics and thus, can not even perform such simple tasks as detecting self-sustaining system states. Therefore, we argue that new tools and techniques are needed to model, monitor and simulate activity dynamics for collaboration networks.

\begin{figure*}[t!]
\captionsetup[subfigure]{captionskip=2pt, farskip=1pt}
\centering
\subfigure[Intrinsic Activity (blue) and Peer Influence (yellow) at time $t_0$]{\label{fig:collab_1}\includegraphics[width=0.24\linewidth]{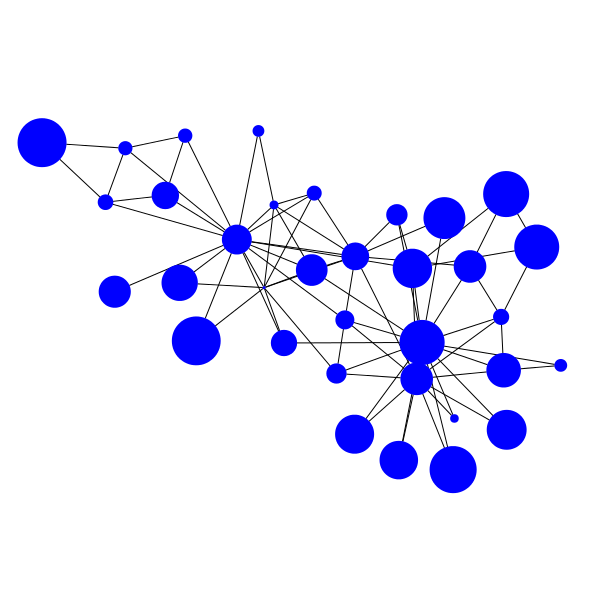}}
\subfigure[Intrinsic Activity (blue) and Peer Influence (yellow) at time $t_1$]{\label{fig:collab_2}\includegraphics[width=0.24\linewidth]{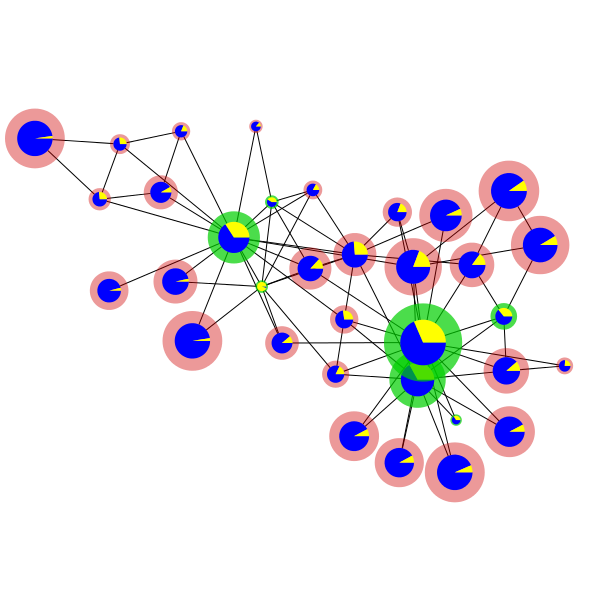}}
\subfigure[Intrinsic Activity (blue) and Peer Influence (yellow) at time $t_2$]{\label{fig:collab_3}\includegraphics[width=0.24\linewidth]{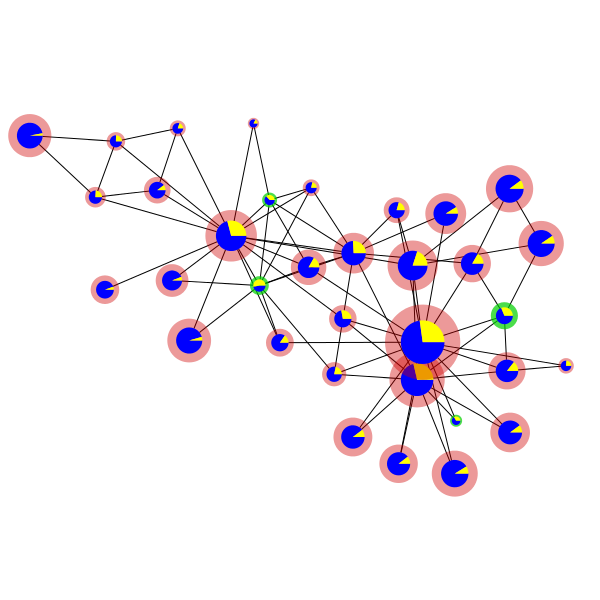}}
\subfigure[Intrinsic Activity (blue) and Peer Influence (yellow) at time $t_3$]{\label{fig:collab_4}\includegraphics[width=0.24\linewidth]{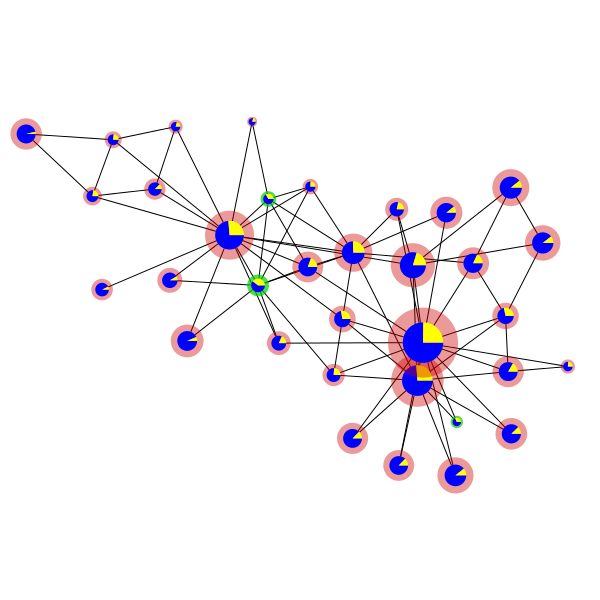}}
\caption{ \textbf{Intrinsic Activity and Positive Peer Influence.} Activity dynamics in collaboration networks, represented by users as nodes, collaboration as edges and activity as node size (Figure~\subref{fig:collab_1}), are based on two opposing principles. The \ldesc postulates the loss of intrinsic activity (blue color of nodes) per user over time. In contrast, the \mdesc follows the intuition, that users in collaboration networks are (positively) influenced by their peers (yellow color of nodes) where more active peers exercise a higher influence than less active peers. We initialize the network at time $t_0$ with random intrinsic activities. Nodes with a green halo at times $t_1$ to $t_3$ represent users that exhibit a gain in their overall activity between two iterations $t_n$ and $t_{n+1}$, as the exercised positive peer influence is higher than the intrinsic loss of activity. Analogously, red halos represent decreases in overall activity. At first, very central (high degree) nodes with smaller activity values manage to increase their overall activity, while very active central nodes already start to lose activity. After $t_3$ or more iterations, due to overall decreasing activities and hence, decreasing peer influences, all nodes in the collaboration network eventually start to lose activity and inevitably converge towards zero activity.  
}
\label{fig:act_dyn_prin}
\end{figure*}

The high-level contributions of this work are two-fold. First, we introduce a model that is capable of simulating activity dynamics for online collaboration networks. Second, we describe in detail how to fit the model to empirical datasets, simulate trends in activity dynamics and interpret our findings. The proposed model is based on the formalism of continuous deterministic dynamical systems---meaning that activity is modeled by a system of coupled non-linear differential equations. Each user of the system is represented by a single quantity (the current activity), and the social ties between users define the coupling of variables. \revone{In general, when using dynamical systems on networks, we define the (micro-)behavior of each user to observe and gather new insights into the (macro-)behavior of the system. For a more detailed introduction to dynamical systems see Section~\ref{sec:related work} and \citet{newman2010networks}.}
For simplicity, we do not take individual differences between users into account---the dynamics and its parameters are the same for each user in the population. This allows us to configure the model with a single parameter, which is a ratio of the following two parameters, representing two basic activity mechanisms (cf. Figure~\ref{fig:act_dyn_prin}) in online collaboration networks:
\begin{enumerate}[label=(\roman*)]
\item \ldesc $\lambda$, which postulates how fast users lose interest to contribute,
\item \mdesc $\mu$, postulating to what extent users are influenced by the actions taken by their peers.
\end{enumerate}

A first analysis of the model shows that activity dynamics in collaboration networks have an obvious and natural fixed point---the point of complete inactivity---where all contributions of the users have seized. However, by slightly manipulating the parameters in our model we show that it is possible to destabilize the fixed point, resulting in a potential increase of activity. We then outline the process of calculating the \ldesc and \mdesc for existing collaboration networks, simulate their corresponding activity dynamics and expand our understanding of critical mass---via the notion of \textit{System Mass} and \textit{Activity Momentum}---in collaboration networks by interpreting our findings.

The remainder of this paper is structured as follows: In Section~\ref{sec:modeling activity dynamics} we introduce and examine our model analytically. We then continue with the model illustration by simulating activity dynamics for a synthetic dataset and discuss different evolution scenarios of our parameters and their implications.  In Section~\ref{sec:illustrative examples} we outline the process of applying our model on empirical datasets. In Section \ref{sec:momentum} we introduce the notion of \emph{System Mass} and \emph{Activity Momentum}, review related work in Section~\ref{sec:related work} and summarize our findings and discuss limitations and implications for future work in Section~\ref{sec:discussion}.

\section{Modeling Activity Dynamics}
\label{sec:modeling activity dynamics}

We model activity dynamics in an online collaboration network as a dynamical system on a network. Hereby, the nodes of a network represent users of the system and links represent the fact that the users have collaborated in the past. We represent the network with an $n \times n$ adjacency matrix $\bm{A}$, where $n$ is the number of nodes (users) in the network. We get $A_{ij} = 1$ if nodes $i$ and $j$ are connected by a link and $A_{ij} = 0$ otherwise. Since collaboration links are undirected, the matrix $\bm{A}$ is symmetric, thus $A_{ij} = A_{ji}$, for all $i$ and $j$. We denote the total number of links in the network with $m$, and thus we have $2m=\sum_{ij}A_{ij}$.

We model activity as a continuous real-valued variable $a_i$ evolving on node $i$ of the network in continuous time $t$. The general time evolution equation can be written as follows (see also \citet{newman2010networks}):
\begin{equation}
 \frac{da_i}{dt} = \underbrace{f_i(a_i)}_\text{\parbox{2.2cm}{\centering Intrinsic Activity Evolution of i}} + \overbrace{\sum_j A_{ij} \underbrace{g_i(a_i, a_j)}_\text{Influence of j on i}}^\text{Peer Influence},
\label{eq:gtee}
\end{equation}

where $f(a_i)$ specifies the intrinsic activity evolution of node $i$ and $g(a_i, a_j)$ describes the influence of neighbor $j$ on node $i$.  To simplify, we assume that the intrinsic activity dynamics as well as the influence of node neighbors are the same for each node $i$ and for each neighbor pair $(i, j)$. This means that we have a single intrinsic activity function $f(a_i)$ for all nodes $i$, as well as a single peer influence function $g(a_i, a_j)$ for all node pairs $(i, j)$.

In addition, we make the following assumptions:

\noindent\textbf{Intrinsic Activity Decay.} Without external incentives or without positive influence from their social connections, each user has a tendency to slowly reduce activity. For example, people slowly lose interest to participate in collaborative networks or exhaust their resources. An observation that specifically reflects this inherent exhaust of activity over time has been made by \citet{Danescu-Niculescu-Mizil:2013:NCO:2488388.2488416} for different online communities. We model this situation by using a linear function for $f(a_i)$:
\begin{equation}
 f(a_i) = -\lambda a_i, \lambda > 0
\end{equation}

\begin{figure*}[t!]
\captionsetup[subfigure]{captionskip=2pt, farskip=1pt}
\centering
\subfigure[Intrinsic Activity Decay]{\label{fig:fx}\includegraphics[width=0.48\linewidth]{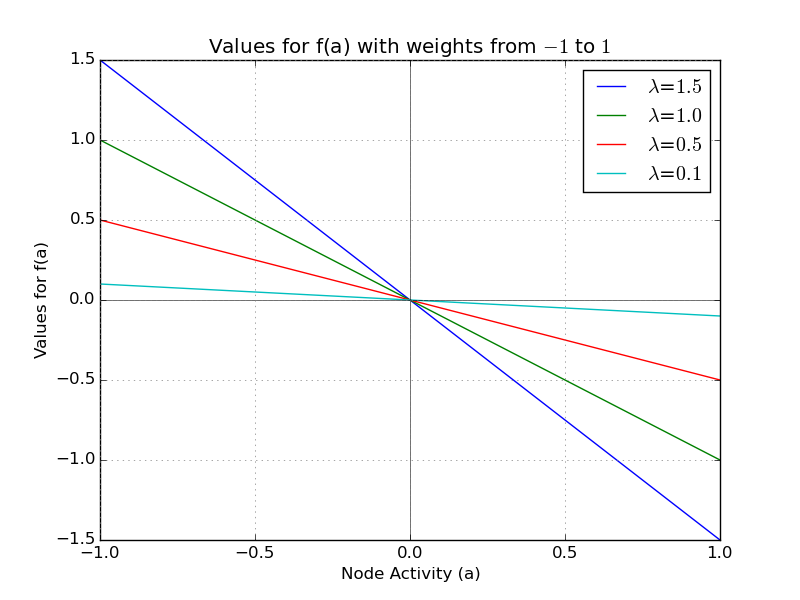}}
\subfigure[Extrinsic Peer Influence]{\label{fig:gx}\includegraphics[width=0.48\linewidth]{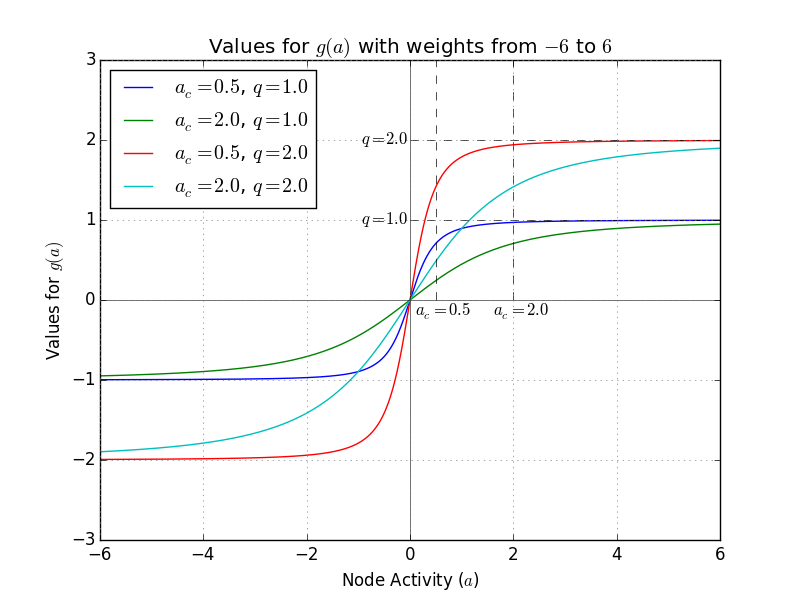}}
\caption{ \textbf{Intrinsic Activity Decay} is the rate at which users reduce their activity per unit time and is represented as a linear function in the form of $f(a)=-\lambda a$, which results in an exponential decay in activity that converges towards zero. \textbf{Extrinsic Positive Peer Influence} describes to what extent users are influenced by the actions taken by their peers, and is represented as a monotonically increasing function of a users activity in the form of $g(a) = (qa)/\sqrt{a_c^2 + a^2}$. It naturally saturates at \qdesc $q$ as activity reaches infinity and, in our simulations, can never be negative per definition (see Equation~\ref{eq:gx}). When the user activity passes the point of the \acdesc $a_c$, peer influence gains notable weight and influences neighbors to ``do something'' (become active).
}
\label{fig:Functions}
\end{figure*}

We call parameter $\lambda$ the \ldesc---the rate at which users reduce their activity per unit time, given a complete absence of other (positive) incentives. The specific form of $f(a_i)$ results in an exponential decay ($a_i(t) = a_i(t_0)e^{-\lambda t}$, with $a_i(t_0)$ being the initial activity of node $i$ at time $t_0$) of activity without any external influence. Thus, without other positive impulses the activity of every user will decay over time (see Figure~\ref{fig:fx}).

\noindent\textbf{Positive Peer Influence.} People tend to copy their friends \cite{christakis2008collective, aral2012identifying, wagner2012social}, meaning that if neighbors of a node $i$ are active they will positively influence node $i$ to become active as well. The magnitude of the influence, or the ``speed'' at which the influence is transferred from an active node to its neighbors will depend on two quantities (cf. Figure~\ref{fig:Functions}):
\begin{enumerate}[label=(\roman*)]
 \item \acdesc $a_c$, which represents a soft threshold of activity that marks the point when users have an activity potential, that notably exercises influence on their peers. Note that influence is exercised at all levels of $a_c$. However, once $a_c$ is reached, the influence is determined as ``notable'' (e.g., a level of activity that is above the average activity per user) for the corresponding peers. Hence, this critical level of activity is a system-dependent quantity. One can imagine that in a system with high user activity (e.g., a large number of changes per user) the critical activity is higher than in a system with lower levels of activity. For example, in the latter case the users will sooner notice a neighbor who became active recently. We model the \acdesc as a continuous threshold. Meaning that active users will always influence their neighbors, but will exercise more influence after they have passed the critical level of activity.
\item \qdesc $q$ represents the maximum activity flow per unit time from users to each of their neighbors. This maximum flow is reached as user activity approaches infinity. However, substantial amounts of the maximum flow are already reached whenever the user activity passes the level of the critical activity $a_c$.
\end{enumerate}

Thus, to model peer influence, we resort to a monotonically increasing function, where more active neighbors are always more influential than less active ones. Additionally, the function $g(a_j)$ saturates for sufficiently large values of activity, inducing a natural limit on how much users can be influenced by their neighbors. We model this by setting $g(a_i, a_j) = g(a_j)$ and choosing an algebraic sigmoid function with:
\begin{equation}
\label{eq:gx}
 g(a_j) = \frac{qa}{\sqrt{a_c^2 + a_j^2}}, q, a_c > 0.
\end{equation}

Peer influence can also be analyzed in terms of the growth rate of $g(a)$, in the form of the derivative $dg/da$ of the function $g(a)$. After simplifying and rearranging, the growth rate can be calculated as:
\begin{equation}
 \frac{dg}{da} = \frac{qa_c^2}{(a_c^2 + a^2)^{3/2}}.
\end{equation}

In the limit of large activity $a$ the derivative of $g(a)$ tends towards zero, thus peer influence saturates at $q$. On the other hand, the maximum change in influence is observed when $a = 0$---neighbors who suddenly become active will be noted most, in terms of activity, by their peers.

\subsection{Dynamics Equation}

With $f(a_i)$ and $g(a_j)$ defined, the activity dynamics equation becomes:
\begin{equation}
\frac{da_i}{dt} = -\lambda a_i + \sum_j A_{ij} \frac{qa}{\sqrt{a_c^2 + a_j^2}}.
\end{equation}

The different parameters of the equation have dimensions. For example, $a_i$ and $a_c$ have activity as unit, $t$ has seconds as unit, $\lambda$ is a rate and has inverse seconds as unit, and $q$ has activity per second as unit. Further, the equation has three free parameters, which span a huge parameter space that is difficult to explore in detail. Therefore, our first step is to simplify the equation and express it in a dimensionless form, which typically also has a smaller number of parameters as only their relative ratios, rather than their absolute values, are of importance. Another advantageous side-effect of a dimensionless formulation is that it eliminates the absolute values of the properties under investigation, in our case user activity, which can be difficult to interpret.

There are many ways to eliminate dimensions from such equations \cite{lin1988mathematics}. A useful heuristic is to try to first eliminate the dimensions from the most non-linear term in the equation, which in our case is $g(a_j)$. Thus, we begin by defining a relative activity $x$ as the ratio between the activity $a$ and the critical activity $a_c$:
\begin{equation}
 x = \frac{a}{a_c}.
\end{equation}

The variable $x$ is dimensionless now, and it is easy to interpret. For example, the fact that $x=5$ means that users exercises a strong influence on their neighbors, since the level of activity is five times the critical activity $a_c$. In fact, the influence in this case is $g(5a_c)=(5q)/\sqrt{26} \approx 0.98q$. On the other hand if $x\ll1$ (e.g., $x=0.1$), this then means that the influence of users on their neighbors is much smaller as $g(0.1a_c)=(0.1q)/\sqrt{1.01} \approx 0.1q$.

\revone{By rearranging, substituting $x$ for $a$ and simplifying ($a_c$ cancels in the second term) our activity dynamics equation reduces to:

\begin{equation}
 a_c\frac{dx_i}{dt} = -\lambda a_c x_i + \sum_j A_{ij} \frac{q x_j}{\sqrt{1 + x_j^2}}.
\end{equation}
}

To eliminate the dimensions from the second term we divide both sides with $q$:
\begin{equation}
 \frac{a_c}{q}\frac{dx_i}{dt} = -\lambda \frac{a_c}{q} x_i + \sum_j A_{ij} \frac{x_j}{\sqrt{1 + x_j^2}}.
\end{equation}

The term $q/a_c$ is the growth rate of the function $g(a)$ evaluated at zero:
\begin{equation}
 \left.\frac{dg}{da}\right|_{a=0} = \left.\frac{qa_c^2}{(a_c^2 + a^2)^{3/2}}\right|_{a=0}=\frac{q}{a_c}.
\end{equation}

This quantity gives the rate at which the influence on the peers grows if the user activity experiences a small displacement from the point of zero activity. Let us now define this quantity as \mdesc and denote it with $\mu = q/a_c$ since this will simplify the algebra and will make the model interpretation more intuitive. Thus, the last equation can then be written as:
\begin{equation}
 \frac{1}{\mu}\frac{dx_i}{dt} = -\frac{\lambda}{\mu} x_i + \sum_j A_{ij} \frac{x_j}{\sqrt{1 + x_j^2}}.
\label{eq:dd}
\end{equation}

Finally, we also want to scale time $t$ and express the equation in terms of dimensionless time $\tau$. This last reformulation will further simplify the equation and allows us to interpret and compare activity dynamics over time across various systems. The latter is possible due to the usage of dimensionless time $\tau$ to scale and compare the time evolution of different systems relative to each other. Let us make the following substitution:
\begin{equation}
 \tau = \mu t.
 \label{eq:mu}
\end{equation}

\revone{By substituting $\tau$ for $t$ in the term on the left hand side in Equation \ref{eq:dd} we arrive at the dimensionless dynamics equation:
\begin{equation}
 \frac{dx_i}{d\tau} = -\frac{\lambda}{\mu} x_i + \sum_j A_{ij} \frac{x_j}{\sqrt{1 + x_j^2}}.
\label{eq:fd}
\end{equation}}

Now, there is only one parameter in our dynamics equation, namely the ratio $\lambda/\mu$. This is a dimensionless ratio of two rates:
\begin{inparaenum}[(i)]
 \item The \ldesc $\lambda$, which is the rate at which a user loses activity, and
\item the \mdesc $\mu$, which is the rate at which a user gains activity due to the influence of a \emph{single} neighbor.
\end{inparaenum}

The ratio between those two rates is the ratio of how much faster users lose activity due to the decay of intrinsic activity (or interest) than they can gain due to positive peer influence of a single neighbor. For example, a ratio of $\lambda/\mu = 100$ would mean that the users intrinsically lose activity $100$ times faster than they potentially can get back from one of their neighbors. If we would set $\lambda/\mu = 1$, it would mean that users would lose activity as fast as they can regain it from one of their peers. For a short description of all parameters of the activity dynamics model see Table~\ref{tab:model params}.

\begin{table}[!b]
\centering
\caption{\textbf{Model and model parameters.} The activity dynamics equation is in a dimensionless form and scales over relative time $\tau$. All properties, as well as the single parameter of the model, are briefly described under $Properties$ and $Parameters$.}
\centering
\begin{tabular}{l  p{6.5cm}}
\toprule
\multicolumn{1}{c}{Equation} & \multicolumn{1}{c}{Name} \\ \midrule
$ \frac{dx_i}{d\tau} = -\frac{\lambda}{\mu} x_i + \sum_j A_{ij} \frac{x_j}{\sqrt{1 + x_j^2}}$ & Activity Dynamics Equation\\
\midrule
\multicolumn{1}{c}{Properties} & \multicolumn{1}{c}{Name} \\ \midrule
\multicolumn{1}{c}{$\lambda$} & \ldesc \\
\multicolumn{1}{c}{$q$} & \qdesc \\
\multicolumn{1}{c}{$a_c$} & \acdesc \\
\multicolumn{1}{c}{$\mu = \frac{q}{a_c}$} & \mdesc \\ 
\multicolumn{1}{c}{$\tau$} & Relative Time Scale \\ \midrule
\multicolumn{1}{c}{Parameter} & \multicolumn{1}{c}{Name} \\ \midrule
\multicolumn{1}{c}{\multirow{4}{*}{$\frac{\lambda}{\mu}$}} & The ratio, describing how fast users intrinsically loses activity compared to how fast they get it back from (one of) their neighbors. \\
\bottomrule
\end{tabular}
\label{tab:model params}
\end{table}

\subsection{Linear Stability Analysis}
\label{subsec:linear stability analysis}
In general, Equation \ref{eq:fd} is a coupled set of $n$ ($n$ being the number of nodes or users in the network) non-linear differential equations, for which, in a typical case, no closed form solution can be found. Therefore, we turn our attention to the properties of so-called fixed points. A fixed point $\bm{x^*}$ represents all the values for $x_i^*$ for which the system does not change in time:
\begin{equation}
 \frac{dx_i}{d\tau} = -\frac{\lambda}{\mu} x_i + \sum_j A_{ij} \frac{x_j}{\sqrt{1 + x_j^2}} = 0, \forall i.
\label{eq:fp}
\end{equation}

Suppose that we are able to find a fixed point $\bm{x^*}$ by solving Equation \ref{eq:fp}. One obvious fixed point in our model is $\bm{x^*} = \bm{0}$, meaning that $x_i^*$ has the same value for every $i$: $x_i^*=x^*=0$, representing a simple special case: a symmetric fixed point. We can easily check that $x^*=0$ is indeed a fixed point since $f(x^*) = g(x^*) = 0$, and this also gives $f(x^*) + \sum_j A_{ij} g(x^*) = 0, \forall i$.

\revone{We are investigating this specific fixed point, as it also has a particular interpretation in our model. At this fixed point all users have zero activity, which means that they are completely inactive and the system is in an inactive or ``dead'' state.} If the system is in such a state and no external incentives are provided, nothing will ever change and the system will remain inactive indefinitely.

Typically, we are interested in the implications on the system if we provide a small enough impulse to leave such a steady (inactive) state. In our context, the most interesting question is if the system will move from an inactive state towards a state of lively activity or if it will just revert to the inactive state. Technically, we are interested in the stability of the fixed point. In particular, we want to know if the fixed point is attracting (meaning that the system's activity in the proximity of the fixed point will be attracted to it) or repelling (meaning that the system's activity close to the fixed point will be pushed away from it).

To answer this question we linearize the functions in the proximity of a fixed point. We represent the value of $x_i$ close to the fixed point with $x_i = x^* + \epsilon_i$, where $\epsilon_i$ is sufficiently small. To simplify the calculations, we concentrate on the case of a symmetric fixed point, such as $\bm{x^*} = \bm{0}$. Next, we perform a Taylor expansion about the fixed point and linearize by neglecting the terms of second and higher orders. After simplification we obtain (for details see e.g. \citet{newman2010networks}):
\begin{equation}
\frac{d\epsilon_i}{d\tau} = -\frac{\lambda}{\mu}\epsilon_i + \sum_j A_{ij} \epsilon_j,
\label{eq:epsilon}
\end{equation}
where $\epsilon_i$ is the displacement of $x_i$ from the fixed point $x^*$.

We can also write Equation \ref{eq:epsilon} in matrix form, which gives:
\begin{equation}
\frac{d\bm{\epsilon}}{d\tau} = (-\frac{\lambda}{\mu}\bm{I} + \bm{A})\bm{\epsilon},
\label{eq:epsilon_time}
\end{equation}
where $\bm{I}$ is the identity matrix and $\bm{A}$ is the adjacency matrix.

We can solve the last equation by writing $\bm{\epsilon}$ as a linear combination of eigenvectors $\bm{v}_r$ of the symmetric real matrix $(-(\lambda/\mu)\bm{I} + \bm{A})$:
\begin{equation}
\bm{\epsilon}(\tau) = \sum_r c_r(\tau)\bm{v}_r.
\end{equation}

Equation \ref{eq:epsilon_time} then becomes:
\begin{equation}
\sum_r \frac{dc_r}{d\tau} \bm{v}_r = (-\frac{\lambda}{\mu}\bm{I} + \bm{A}) \sum_r c_r(\tau)\bm{v}_r = \sum_r c_r(\tau)(-\frac{\lambda}{\mu} + \kappa_r) \bm{v}_r,
\label{eq:coef}
\end{equation}
where $\kappa_r$ are the eigenvalues of the graph adjacency matrix $\bm{A}$. We also used the fact that the matrix $(-(\lambda/\mu)\bm{I} + \bm{A})$ has the same eigenvectors as $\bm{A}$, but with the eigenvalues $-\lambda/\mu + \kappa_r$.

The solution of the last equation for the coefficients of the linear combination is then:
\begin{equation}
 \frac{dc_r}{d\tau} = (-\frac{\lambda}{\mu} + \kappa_r) c_r(\tau) \implies c_r(\tau) = c_r(t_0) e ^{(-\frac{\lambda}{\mu} + \kappa_r)\tau}.
\end{equation}

Now, the displacement from the fixed point will decay in time towards $0$ if the exponents for the coefficients $c_r(\tau)$ are all negative. Thus, we arrive at the master stability equation for the special case of a dynamical system that we defined as:

\begin{equation}
 -\frac{\lambda}{\mu} + \kappa_r < 0, \forall r,
\label{eq:stab}
\end{equation}

\begin{figure*}[h!]
\captionsetup[subfigure]{captionskip=2pt, farskip=1pt}
\centering
\subfigure[\kcaption]{\label{fig:karate_club_graph}\includegraphics[width=0.3\linewidth]{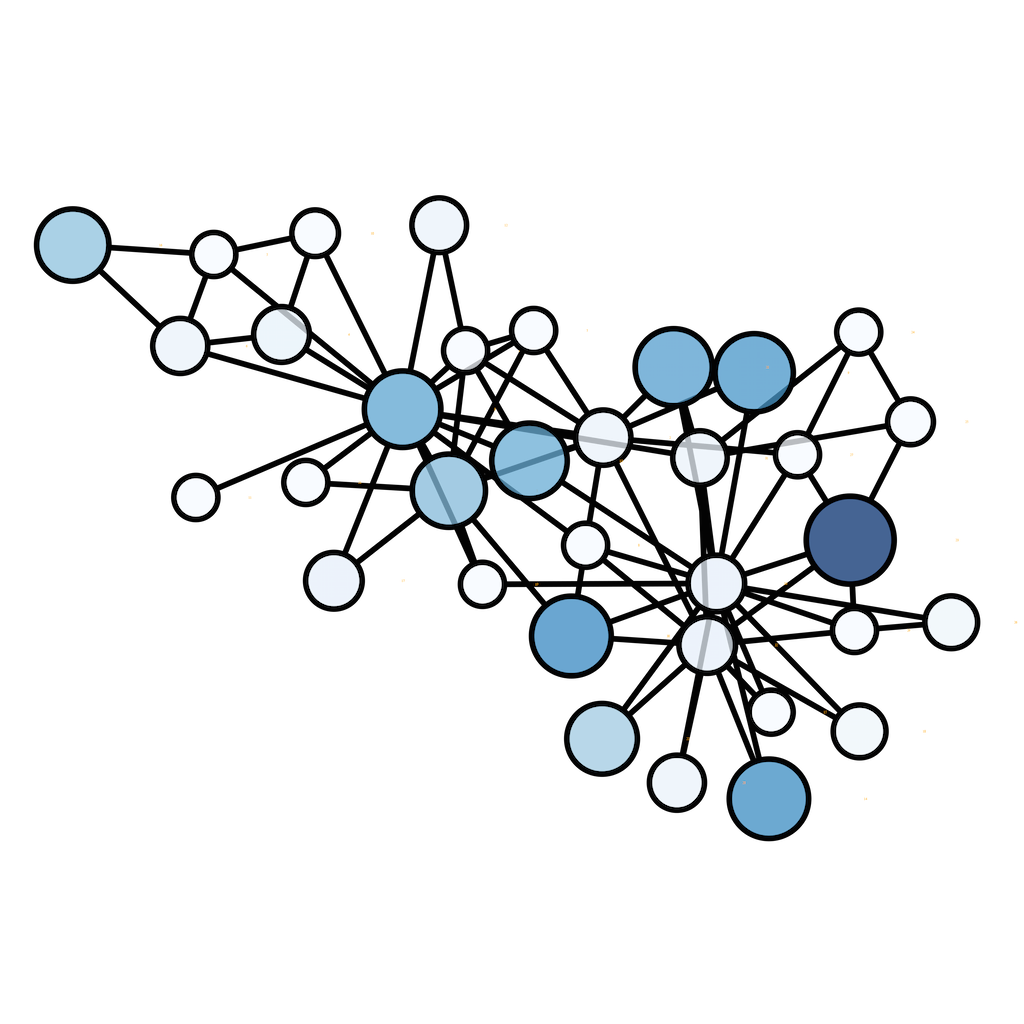}}
\subfigure[Adjacency Spectrum ($\kappa_1 = 6.726$)]{\label{fig:karate_club_eigenvalues}\includegraphics[width=0.3\linewidth]{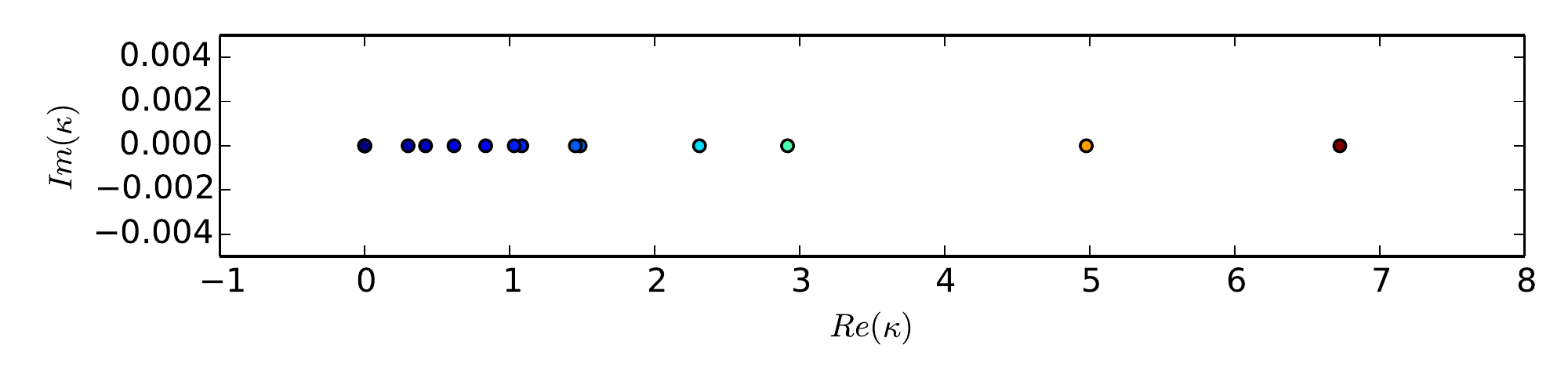}}

\subfigure[Activity Evolution $\frac{\lambda}{\mu}>\kappa_1$]{\label{fig:karate club valid}\includegraphics[width=0.3\linewidth]{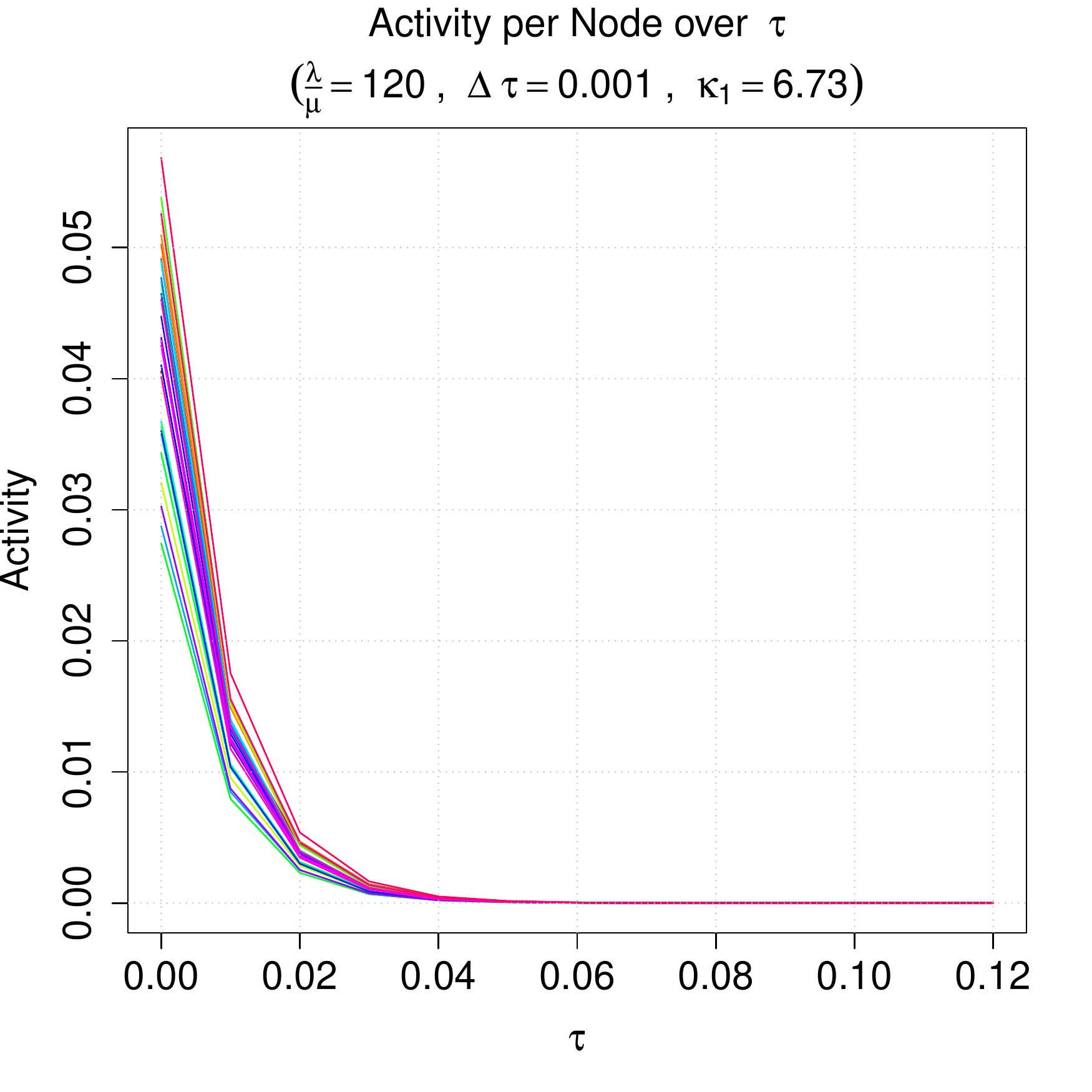}}
\subfigure[Activity Evolution $\frac{\lambda}{\mu}<\kappa_1$]{\label{fig:karate club invalid}\includegraphics[width=0.3\linewidth]{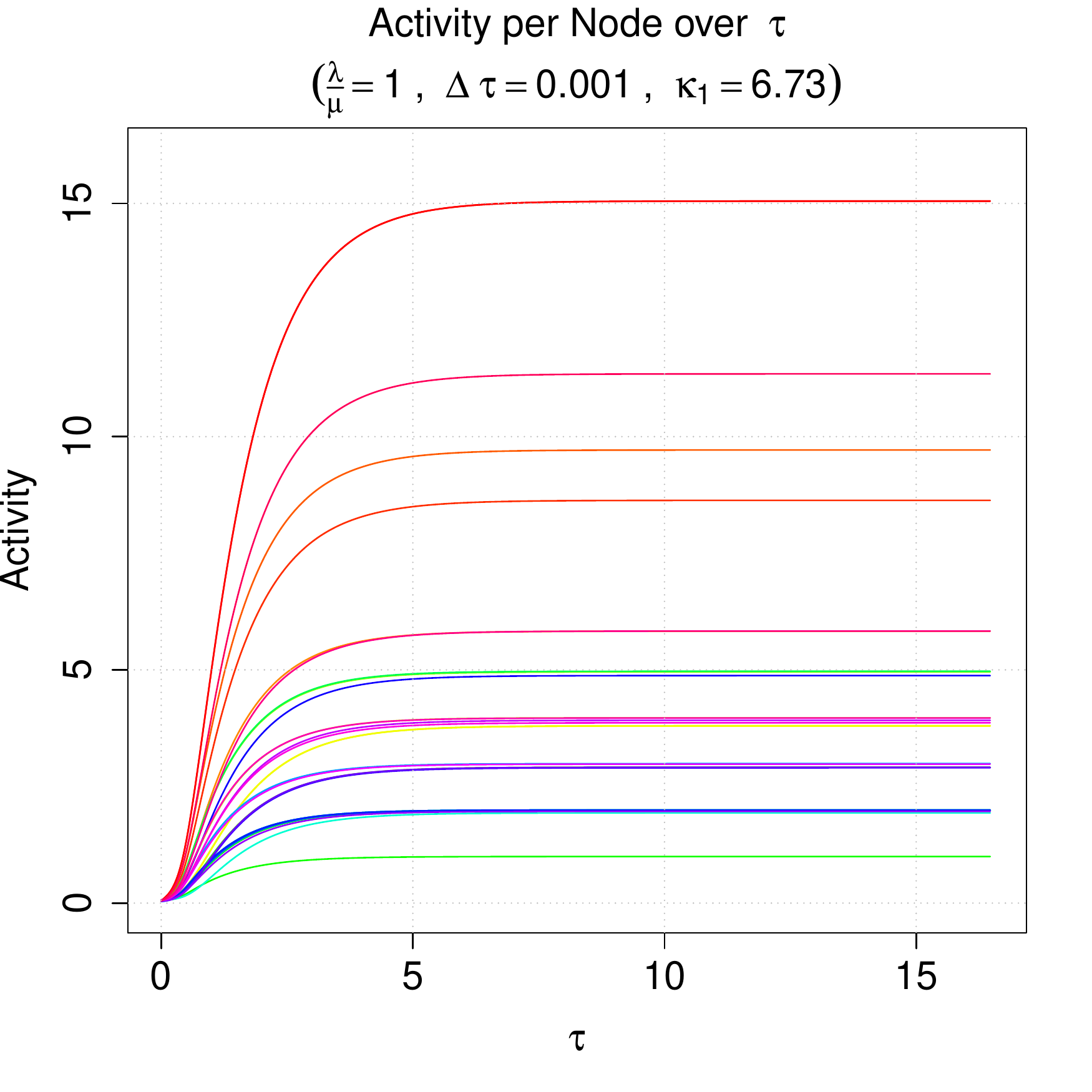}}
\caption{ \textbf{Illustrative example.}  \textbf{Top Left (a):} Visualization of Zachary's Karate Club. The size and color of a node represent random activity values between $0.0$ and $0.1$ of the corresponding nodes (bigger and darker equals higher values). \textbf{Top Right (b):} Eigenvalue spectrum of Zachary's Karate Club network. 
The highest eigenvalue is $6.726$. \textbf{Bottom (c and d):} Evolution of activity with random initial activities (averaged over 10 runs).
\textbf{Bottom Left (c):} Activity dynamics with parameters satisfying the master stability condition $\kappa_1 < \lambda/\mu$. Each line represents one node; all activities converge to the state of zero activity. \textbf{Bottom Right (d):} Invalidation of the master stability condition $\kappa_1 < \lambda/\mu$, activity converges towards a new and permanently active fixed point.}
\label{fig:Karate}
\end{figure*}

Since the adjacency matrix has both positive and negative eigenvalues, a necessary stability condition is $\lambda/\mu > 0$, which is satisfied by definition.
Thus, we can rearrange Equation~\ref{eq:stab} and obtain the following inequality:

\begin{equation}
 \kappa_1 < \frac{\lambda}{\mu}.
\end{equation}
where $\kappa_1$ is the largest positive eigenvalue of the graph adjacency matrix. Note that this inequality separates the network structure ($\kappa_1$) from the activity dynamics  ($\lambda/\mu$).

If this stability condition is satisfied, the fixed point $x^*= 0$, in which there is no activity at all (``inactive'' system), represents a stable fixed point. This also means that small changes in activity only cause the system to momentarily leave the (attracting) fixed point until it becomes inactive again. 

For illustration, we initialized Zachary's Karate Club Network (cf. Figures \ref{fig:karate_club_graph} and \ref{fig:karate_club_eigenvalues}) with random activities between $0$ and $0.1$ per node and simulate activity with our model. If the master stability equation holds (Figure~\ref{fig:karate club valid}), activity converges towards zero. However, when invalidating the master stability equation (Figure~\ref{fig:karate club invalid}), activity converges to a new and permanently active fixed point.

\revone{
\begin{figure*}[!t]
\centering
\includegraphics[width=0.8\linewidth]{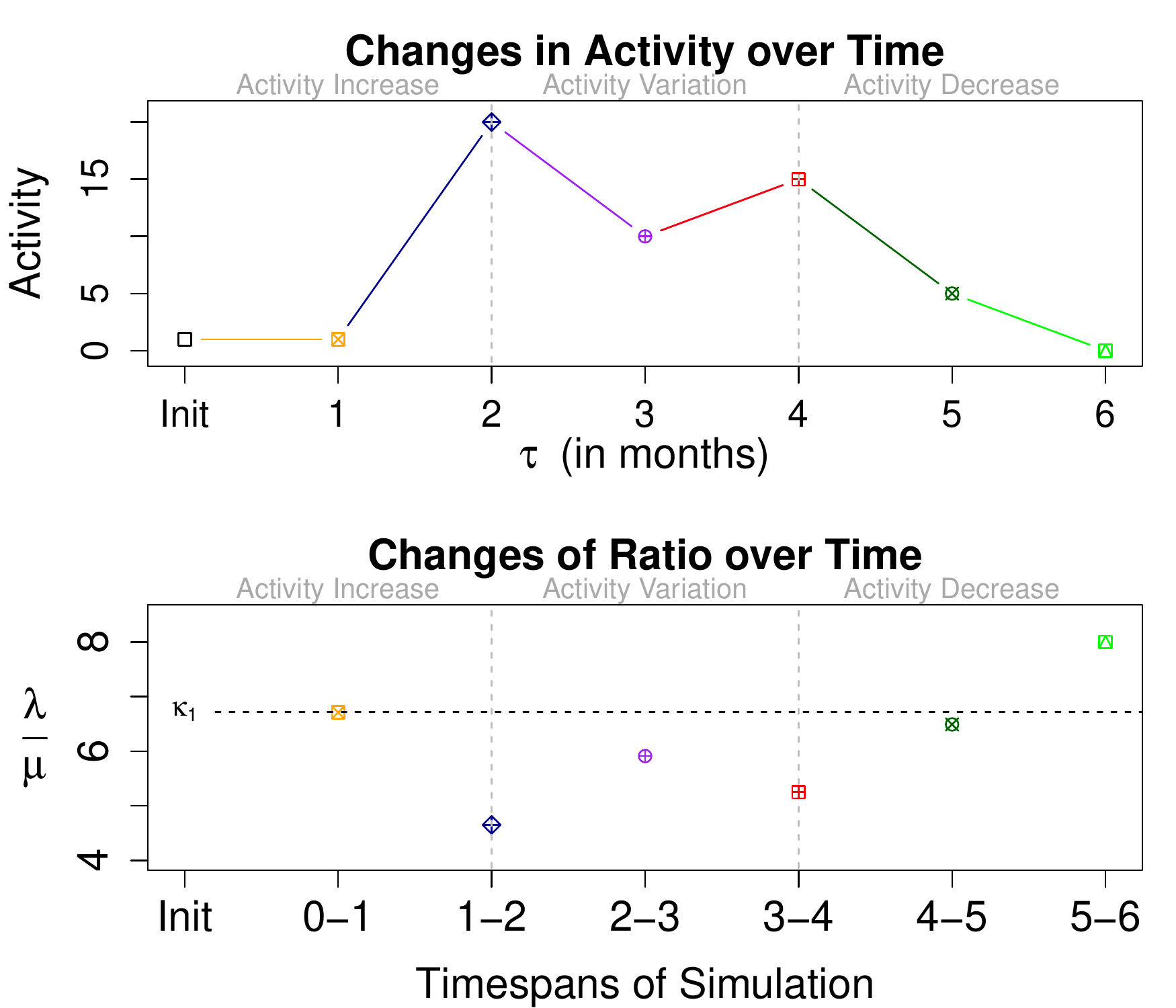}
\caption{ \revone{\textbf{Coupled evolution of activity and \ratio.} The \textbf{top} Figure depicts the evolution of activity ($y$-axis) over time ($x$-axis; in months) for Zachary's Karate Club network with synthetically created (random) activities. The ratios, which correspond to the activity evolutions over time in the top Figure, are depicted in the \textbf{bottom} Figure (same symbol and color), with the $y$-axis representing the value of the ratio, while the different timespans are depicted on the $x$-axis. As long as $\lambda/\mu<\kappa_1$ the network converges towards a state of immanent activity, yet decreases in activity are possible (see timespans $2-4$ of\textit{Activity Variation} sections in \textbf{top} and \textbf{bottom}). If $\lambda/\mu>\kappa_1$ the network converges towards an inactive state.
}}
\label{fig:empirical_ratios_karate}
\end{figure*}
}

\revone{
In practice, additional system configurations are imaginable.
Whenever the ratio is below \kone, the system becomes unstable leaving the inactive state. However, due to the special form of the peer influence function, which saturates for large values of activity, the system will converge towards another stable state of immanent activity (i.e., ratios for periods $1-5$ of Figure~\ref{fig:empirical_ratios_karate}). 

Thus, if the system is in the state where $\kappa_1 > \lambda/\mu$, we can think of \textbf{three different activity evolution scenarios}, depending on the current levels of activity present in the network: 
\begin{enumerate}
\item If the levels of activity are lower than the ones the network converges towards with the new ratio, we will see an increase in activity (e.g., timespans $1-2$ of \textit{Activity Increase} in Figure~\ref{fig:empirical_ratios_karate}). 
\item If the new ratio lets the system converge towards lower levels of activity than currently present, activity will decrease, even though $\kappa_1 > \lambda/\mu$ (e.g., see timespans $2-3$ or $4-5$ of \textit{Activity Variation} and \textit{Activity Decrease} of Figure~\ref{fig:empirical_ratios_karate}). 
\item Lastly, the levels of activity have already converged towards their fixed point and \ratio is left unchanged, retaining the levels of activity from the past (e.g., see timespans $0-1$ of \textit{Activity Increase} in Figure~\ref{fig:empirical_ratios_karate}).
\end{enumerate}

If $\kappa_1 < \lambda/\mu$ holds, the system is stable and activity converges towards the attracting fixed point at zero activity (see timespans $5-6$ of \textit{Activity Decrease} in Figure~\ref{fig:empirical_ratios_karate}).}

\smallskip\noindent\textbf{Summary of system stability analysis.} In order to permanently leave the stable state of complete inactivity we are interested in making the system unstable. To be able to leave the attracting force of the fixed point at zero activity we have the following two options:
\begin{enumerate}[label=(\roman*)]
 \item We \textbf{provide (continuous) external impulses} to the system, for example, in the form of incentives for users to increase their activity, pushing the system far away from the fixed point of no activity (and hope that it will be attracted by another fixed point where activity is not zero). 
\item We \textbf{compromise the stability condition} by either manipulating:
\begin{enumerate}
\item the network structure (i.e., \emph{making $\kappa_1$ larger}) or
\item the activity dynamics (i.e., \emph{making $\lambda/\mu$ smaller}).
\end{enumerate}
\revone{Structurally, we can manipulate the size of $\kappa_1$ by creating or removing links (and nodes) in our network (for more information on how to manipulate $\kappa_1$ see \cite{newman2010networks}).} Dynamically, $\lambda/\mu$ becomes smaller if either $\lambda$ becomes smaller, meaning that the intrinsic user activity decays at a slower pace or $\mu$ becomes larger, meaning that people copy their friends more and faster, or both.
\end{enumerate}

\revone{\subsection{Discussion on Parameter Evolution}
\label{subsec:parameter evolution}}

At this time, we leave the investigation of the manipulation of the activity dynamics ratio \ratio as well as the manipulation of the network structure to invalidate the master stability equation open for future work. Nevertheless, before illustrating how our proposed activity dynamics model can be applied to empirical datasets, we discuss potential system evolution scenarios and their implications for activity.

\revone{\smallskip\noindent\textbf{\ldesc.} Technically, if \lam increases, the ratio \ratio increases as well, resulting in higher (faster) losses of activity per timespan. Once the system satisfies the master stability equation ($\kappa_1 < \lambda/\mu$) it will inevitably become inactive. To be precise, the larger \lam for a stable system, the faster activity will converge towards zero.
Essentially, an increase in \lam represents an increased intrinsic loss of activity for all users (e.g., due to a lack of interest to contribute) while a decrease of \lam can be interpreted as an increase of interest (more precisely, slower loss of interest) and thus higher levels of activity. 

\smallskip\noindent\textit{Evolution scenarios of} \ldesc. We would expect to see an increase in \lam on websites with low levels of user interaction and activity (i.e., meaning that individual contributions are not valued, as no feedback is provided). On the other hand, websites that engage with their users and provide steady updates (e.g., new content or functionality) will likely see a consistent or even decreasing \lam. In general, practitioners can influence \lam by, for example, providing incentives for users to contribute, such as badges, barn stars, likes, reputation systems, or monetary incentives.}

\revone{
\begin{figure*}[!t]
\centering
\subfigure[Evolution of \acdesc]{\label{fig:scenario_peer_noise}
\includegraphics[width=0.48\linewidth]{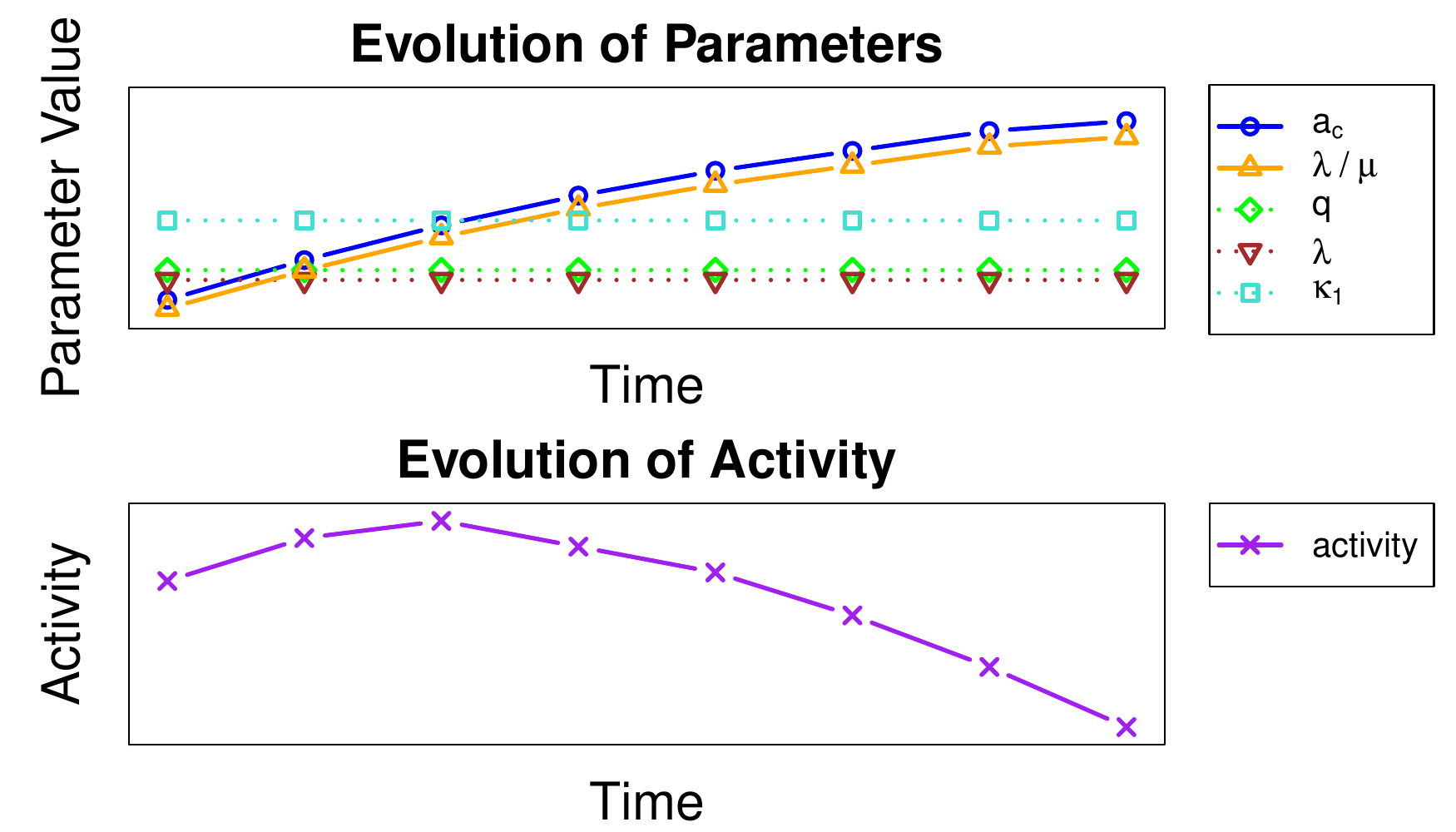}}
\subfigure[Coupled Evolution of Parameters]{\label{fig:scenario_all}
\includegraphics[width=0.48\linewidth]{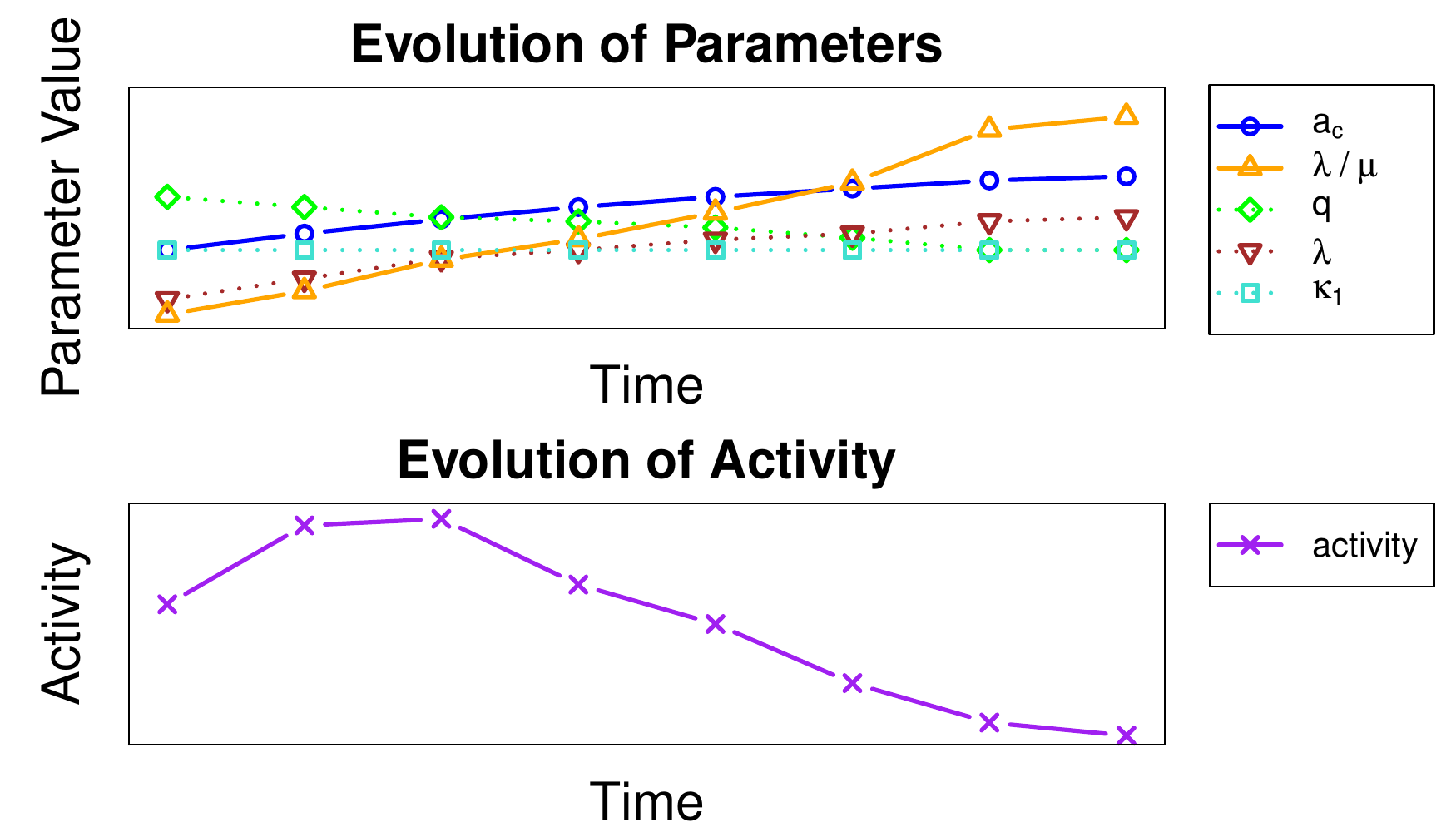}}
\caption{ \revone{\textbf{Parameter Evolution Scenarios.}
In a system with (at first) increasing overall levels of activity and fixed values for $q$ and \lam for all users, we expect $a_c$ to slowly increase (see \textbf{(a)}), as individual contributions are indistinguishable due to a flood of newly added content (activity). As a consequence, more posts and replies are required from all users to exercise the same amount of peer influence---represented by increasing values for $a_c$ over time. After a certain point in time, $a_c$ will reach a threshold and activity will start to decrease, if not intervened by administrators. In a more realistic scenario (see \textbf{(b)}), again with increasing levels of overall activity, users will---in addition to increasing values of $a_c$---start to lose interest in contributing to the system, represented by increasing values for \lam. As a consequence, activity will decrease at a faster pace.
}}
\label{fig:evolution_scenarios}
\end{figure*}
}

\revone{\smallskip\noindent \textbf{\mdesc}. With increasing values for \mue the ratio \ratio decreases, resulting either (i) in an overall increase in activity if the system is unstable ($\kappa_1 > \lambda/\mu$), (ii) in prolonged timespans of activity before converging towards inactivity if the system is stable ($\kappa_1 < \lambda/\mu$), (iii) or in an invalidation of the master stability equation if \ratio reaches a tipping point where $\kappa_1 > \lambda/\mu$. 

\smallskip The evolution of \mue directly corresponds to the evolution of the \qdesc and \acdesc.
	
		\revone{\smallskip\noindent{\qdesc.} The parameter $q$ defines the maximum amount of activity (peer influence) that can traverse along the edges of the collaboration network per unit time. If this parameter increases, $\mu = q/a_c$ will increase as well; resulting in an overall increase in activity. In contrast, reducing the value of $q$ results in overall decreasing levels of activity. 

			\smallskip\noindent\textit{Evolution scenarios of} \qdesc. In real-world systems, $q$ is best interpreted as a proxy for the efficiency of the user interface, describing how well information (or influence) is transported (e.g., highlighted or visualized) across users. For example, practitioners can influence the \qdesc by adding recommendations for users to collaborate with or by optimizing the presentation of newly added/edited content. 
			Note that with increasing numbers of users and levels of activity it becomes increasingly difficult for practitioners to keep $q$ at its current level, let alone positively influence the parameter due to the vast amount of content and/or activity present in the system.}
	
	\revone{\smallskip\noindent{\acdesc.} The parameter $a_c$ represents a soft threshold, which defines when users start to ``effectively notice'' the actions of their peers and are, as a consequence, ``notably'' influenced (see Figure~\ref{fig:gx}) by them. 
		The larger $a_c$, the more actions (i.e., posts or replies) are required by users to positively influence their peers to copy their actions and increase their activity levels (see Figure~\ref{fig:evolution_scenarios}). 
		
		\smallskip\noindent\textit{Evolution scenarios of} \acdesc. In practice, we would expect to see an increasing $a_c$ with an increasing number of active users and levels of activity. 
		For example, in a system with low activity and a small number of users, each action by a particular user will be noticed immediately by all others---meaning that the level of $a_c$ is low. However, with increasing numbers of users and an increase in activity, users have to increase their number of posts and replies to be noticed by their peers. 
		Hence, the more active users are present in a system, the harder it becomes for users to specifically notice each contribution of their peers individually. In a worst case, users are confronted with an activity overload that might even result in decreasing levels of (positive) peer influence. In particular, an initial increase in activity likely leads to an increase in $a_c$, which in turn decreases activity in the system. Thus, evolution of $a_c$ represents a \textit{negative feedback loop} in the system.
		In contrast to $q$, which serves as a proxy for the user-interface, $a_c$ represents an intrinsic parameter of the users of a system. Administrators of such networks and websites can influence $a_c$ by either influencing $q$ (e.g., by adjusting the user interface to better promote each individual action taken by the peers of a user) or by actively avoiding and counteracting the activity overflow by filtering and reducing the amount of new content that is displayed at once. 

		For example, the mechanisms of how Facebook displays posts in its ``News Feed'' can be seen as a measure to filter and limit newly added content; actively avoiding information or activity overloads while maximizing the (peer) influence of each individual contribution.}

}

\revone{\smallskip\noindent\textbf{Summary of evolution scenarios.} If activity increases over time and no adaptations to the system are implemented, activity will inevitably decrease, due to a larger \acdesc (see Figure~\ref{fig:evolution_scenarios}). To counteract this development, website administrator could either try to manipulate \ldesc---an intrinsic property that varies per user---or optimize the user interface, and thus manipulate \qdesc.}

\revone{\section{Empirical Illustration}
\label{sec:illustrative examples}}

We are now interested in modeling and simulating activity dynamics for empirical datasets. In particular, we investigate activity dynamics for an array of different websites, consisting of instances of the StackExchange\footnote{\url{http://www.stackexchange.org/sites}} network as well as multiple Semantic MediaWikis\footnote{\url{http://www.semantic-mediawiki.org}}.

First, we characterize the investigated datasets and outline our methods for the empirical estimation of the required parameters (see Table~\ref{tab:model params}). We then fit our model to the collaboration networks and present the results of the activity dynamics simulation.

\begin{figure*}[t!]
\centering
\subfigure[History StackExchange]{\label{fig:history_stackexchange_graph}\includegraphics[width=0.24\linewidth]{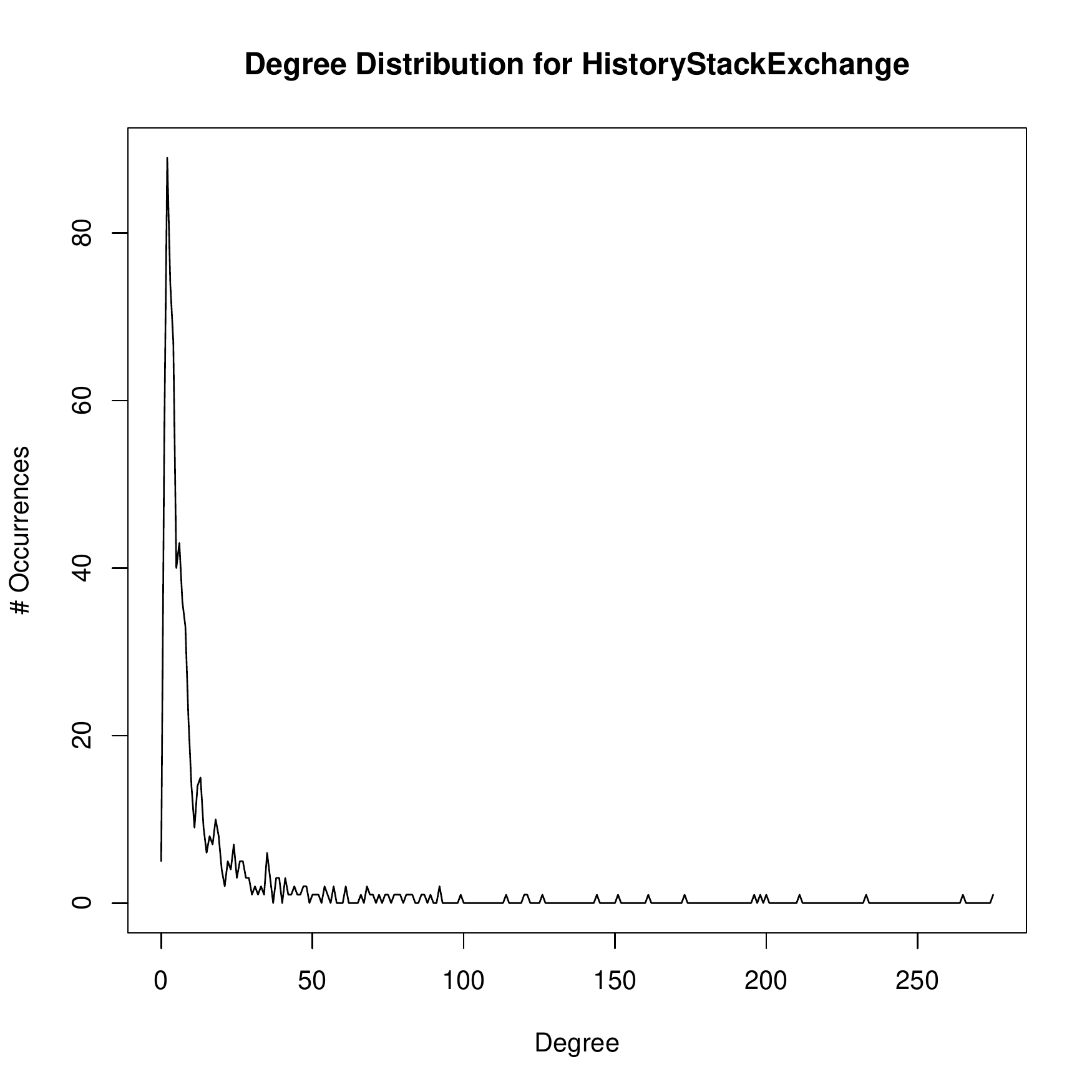}}
\subfigure[Bitcoin StackExchange]{\label{fig:stackoverflow_graph}\includegraphics[width=0.24\linewidth]{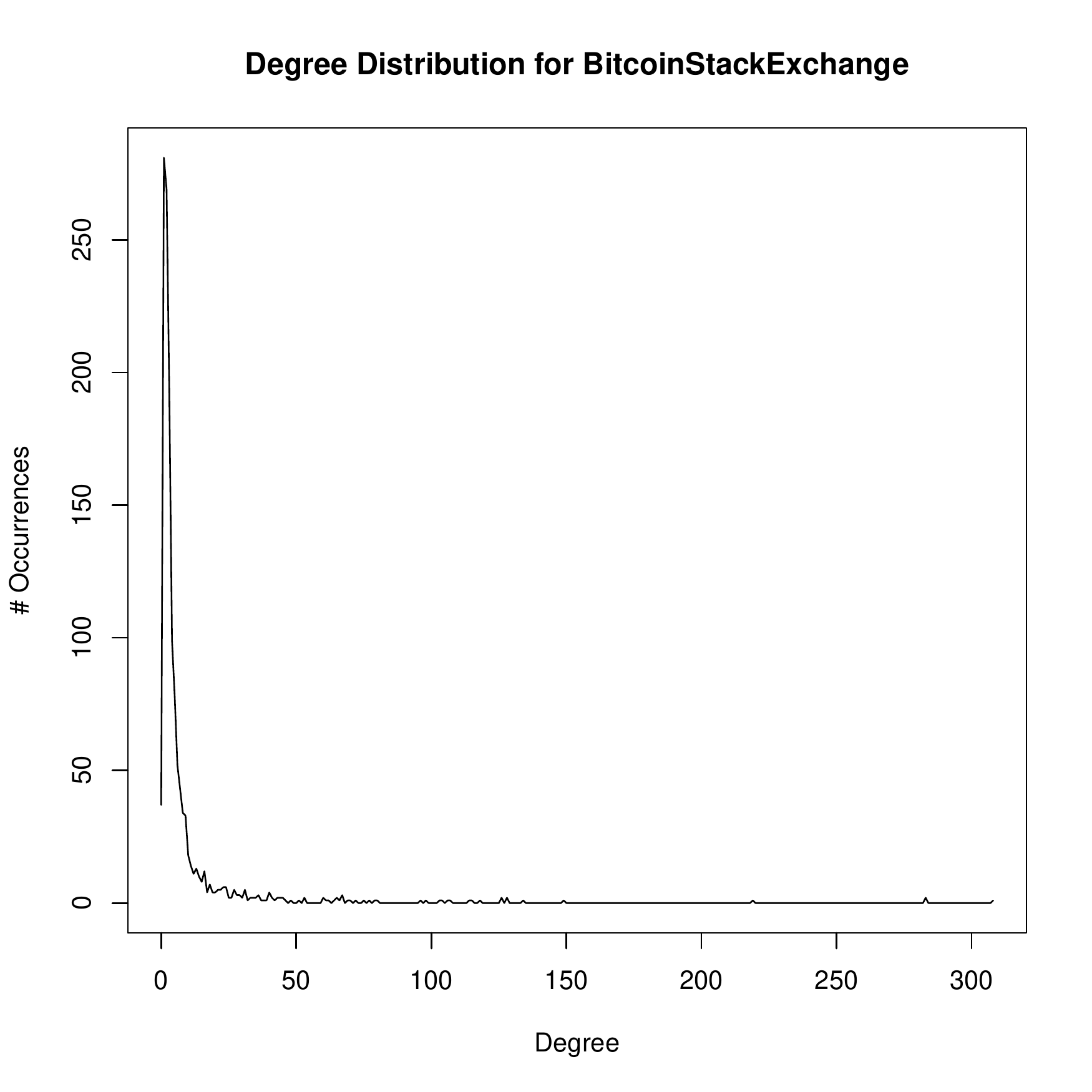}}
\subfigure[English Language \& Use  StackExchange]{\label{fig:english_stackexchange_graph}\includegraphics[width=0.24\linewidth]{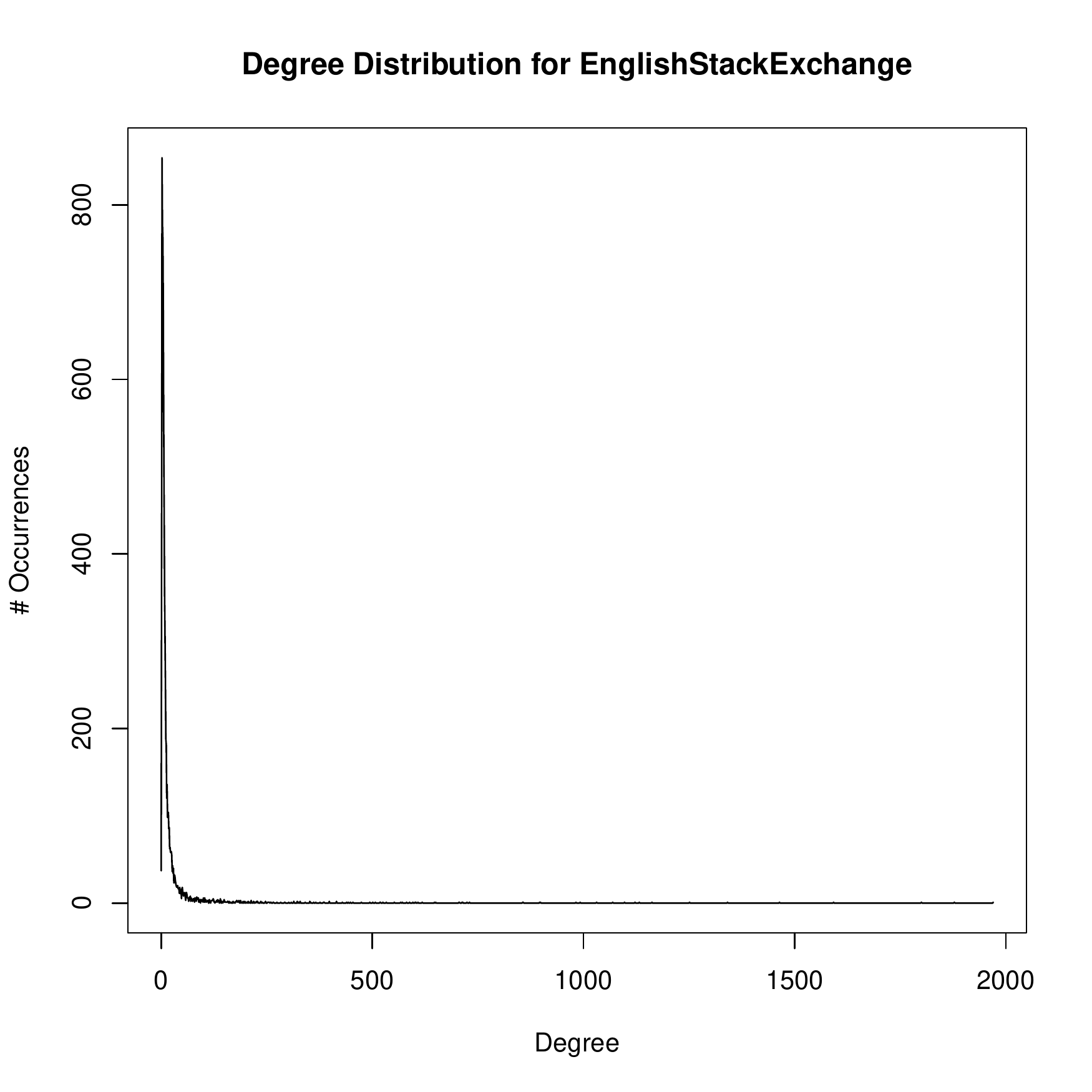}}
\subfigure[Mathematics StackExchange ]{\label{fig:math_stackexchange_graph}\includegraphics[width=0.24\linewidth]{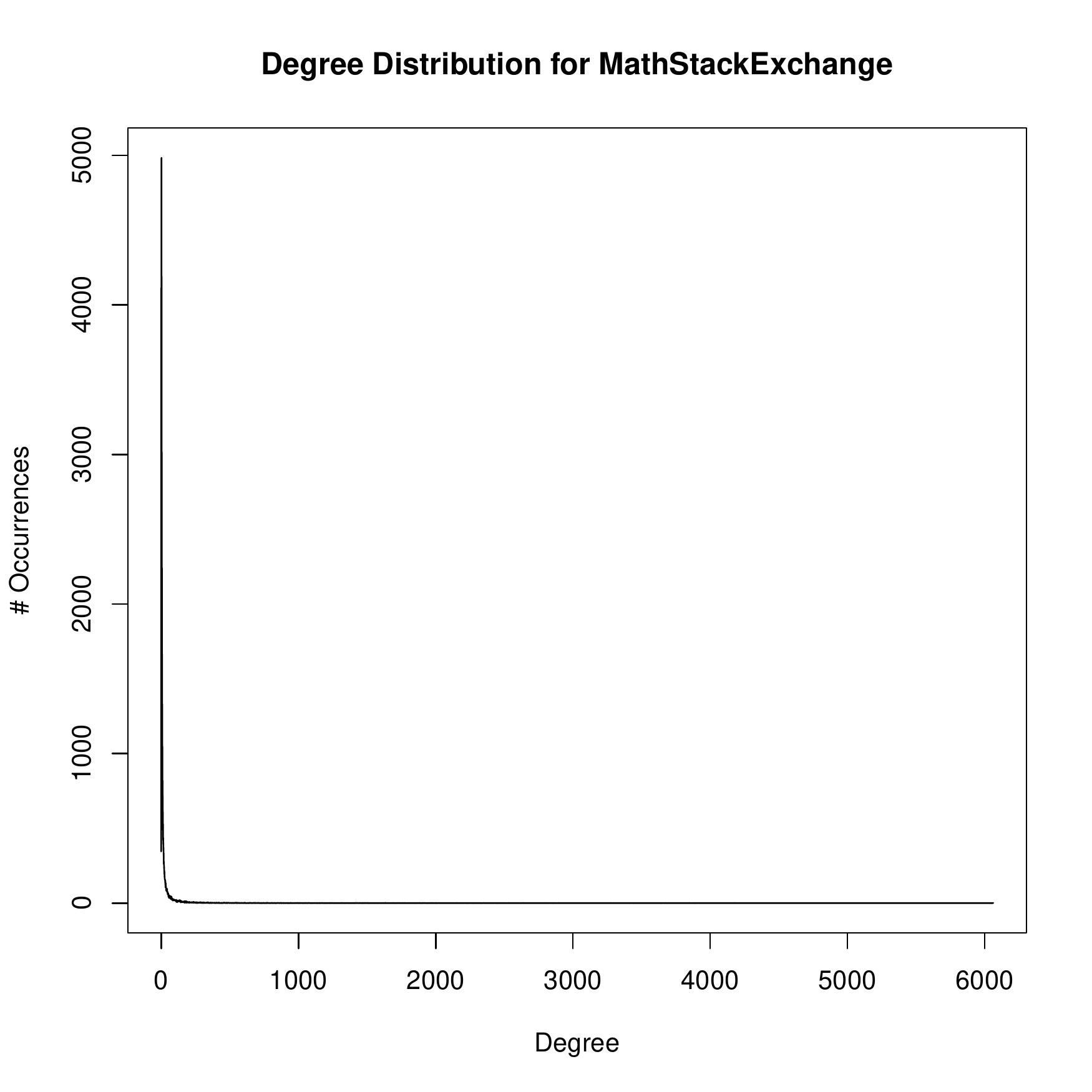}}
\subfigure[Beachapedia Wiki]{\label{fig:beachapedia_graph}\includegraphics[width=0.24\linewidth]{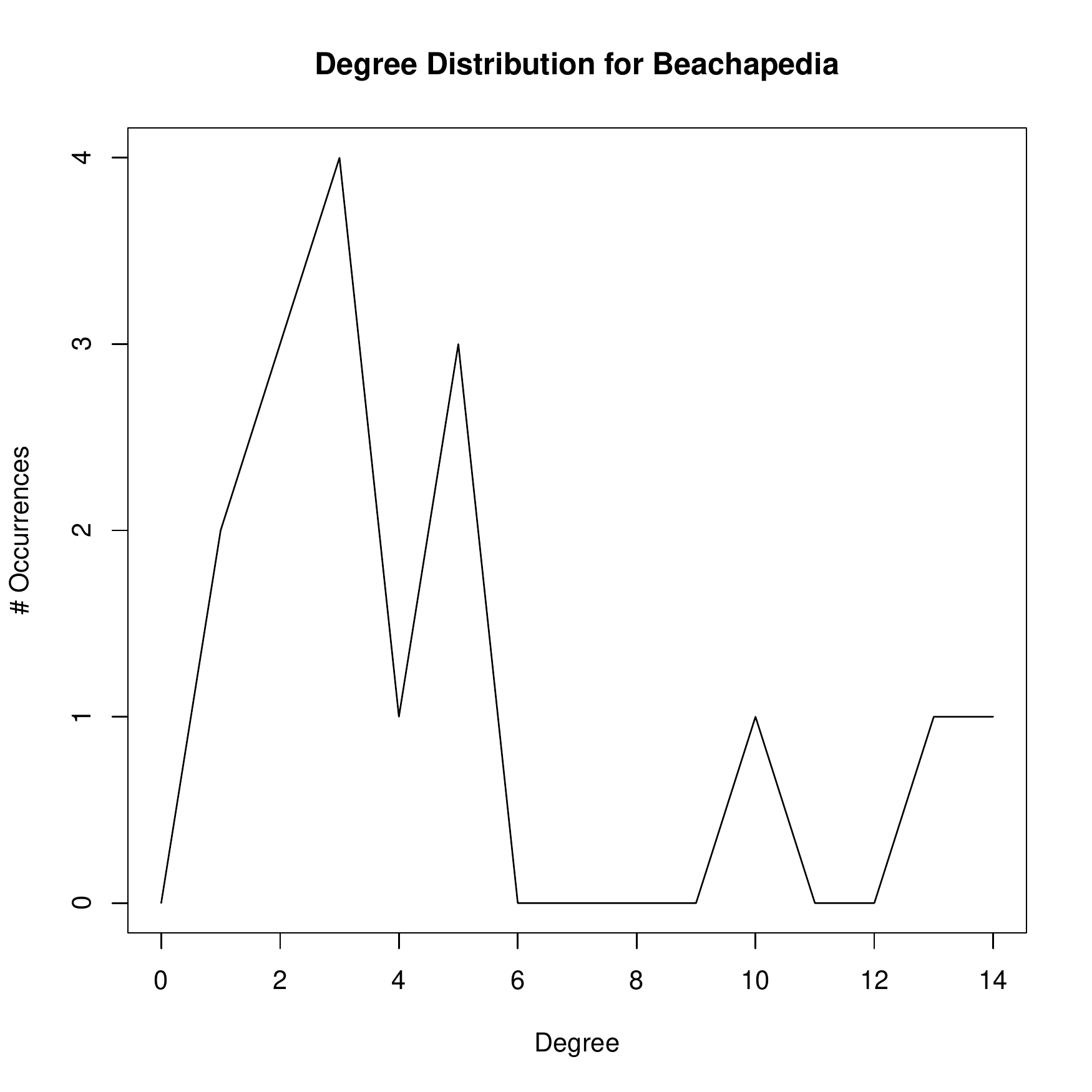}}
\subfigure[Nobbz Wiki]{\label{fig:nobbz_graph}\includegraphics[width=0.24\linewidth]{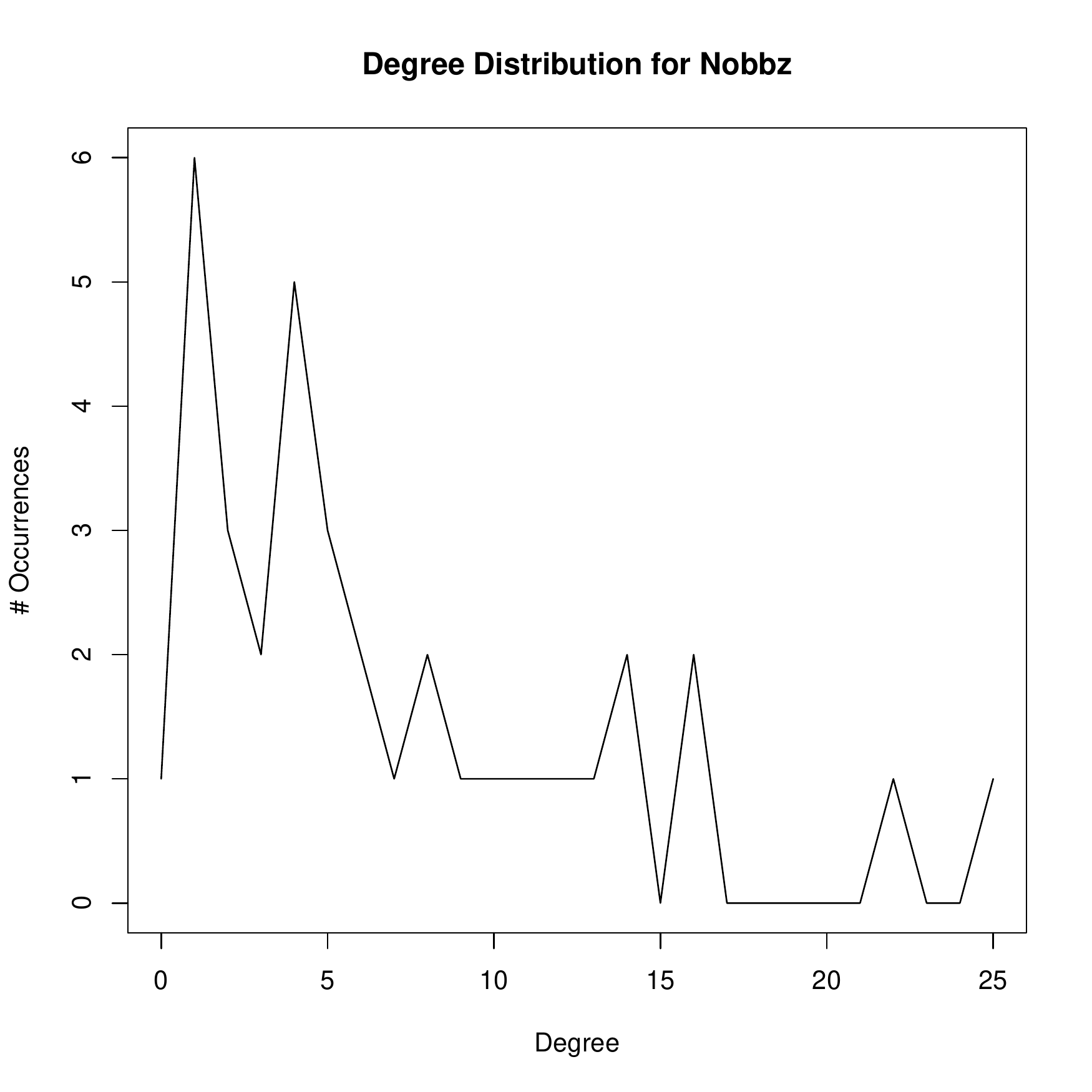}}
\subfigure[NeuroLex Wiki]{\label{fig:neurolex_graph}\includegraphics[width=0.24\linewidth]{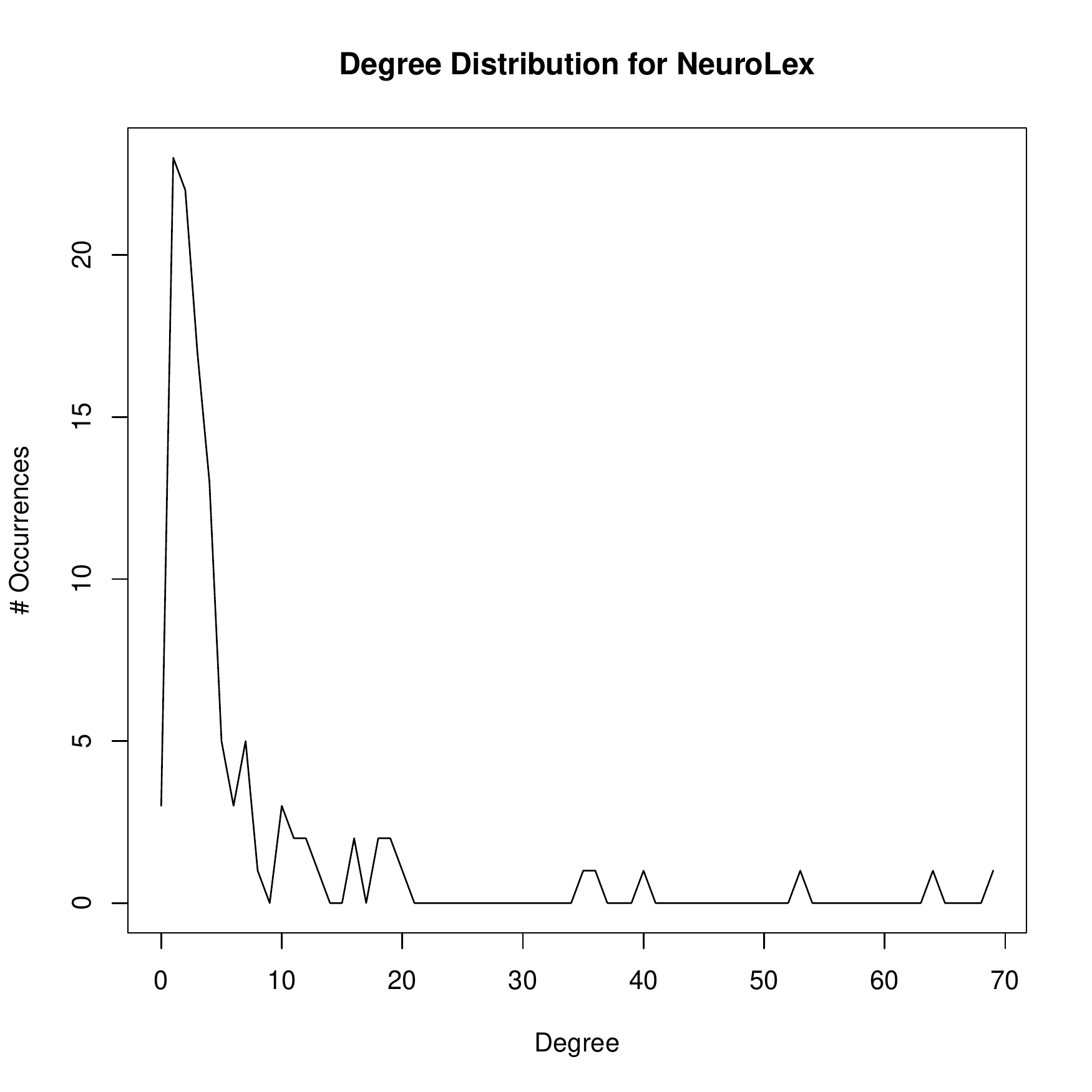}}
\subfigure[15Mpedia Wiki]{\label{fig:15mpedia_graph}\includegraphics[width=0.24\linewidth]{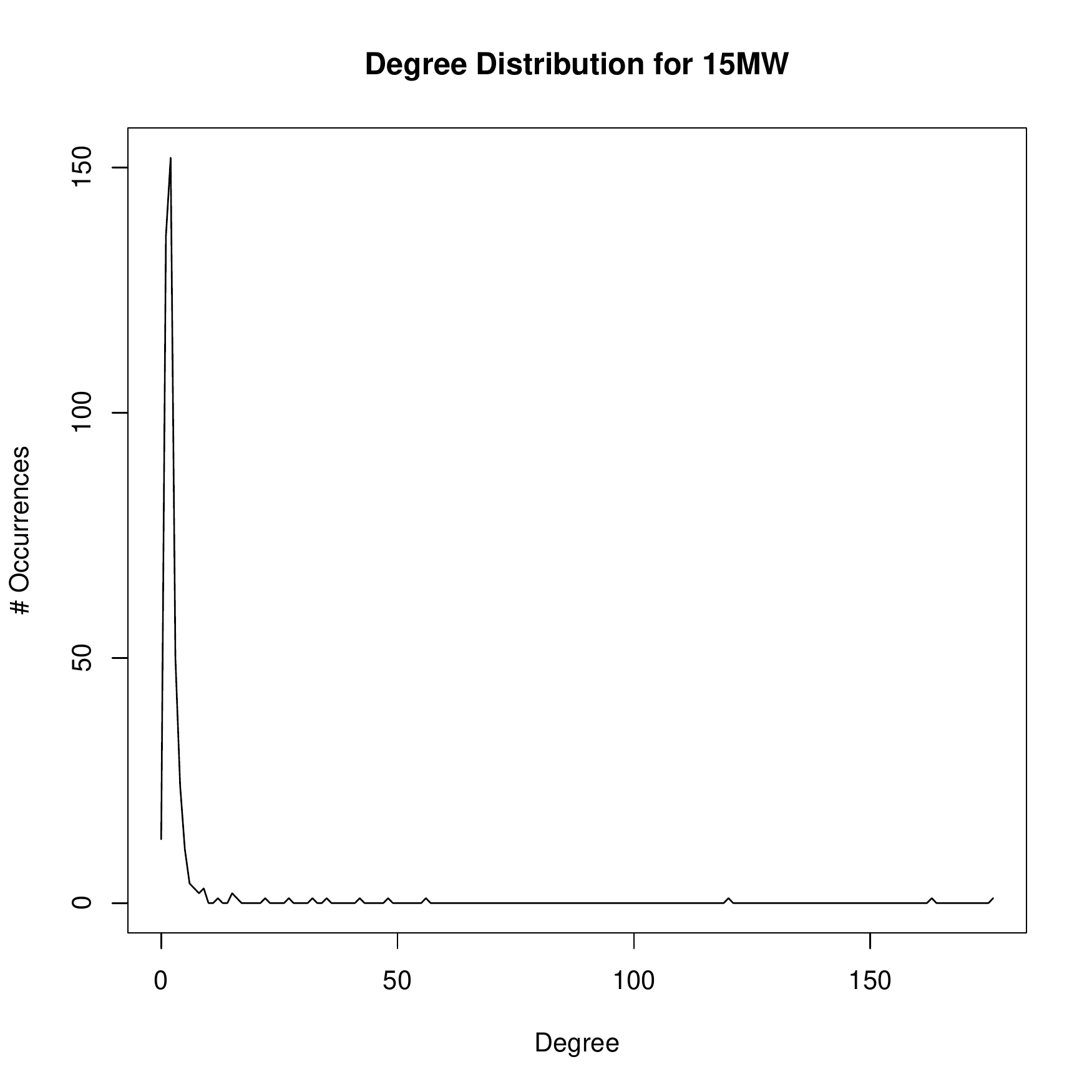}}
\caption{ \revone{\textbf{Degree Distribution of Empirical Collaboration Networks.} Visualization of the degree distribution of all investigated collaboration networks. The \textbf{top row (a to d)} depicts the different StackExchange collaboration networks, while the \textbf{bottom row (e to h)} shows the collaboration network visualizations for the different Semantic MediaWiki instances. The majority of users, across all collaboration networks, exhibits between $0$ and $10$ collaboration edges.}}
\label{fig:emp_graphs}
\end{figure*}

\subsection{Datasets}
\label{subsec:empirical datastes}

We selected a total of four differently sized instances from the StackExchange network as well as four different Semantic MediaWiki instances to model activity dynamics. In particular, we concentrate our efforts on the History StackExchange\footnote{\url{http://history.stackexchange.com}} (HSE), which is the smallest of the StackExchange datasets and allows users to discuss topics and questions related to history and historical events.
\revone{The Bitcoin StackExchange\footnote{\url{http://bitcoin.stackexchange.com}} (BSE) as well as the 
The English Language \& Usage StackExchange\footnote{\url{http://english.stackexchange.com}} (ESE) 
represent two medium-sized websites and are platforms for asking and discussing questions related to everything related to mining, buying and selling of bitcoins and the English language respectively.} On the Mathematics StackExchange\footnote{\url{http://mathematics.stackexchange.com}} (MATHSE) website, which also represents our largest dataset, users can ask and discuss mathematics related questions and topics.

We further investigate activity dynamics for the Beachapedia Wiki\footnote{\url{http://www.beachapedia.org}} (BP), representing the smallest dataset in our activity dynamics analysis, striving to create a structured knowledge base for a variety of topics on beaches in the United States.
The medium-sized german Nobbz Wiki\footnote{\url{http://nobbz.de/wiki}} (NZ) provides a structured knowledge base and discussion platform for the online game ``Die Verdammten''\footnote{\url{http://www.dieverdammten.de/}}.
\revone{The second largest dataset, the NeuroLex Wiki\footnote{\url{http://neurolex.org/}} (NLX), represents a large and semantically enriched lexicon on terms and topics related to neuroscience.} 
Our largest dataset is the 15Mpedia Wiki\footnote{\url{http://wiki.15m.cc/wiki/Portada}} (15MW)---a Spanish Semantic MediaWiki instance that discusses a wide variety of topics related to Spain and its different areas and regions. 

In general, the investigated datasets are very diverse in their characteristics, for example, the number of active users ranges from $35,476$ in MATHSE to a total of $16$ in BP. For the analyses conducted in this paper we focus on the last $52$ weeks of each dataset. For more detailed information see Table~\ref{tab:emp_dataset_details}. The different degree distributions for all collaboration networks are highly heterogeneous (cf. Figure~\ref{fig:emp_graphs}). 
For all investigated datasets, the majority of users exhibit between $0$ and $10$ collaboration edges. However, in all datasets there are a few users with a large number of collaboration edges.

\begin{figure}[t!]
\centering
{\includegraphics[width=0.7\linewidth]{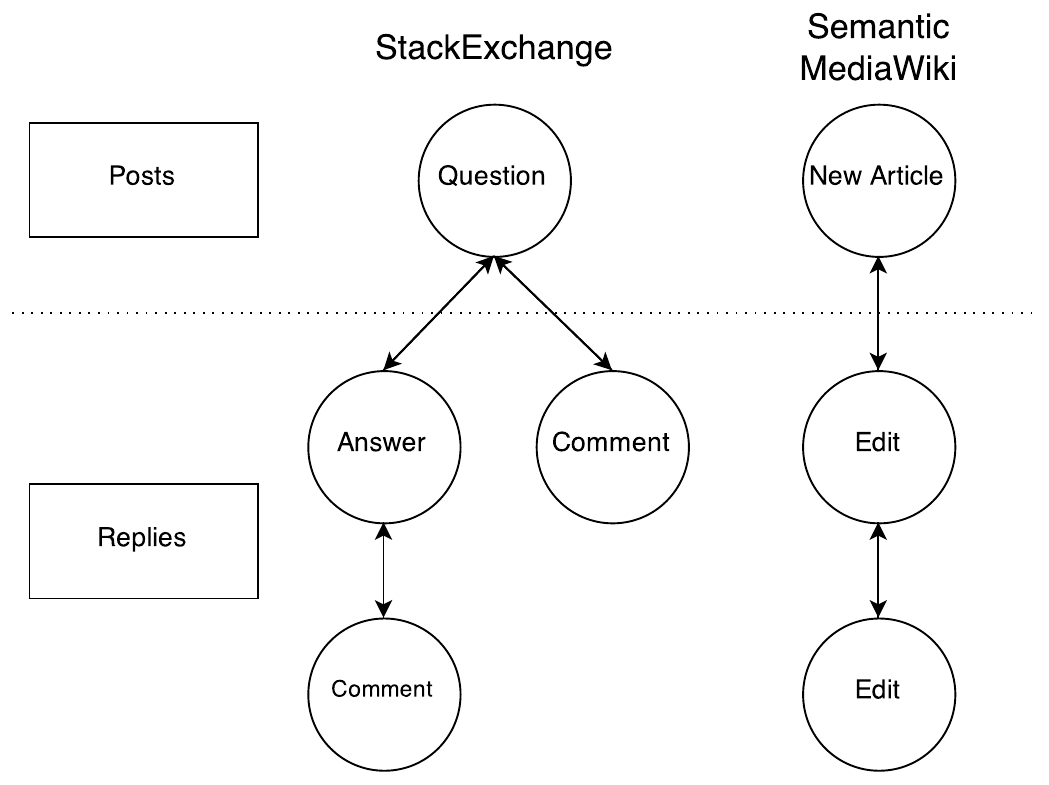}}
\caption{\textbf{Collaboration Network Construction.} This plot depicts the different elements of the StackExchange and Semantic MediaWiki datasets that have been classified as posts and replies (cf. Table~\ref{tab:emp_dataset_details}) as well as the edges that have been drawn between certain entities and change-actions and represent collaboration in our collaboration networks.}
\label{fig:empr_ds_desc}
\end{figure}

From each of these datasets we extracted a collaboration network for the tasks of fitting the model and simulating activity dynamics. Hence, we first parsed the change-logs of all datasets. Each user, who has contributed at least one question, answer or comment for the StackExchange datasets, or created or edited an article for the Semantic MediaWikis is represented as a node in the corresponding collaboration network. Edges between users represent collaboration and are undirected. For the StackExchange datasets, we defined collaboration as either posting an answer to a question or posting a comment on the initial question or an answer. For the Semantic MediaWiki instances, we have created an edge between users who (chronologically and) successively changed the same article (cf. Figure~\ref{fig:empr_ds_desc}). Edges with the same source and target user have been removed in all datasets. 

\revone{Further, users with zero collaboration edges are initialized analogously to all other users and are not specifically filtered from our datasets. However, due to missing positive peer influence, activity will inevitably---as long as $\lambda/\mu>0$---converge towards zero for these users.}

\revone{Note that the presented approach for creating collaboration networks represents just one of many different possibilities to create such networks and is analogous to (undirected) co-authorship networks as presented in \citet{newman2001scientific, barabasi2002evolution}. Given that the created collaboration networks are based on interactions between users, we assume similar characteristics to social networks, particularly with regards to potential peer influence \cite{aral2009distinguishing}.}

\begin{table*}[!b]
\center
\tiny
\caption{\revone{\textbf{Dataset statistics.} Note that all datasets differ in the number of users, collaboration edges and activity. Users refers to the number of unique users that have contributed more than one post or reply to the corresponding datasets within our observation periods. Posts represent newly created questions in the case of the StackExchange network and newly created articles in the case of the Semantic MediaWiki datasets. Replies are either comments or answers for all StackExchange datasets and edits of existing articles for Semantic MediaWikis. $\kappa_1$ denotes the largest eigenvalue of the corresponding collaboration network. For our experiments we limited our observation periods to the last $52+3$ weeks of each dataset.}}
\begin{tabular}{ l | c  c  c  c  c  c  c  c }\toprule
Dataset & HSE & BSE & ESE & MATHSE & BP & NZ & NLX & 15MW  \\\midrule
Users & $682$ & $1,299$ & $7,893$ & $35,476$ 
& $16$ & $36$ & $112$ & $394$ \\
Edges & $5,179$ & $5,528$ & $83,457$ & $477,133$ 
& $38$ & $125$ & $383$ & $772$ \\
$\kappa_1$ & $54.33$ & $43.88$ & $162.04$ & $303.58$ 
& $6.71$ & $11.46$ & $18.4$ & $19.97$\\
\midrule
Posts \& Replies & $12,496$ & $12,295$ & $151,028$ & $986,996$ 
& $2,718$ & $603$ & $33,792$ & $102,521$\\
Weeks & $52+3$ & $52+3$ & $52+3$ & $52+3$ & $52+3$ & $52+3$ & $52+3$ & $52+3$\\
\bottomrule
\end{tabular}
\label{tab:emp_dataset_details}
\end{table*}

\revone{\subsection{Parameter Estimation with Least-Squares}
\label{subsec:model fitting}}

\revone{ To estimate \ratio for (preprocessed) empirical datasets we resort to an output-error estimation method. First, we formulate the estimation of the model parameter as an optimization problem. As objective function we use a well-known least-squares cost function. Second, we solve the optimization problem numerically, using the method of gradient descent in combination with Newton's method to speed up the calculations. Finally (as a proof of concept), we evaluate the accuracy of the ratio estimate by calculating prediction errors on unseen data. Next, we describe these estimation steps in more details. 

\smallskip\noindent\textbf{Preprocessing.} 
First, we aggregate all activities per user per day and apply a rolling mean of $7$ days to smoothen and reduce strong fluctuations in activity, which are likely caused by external influences.
Second, we further aggregate the smoothed activities per user and per (calendar) week.
For an additional noise reduction in our datasets we remove all users that have contributed less than one post or reply in the smoothened dataset during our observation period, as well as the first and last week of our datasets, if they contain less than $7$ days of activity data. Finally, since we only want to illustrate the practical application of our model on the empirical data we extract the last $52 + 3$ to weeks from all our datasets. Note that the $3$ additional weeks are required to calculate a ratio for the simulation of activity for the first week.
}

\revone{\smallskip\noindent\textbf{Formulating estimation as an optimization problem.}
Depending on a particular application of the model we may need to introduce a suitable objective function. For example, we may be interested in applying our model to analyze and simulate the aggregated levels of activity in a system. In other words, we are interested in the overall activity level in a system, rather than in the particular activity distribution over the users (see below for another example involving user activity levels).
Hence, we formulate the objective function (see Equation~\ref{eq:ls:objective_sum_general}) as a least squares cost function, which calculates the error of the sum of activity over multiple data points over a certain period of time $T$:

\begin{equation}
	J(\frac{\lambda}{\mu})=\frac{1}{T}\sum_{k=0}^{T-1}{\left[\sum_{i}^{n}{x_i(k+1)} - \sum_{i}^{n}{\hat x_i(k+1)}\right]^2},
	\label{eq:ls:objective_sum_general}
\end{equation}

where $x_i(k)$ is the empirically observed activity of user $i$ at time $k$, $\hat x_i(k)$ is the estimated activity for user $i$ at time $k$, and $n$ is the total number of users as before. 

To calculate the estimates $\hat x_i(k)$ we numerically integrate the differential equations from our model by applying Euler's method for solving differential equations computationally. 
Thus, we approximate the time evolution of $\hat x_i$ between all time steps $k$ and $k+1$ (for each of these steps we set the total time to $\tau$) by iterating:
\begin{equation}
 \hat x_{i, t+1}(k) = \hat x_{i, t}(k) + \Delta \tau \left[-\hat{\frac{\lambda}{\mu}} \hat{x}_{i,t}(k) + \sum_j A_{ij} \frac{\hat{x}_{j,t}(k)}{\sqrt{1 + \hat{x}_{j,t}(k)^2}}\right],
\end{equation}
where we set $\hat{x}_{i, t=0}(k) = \hat{x}_i(k)$, $\forall i, k$ and use the current estimate for \ratio to perform calculations.
The final equation for $\hat x_i(k+1)$ becomes:
\begin{equation}
 \hat x_i(k+1) = \hat x_i(k) + \Delta \tau \sum_{t=0}^{t=\tau} \left[-\frac{\lambda}{\mu} \hat{x}_{i,t}(k) + \sum_j A_{ij} \frac{\hat{x}_{j,t}(k)}{\sqrt{1 + \hat{x}_{j,t}(k)^2}}\right].
\end{equation}
}

The local approximation error for the Euler's method is of the order $O(\Delta \tau^2)$ and the global of the order $O(\Delta \tau)$. 
\revone{To perform integration between steps $k$ and $k+1$ we need to iterate for $\tau/\Delta\tau$ steps, where $\Delta\tau$ needs to be chosen with care. In general, if we set $\Delta\tau$ too high---meaning that the calculations are less computationally intensive, as we have to run a smaller number of iterations---the accuracy of our simulation (including the estimation of the ratio) will decline, as the potential error per iteration due to our approximations becomes higher. This error can become so large that it could potentially lead to numerical instability, meaning that the overall activity in a system can become negative, which might result in activity to diverge towards $\pm\infty$. With certain combinations of the network structure, $\Delta\tau$ and the calculated ratios, activity can become negative without diverging, oscillating around the fixed point of zero activity until convergence.
In contrast, if we set $\Delta\tau$ too low we end up with a very precise simulation, although the time necessary to compute the simulation will be much higher, as a much larger number of iterations will have to be executed.

\smallskip\noindent\textbf{Numerical solution of the optimization problem.} We solve the optimization problem numerically using the method of gradient descent. The first derivative of the objective function (Equation~\ref{eq:ls:objective_sum_general}) defines the update rule or gradient, which directs if and to what extent we have to increase or decrease \ratio to minimize the error of the sum of activities over all data points during $T$.

Once we calculate the first derivative with the current values of estimated activities we update the ratio by multiplying the derivative with the \textit{learning rate} $\eta$.
Thus, the complete procedure is as follows. First, we initialize our estimation by using \kone for the first simulation. Second, we estimate the activities and calculate the gradient with these estimates. Third, 
we calculate the error between our simulated and empirical values, and adapt the ratio according to the corresponding update function and step size $\eta$. Fourth, we repeat this process until the calculated update for the ratio is smaller than a given convergence criterion (e.g., $10^{-12}$) or if we reach a total of $20,000$ iterations without reaching convergence. Additionally, we have also implemented Newton's method, which in our cases substantially reduces the computation time. In all our experiments we set $T$ to four weeks, meaning that we optimize the objective function by calculating the optimal ratio over a span of four data points (weeks).

}

\begin{figure*}[!t]
\centering
\subfigure{\label{fig:synthetic_example_simulation:inc}\includegraphics[width=0.50\linewidth]{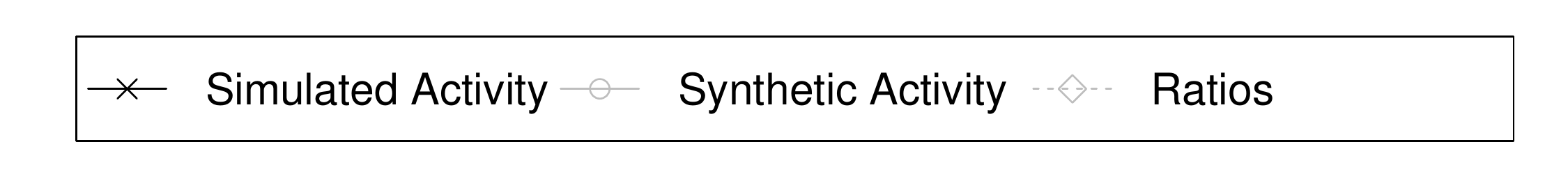}}
\setcounter{subfigure}{0} 

\subfigure[Increasing Activity]{\label{fig:synthetic_example_simulation:inc}\includegraphics[width=0.32\linewidth]{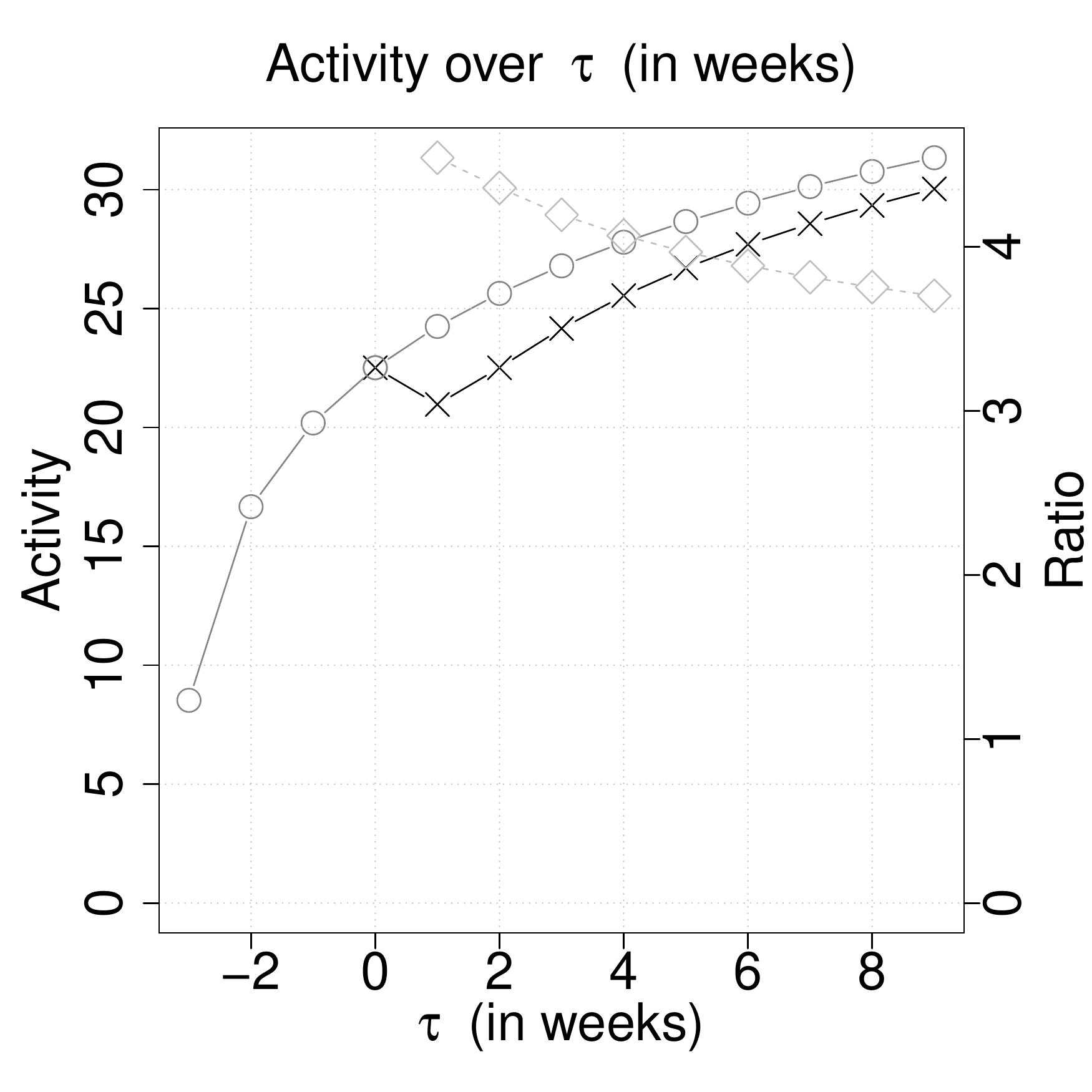}}
\subfigure[Decreasing Activity]{\label{fig:synthetic_example_simulation:dec}
\includegraphics[width=0.32\linewidth]{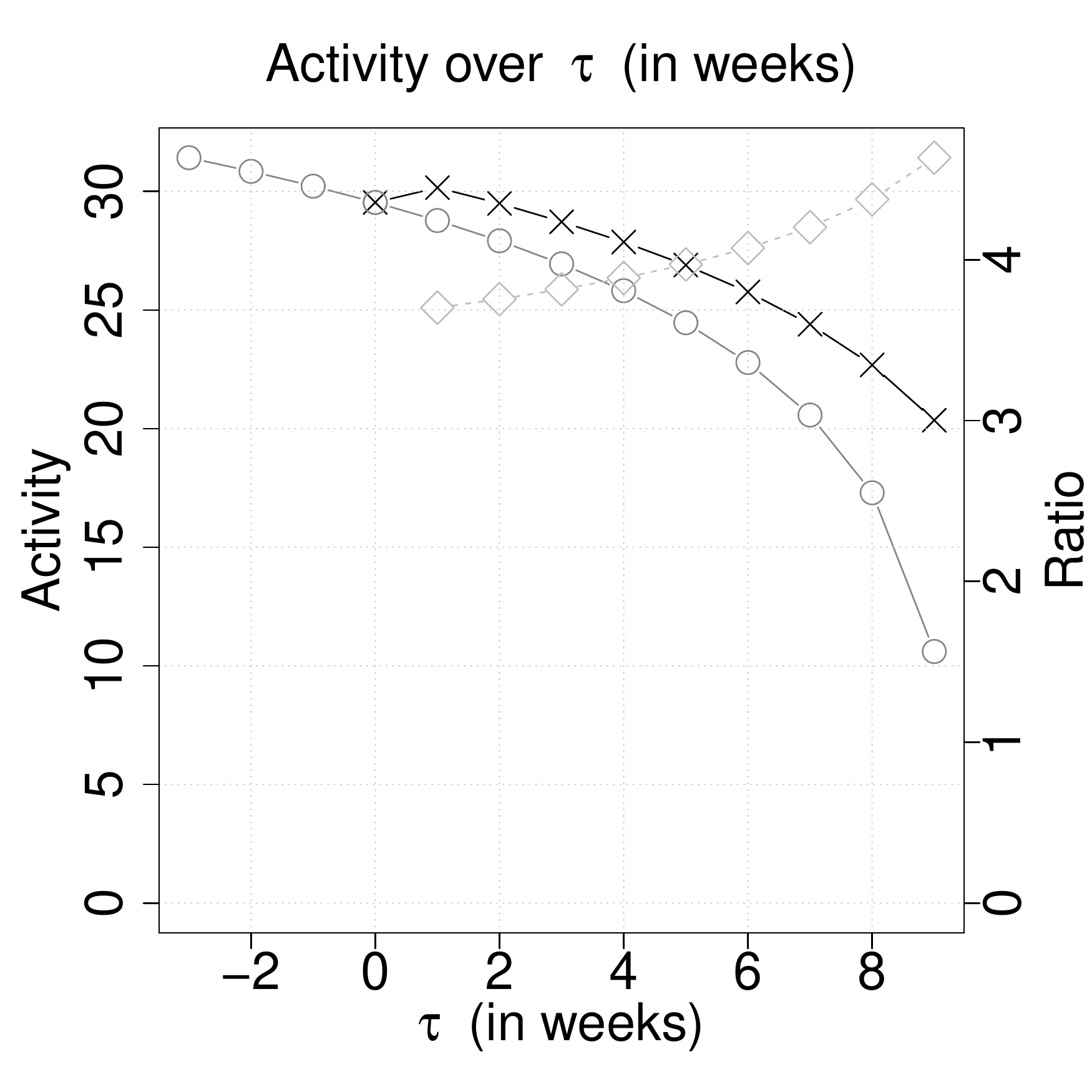}}
\subfigure[Variable Activity]{\label{fig:synthetic_example_simulation:var}\includegraphics[width=0.32\linewidth]{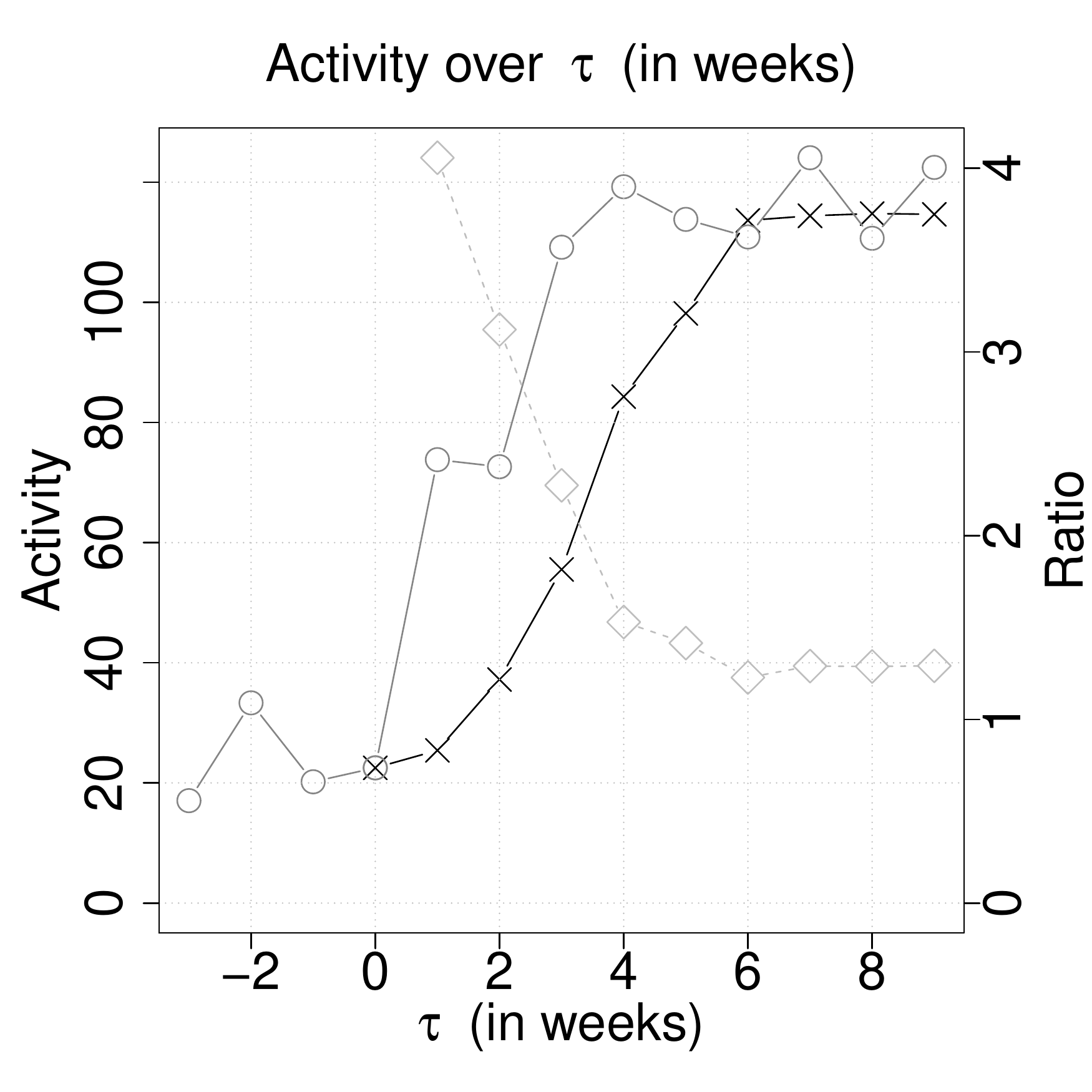}}
\caption{ \revone{\textbf{Illustrations with Synthetic Data.} The plots depict the results of the activity dynamics simulations for Zachary's Karate Club network with synthetic activity values (left $y$-axes) and the corresponding ratios (right $y$-axes). The black solid lines with $x$ markers represent the simulated activity over $t$ (in weeks; $x$-axes). The solid gray lines with circles represent synthetic activities; the gray dotted lines with diamonds represent the ratios corresponding to the simulated activities. With increasing and decreasing activities, the ratios become smaller (see \textbf{(a)}) and larger (see \textbf{(b)}). When setting activity randomly (see \textbf{(c)}) the ratio adjusts analogously.
}}
\label{fig:synthetic_example_simulation}
\end{figure*}

\revone{
\smallskip\noindent\textbf{Evaluation of the parameter estimates.} We evaluate the accuracy of the estimated parameters by cross-validation (leave-one-out method). In particular, we use the estimated ratios over $4$ weeks to simulate activity for the succeeding week. For example, we calculate the optimal \ratio (according to our objective function) for weeks $1$ -- $4$ and predict activity for week $5$. Next, we use the empirical data of weeks $2$ -- $5$ to calculate the ratio to predict activity for week $6$. Hence, we calculate a total of $52$ ratios to simulate activity for a total of $52$ weeks. 

As depicted in Figure~\ref{fig:synthetic_example_simulation}, we have created three synthetic scenarios to test and illustrate the mechanisms of the \textit{Activity Dynamics Model}. 
First, we estimate \ratio (right $y$-axes; gray dotted lines with diamonds) for the three scenarios with synthetically created increasing, decreasing and variable or random activities (left $y$-axes; gray solid lines with circles) over $10+3$ weeks ($x$-axes). In all three scenarios we use Zachary's Karate Club as the underlying collaboration network. Due to our parameter estimation process the simulated levels of activity (left $y$-axes; black solid lines with x markers) exhibit a small lag when activity steadily moves into one direction (i.e., increases or decreases). On the other hand, small fluctuations (see weeks $6$ -- $9$ in Figure~\ref{fig:synthetic_example_simulation:var}) are mitigated. The ratios (right $y$-axes), which correspond to the simulated levels of activity in the same week, are depicted as well.

\smallskip\noindent\textbf{Discussion on parameter estimation method.}
To validate the correctness of our implementation of the method of least squares, we have simulated activity for datasets with a preset ratio (and random weights for initialization) for $3$ weeks. We then used the random activity initialization values, as well as the activity values for each of the $3$ weeks as input for the calculation of the ratio with the method of least squares. Using this approach, we were able to estimate previously set ratios with negligibly small errors. When adding noise to the simulated activity values, the obtained ratios were less accurate accordingly.

Note that the estimation and validation method that we apply is only one of many possible methods. In this paper, we want to illustrate the general applicability of our method as well as its potential to gather new insights into the intricate dynamics of activity in online collaboration networks. We measure the accuracy of the prediction only as a general proof of concept of our model and leave further investigations of the predictive power of our method open for the future work. Following up on this notion, we now shortly discuss some alternative approaches for formulating the objective function and their implications.
}

\revone{\smallskip\noindent\textbf{Alternative objective functions.}}
\revone{To demonstrate the versatility of our model---if we are interested in answering questions about the distribution of the activities over users---we may change the formulation of the objective function to calculate ratios that minimize the error of activity per user and per data point (see Equation~\ref{eq:ls:objective_user_general}). Note that when optimizing towards aggregated levels of activity, we obtain ratios that characterize the systems. In contrast, with the adapted objective function, we are interested in learning more about the users of such systems. The alternative objective function may be defined as follows:

\revone{\begin{equation}
	J(\frac{\lambda}{\mu})=\frac{1}{T}\sum_{k=0}^{T-1}{\left[\bm{x}(k+1)- \hat{\bm{x}}(k+1)\right]^2},
	\label{eq:ls:objective_user_general}
\end{equation}}
where $\bm{x}$ and $\hat{\bm{x}}$ are now $n$-dimensional vectors storing the activities of all $n$ users.
}
\revone{Thus, this objective function represents the sum of squared errors calculated for each of the $n$ users of the corresponding systems over a total of $T$ data points. 

We have estimated \ratio and simulated activity for HSE using this objective function. In contrast to the aggregated levels of activity, we obtain a more accurate distribution of activities across all users, as was intended. However, each of the $4$ data points in $T$ now corresponds to a vector of $n$ users, as opposed to a single value (the aggregated activities), resulting in either much higher computation times, a larger error for the prediction tasks or both.

Additionally, to tackle the prediction problem and to avoid overfitting we may introduce a \textit{regularization term} to the objective function. For example, we might be interested in keeping the ratio or the difference between the ratio and \kone small. In the latter case we would add a term such as $\gamma(\kappa_1 - \lambda/\mu)^2$ to our objective function, where $\gamma$ represents the strength of regularization.

We leave a detailed analysis and comparison of different objective functions open for future work. The ratios calculated to minimize the error for aggregated activity levels exhibit higher accuracy in our simulations (in terms of overall activity per month). The trade-off for a more accurate distribution of activities over users with the changed objective function are worse results for the simulation of activity, as not only the aggregated activity levels are considered, but the vector of activities of all user in our datasets over multiple points in time. However, these ratios provide a better overall correlation between simulated and empirical activities per contributor of our system.
}

\revone{\subsection{Illustration on Empirical Datasets}}
\label{subsec:empirical illustration}

After calculating \ratio and setting $\Delta\tau$ we simulate activity in our collaboration networks. Due to our chosen approximations, the main goal of the presented illustration is not to predict activity in collaboration networks. Rather, we are interested in demonstrating that our assumptions regarding the \ldesc and the \mdesc hold and allow us to simulate trends in activity dynamics for given and real values. Further, by modeling and simulating activity dynamics for empirical datasets we not only deepen our understanding of the model but we also---depending on the values of the parameters---potentially obtain new insights into the systems under investigation.

\begin{figure*}[!t]
\centering
\subfigure{\label{fig:leg}\includegraphics[width=0.35\linewidth]{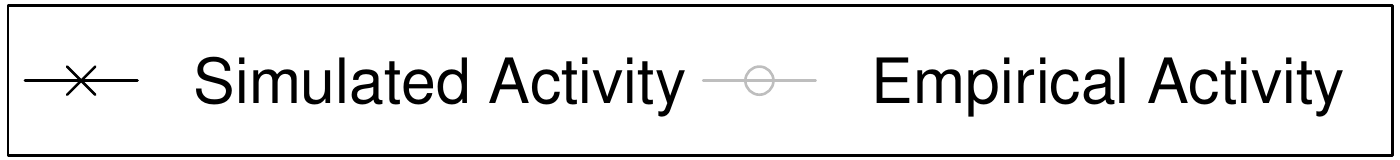}}
\setcounter{subfigure}{0} 

\subfigure[History StackExchange Activity]{\label{fig:hse_act}\includegraphics[width=0.24\linewidth]{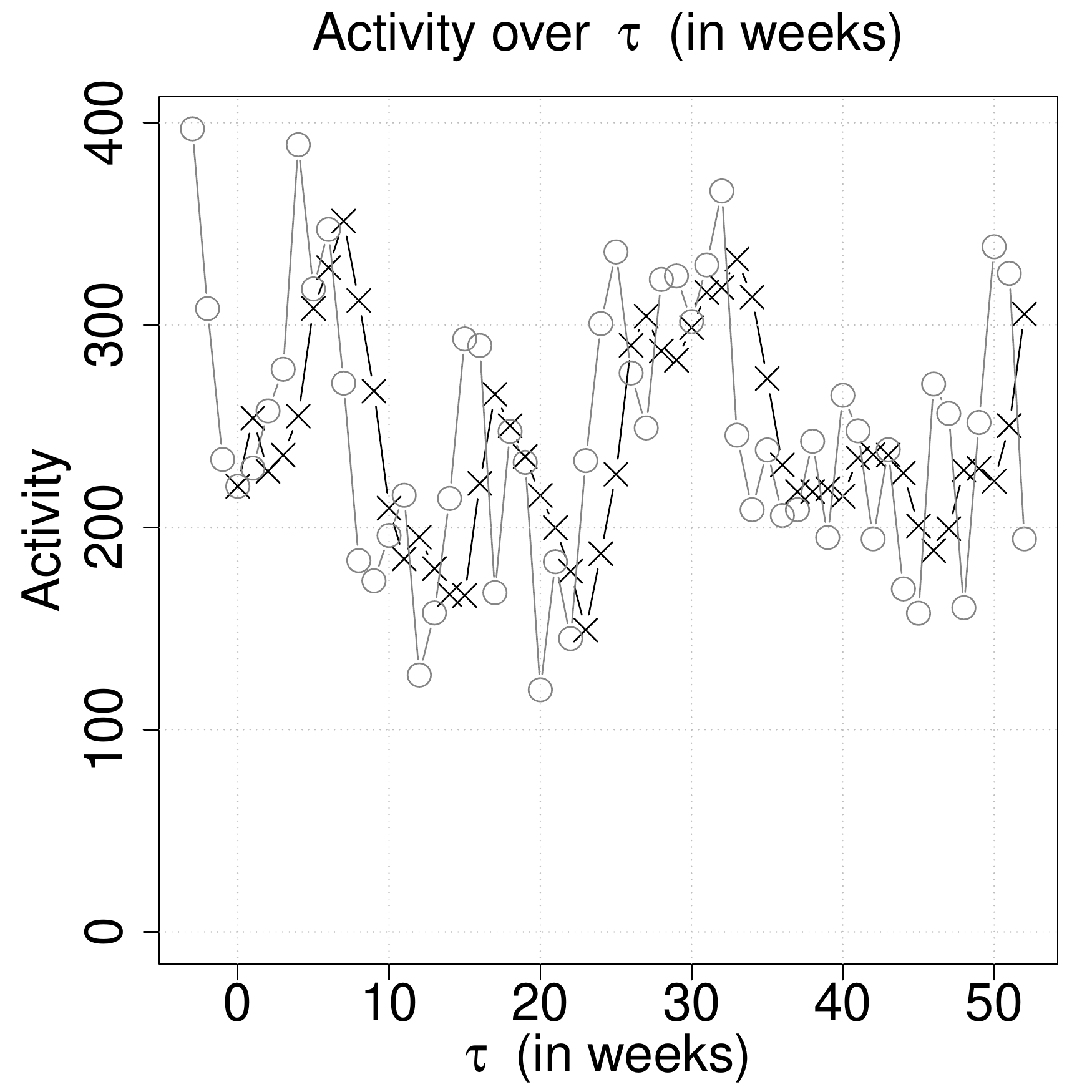}}
\subfigure[Bitcoin StackExchange Activity]{\label{fig:bse_act}
\includegraphics[width=0.24\linewidth]{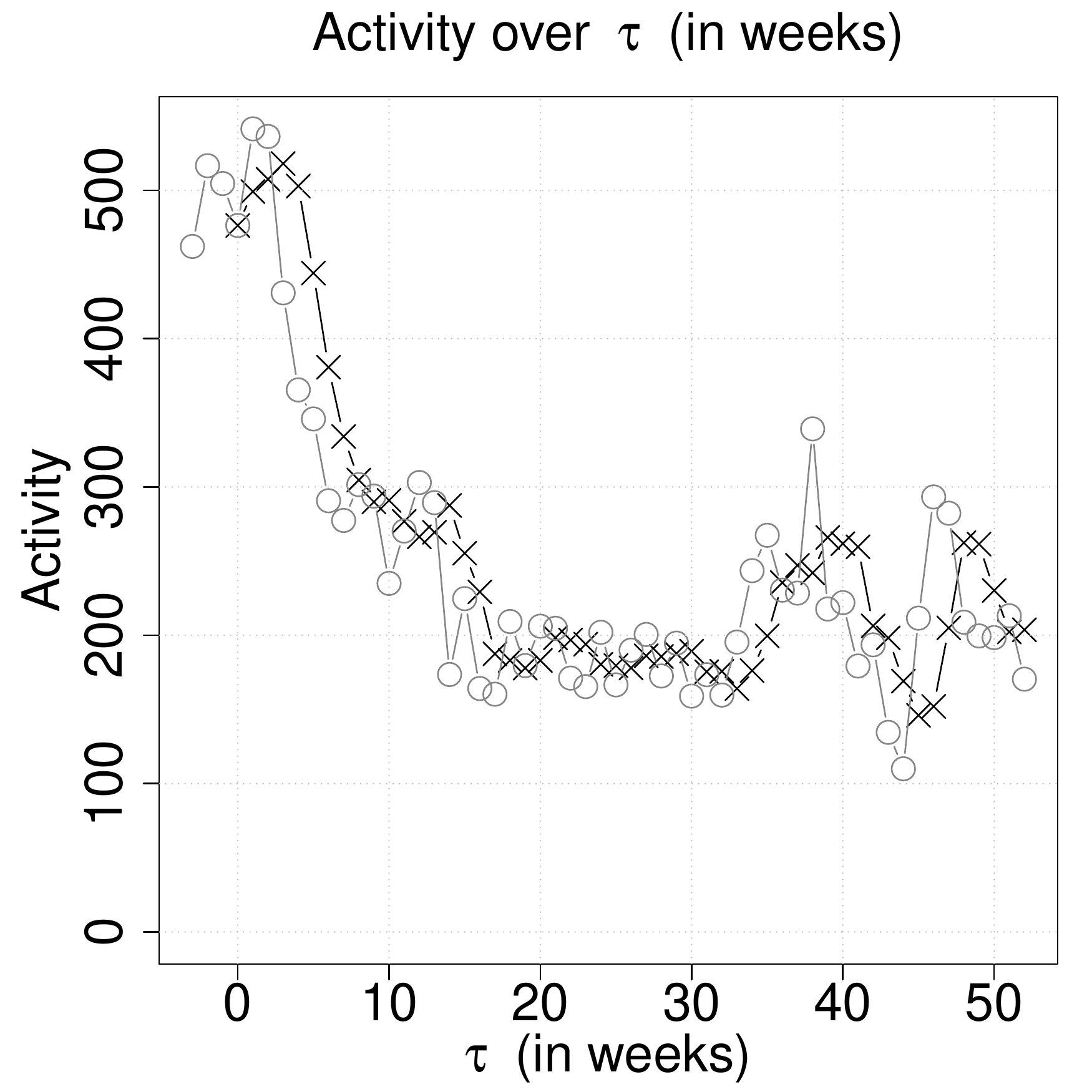}}
\subfigure[English StackExchange Activity]{\label{fig:ese_act}\includegraphics[width=0.24\linewidth]{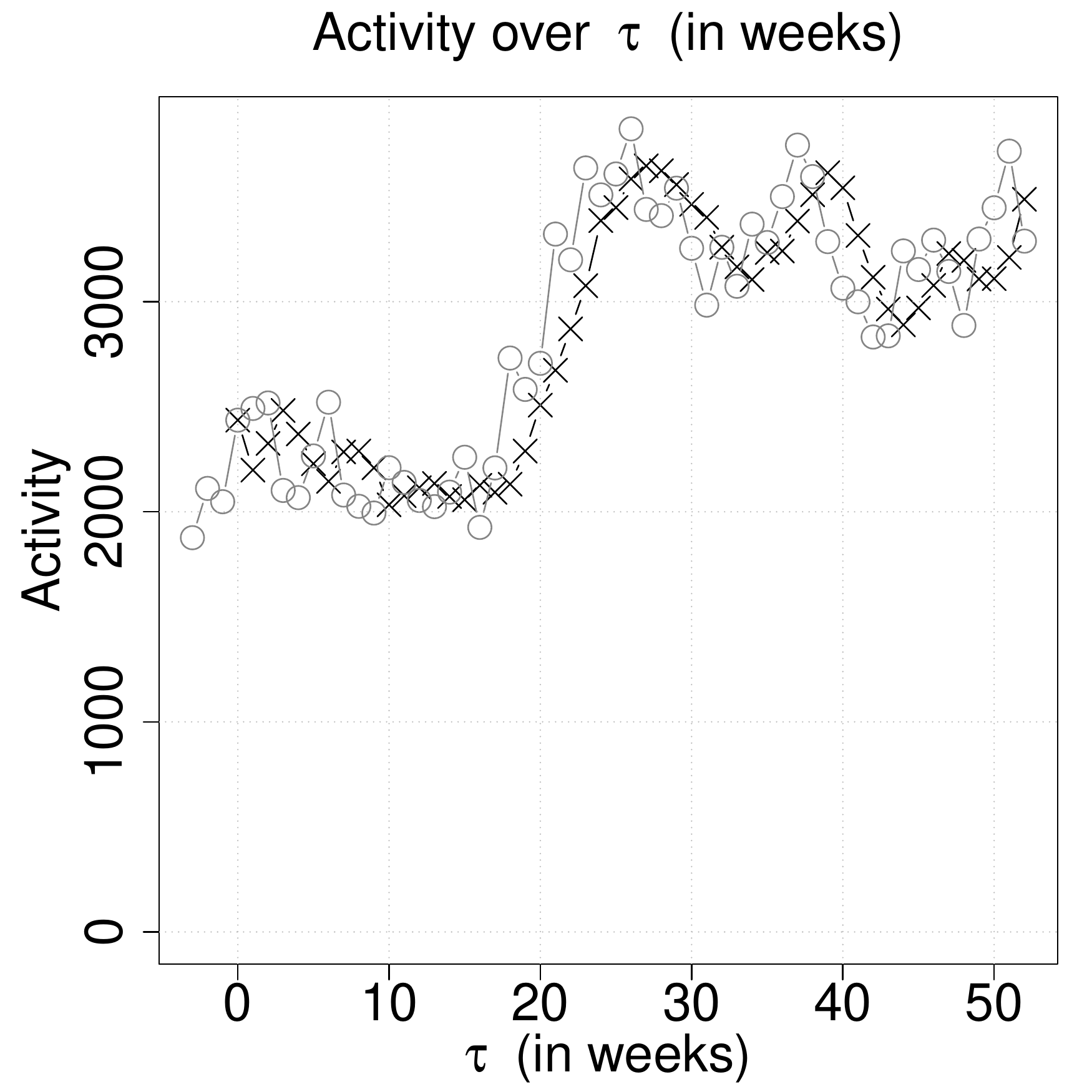}}
\subfigure[Mathematics StackExchange Activity]{\label{fig:mse_act}\includegraphics[width=0.24\linewidth]{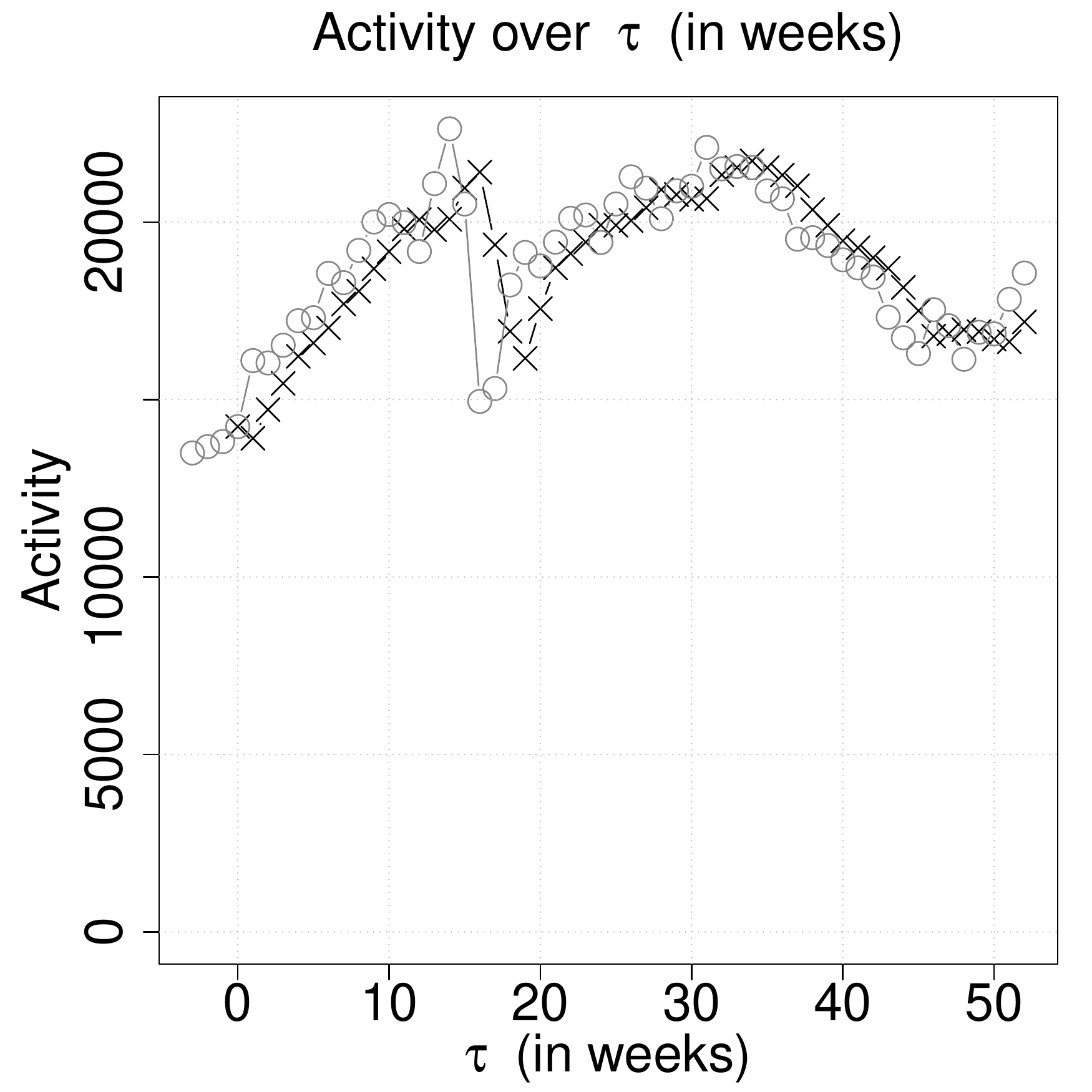}}
\subfigure{\label{fig:leg}\includegraphics[width=0.35\linewidth]{legend_emp_act.pdf}}
\setcounter{subfigure}{4}
 
\subfigure[Beachapedia Activity]{\label{fig:bp_act}\includegraphics[width=0.24\linewidth]{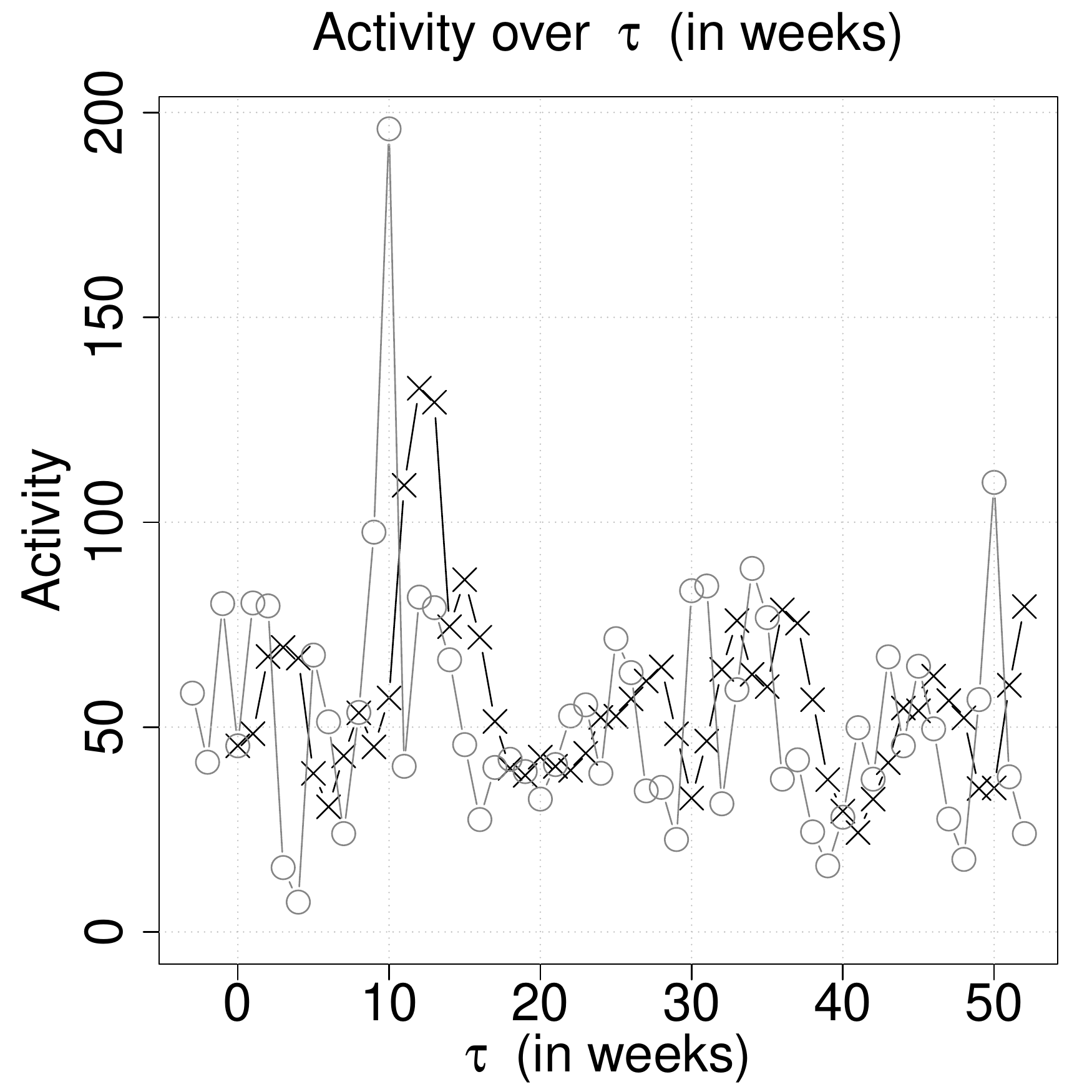}}
\subfigure[NOBBZ Activity]{\label{fig:nbz_act}\includegraphics[width=0.24\linewidth]{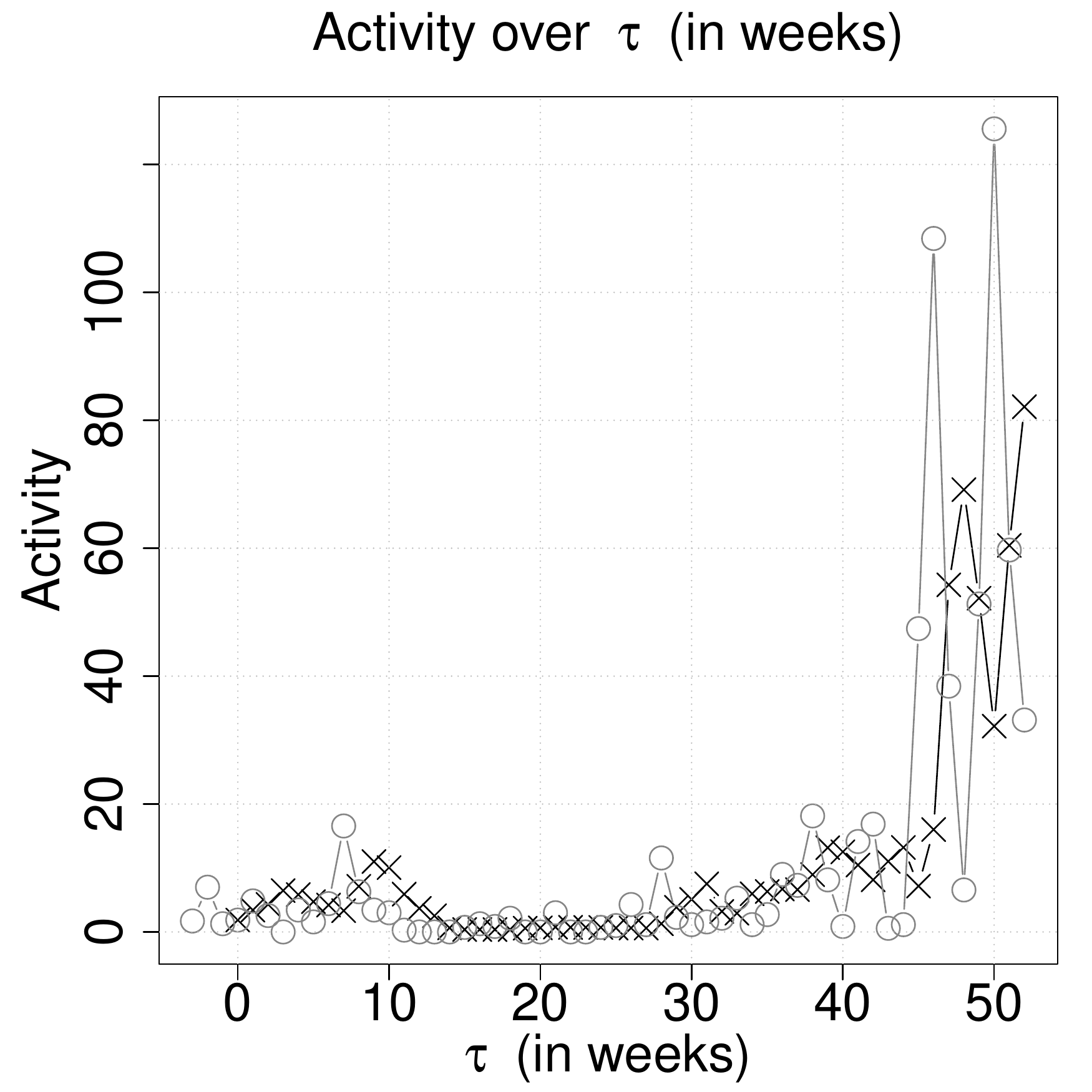}}
\subfigure[NeuroLex Activity]{\label{fig:nlx_act}\includegraphics[width=0.24\linewidth]{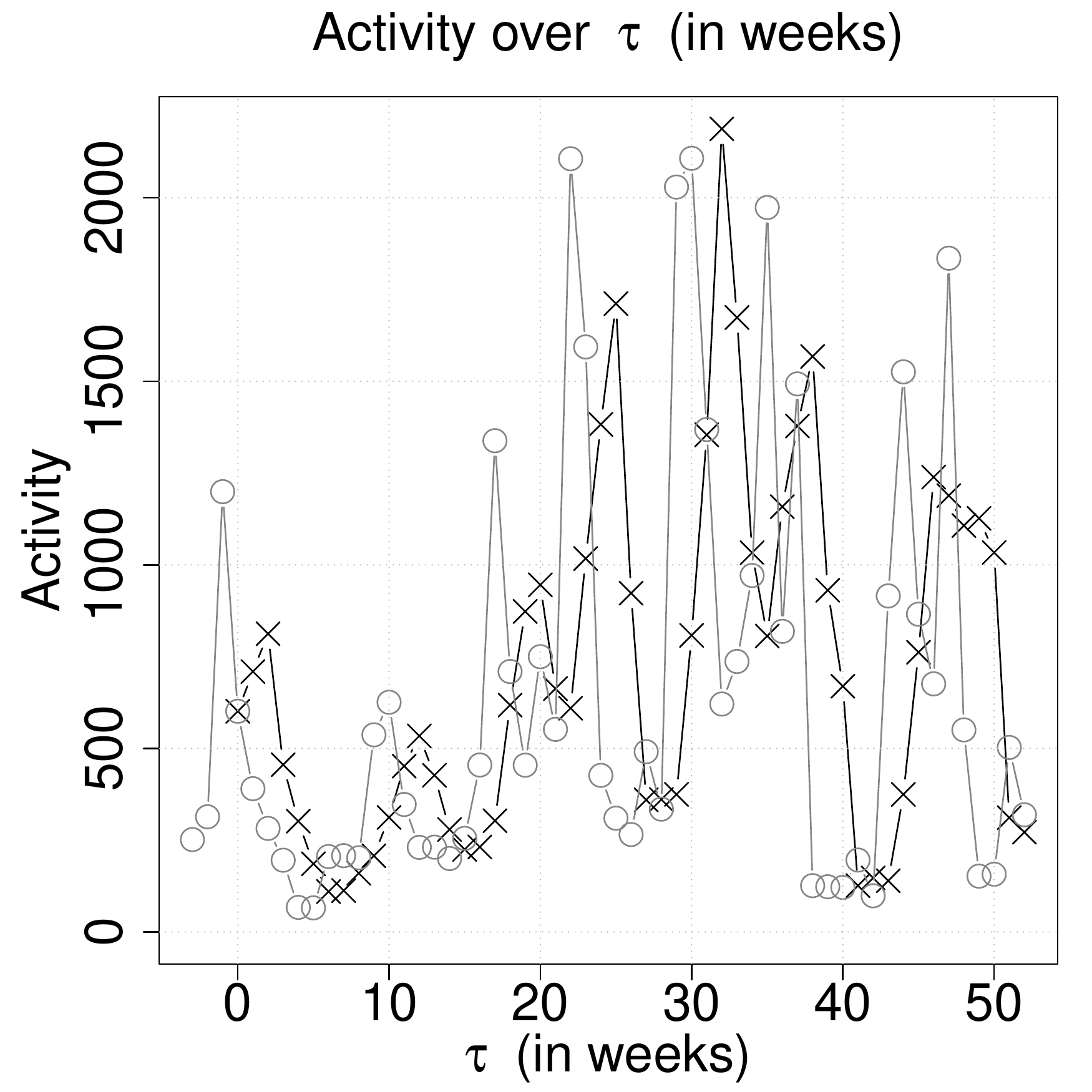}}
\subfigure[15MW Activity]{\label{fig:w15_act}\includegraphics[width=0.24\linewidth]{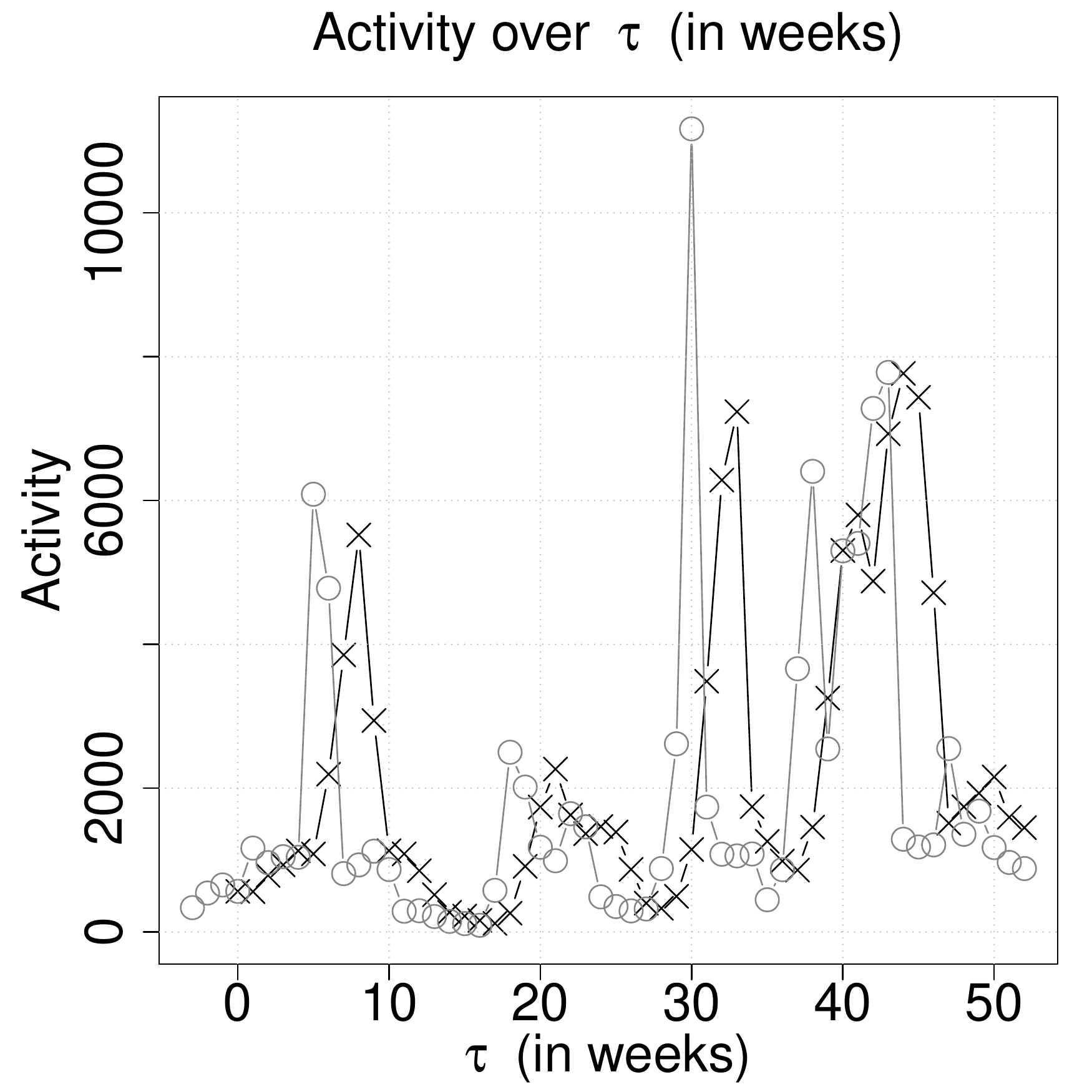}}
\caption{ \revone{\textbf{Results for the activity dynamics simulation.} The plot depicts the results of our activity dynamics simulation for the StackExchange datasets (\textbf{top row}) and Semantic MediaWiki instances (\textbf{bottom row}). The solid gray lines with circles represent the empirical (observed) activity over $t$ (in weeks; $x$-axes), while the solid black lines represent the simulated activity dynamics ($y$-axes). In all of our analyzed datasets, the simulated activity dynamics exhibit a notable resemblance to the empirical activity.}}
\label{fig:empirical_illustration}
\end{figure*}

\begin{table*}[b!]
\center
\tiny
\caption{\revone{\textbf{RMSE.} The table depicts root mean-squared errors (RMSE) of our activity dynamics simulation per user and week for all datasets. Our simulation yields a small RMSE for all StackExchange datasets. RMSE for the Semantic MediaWiki datasets is slightly higher, which is likely due to the lower number of active users (listed in the Users column).}}
\begin{tabular}{ l | c c c c c c c c }\toprule
\textbf{Dataset} & HSE & BSE & ESE & MATHSE & BP & NZ & NLX & 15MW \\\midrule
\textbf{Activity} & $12,496$ & $12,295$ & $151,028$ & $986,996$ & $2,718$ & $603$ & $33,792$ & $102,521$\\
\textbf{Users} & $682$ & $1,299$ & $7,893$ & $35,476$ & $16$ & $36$ & $112$ & $394$ \\
\textbf{RMSE} & $0.076$ & $0.031$ & $0.029$ & $0.030$ & $1.755$ & $0.274$ & $4.397$ & $4.043$ \\
\bottomrule
\end{tabular}
\label{tab:rmse}
\end{table*}

Figure~\ref{fig:empirical_illustration} depicts the results of the activity dynamics simulation. The root mean-squared errors (RMSEs) of the simulations are listed in Table~\ref{tab:rmse}.

Overall, the results gathered from the activity dynamics simulation exhibit a notable resemblance to the real activities of the corresponding datasets. Due to the chosen approximations and simplifications when estimating \ratio for our model (i.e., static network structure and average model parameters over weeks and users), the simulated activity is naturally limited in its accuracy. These limitations are particularly visible whenever there are large and sudden increases of activity in the collaboration networks. \revone{Note that \ratio will only be higher than $\kappa_1$ if activity in our datasets is either zero or the relative difference in activity between two months is extremely high, which is never the case for our smoothed empirical datasets.}

Further, the assumption of a fixed network structure of our investigated collaboration networks also (negatively) influences the obtained results of our simulation. For example, it is possible for our simulation to yield higher increases in activity (e.g., Figure~\ref{fig:bse_act}), as users might be influenced by peers, who would join the collaboration network only at a later point in time.

\begin{figure*}[!t]
\centering
\subfigure[History StackExchange Ratios]{\label{fig:hse_ratio}\includegraphics[width=0.24\linewidth]{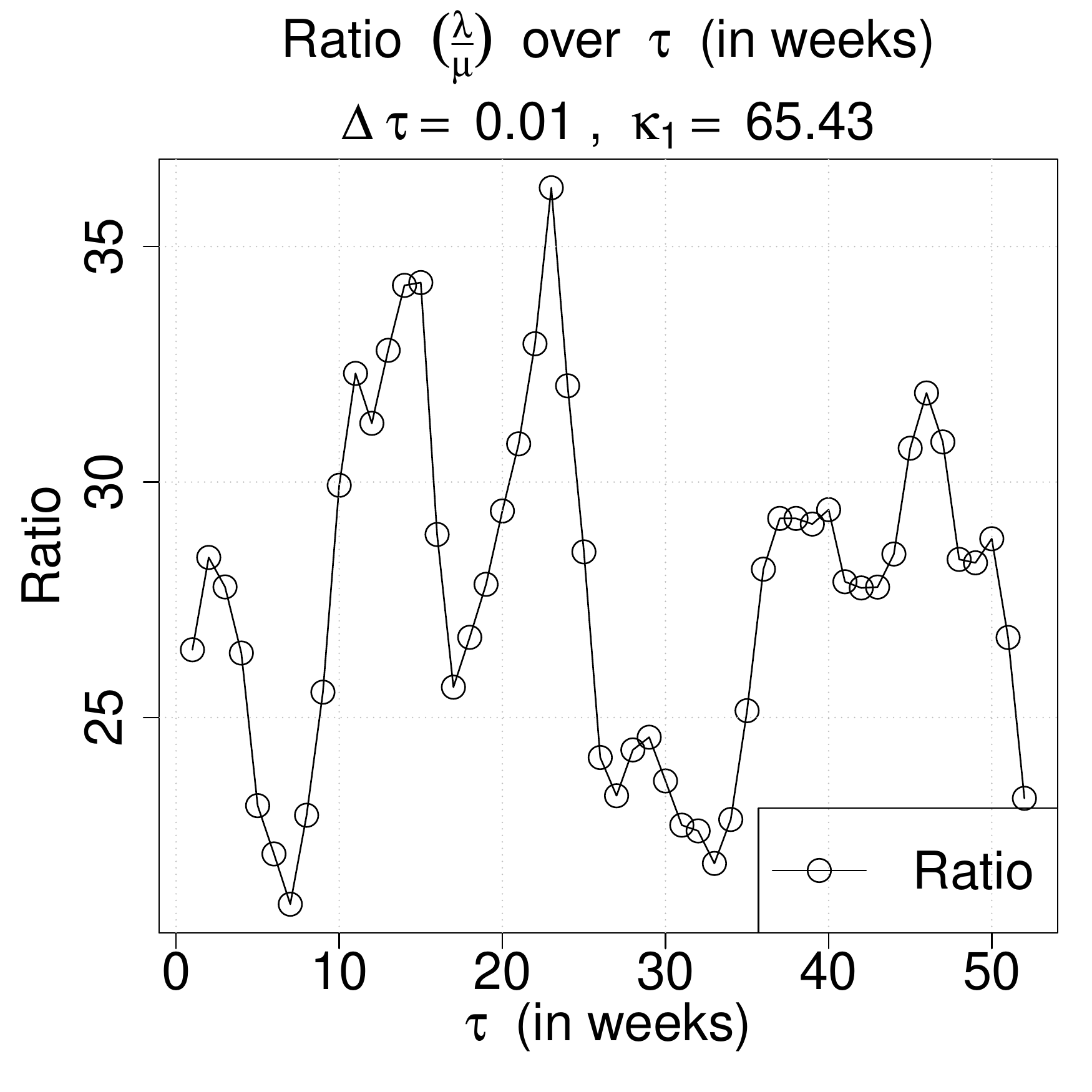}}
\subfigure[Bitcoin StackExchange Ratios]{\label{fig:bse_ratio}
\includegraphics[width=0.24\linewidth]{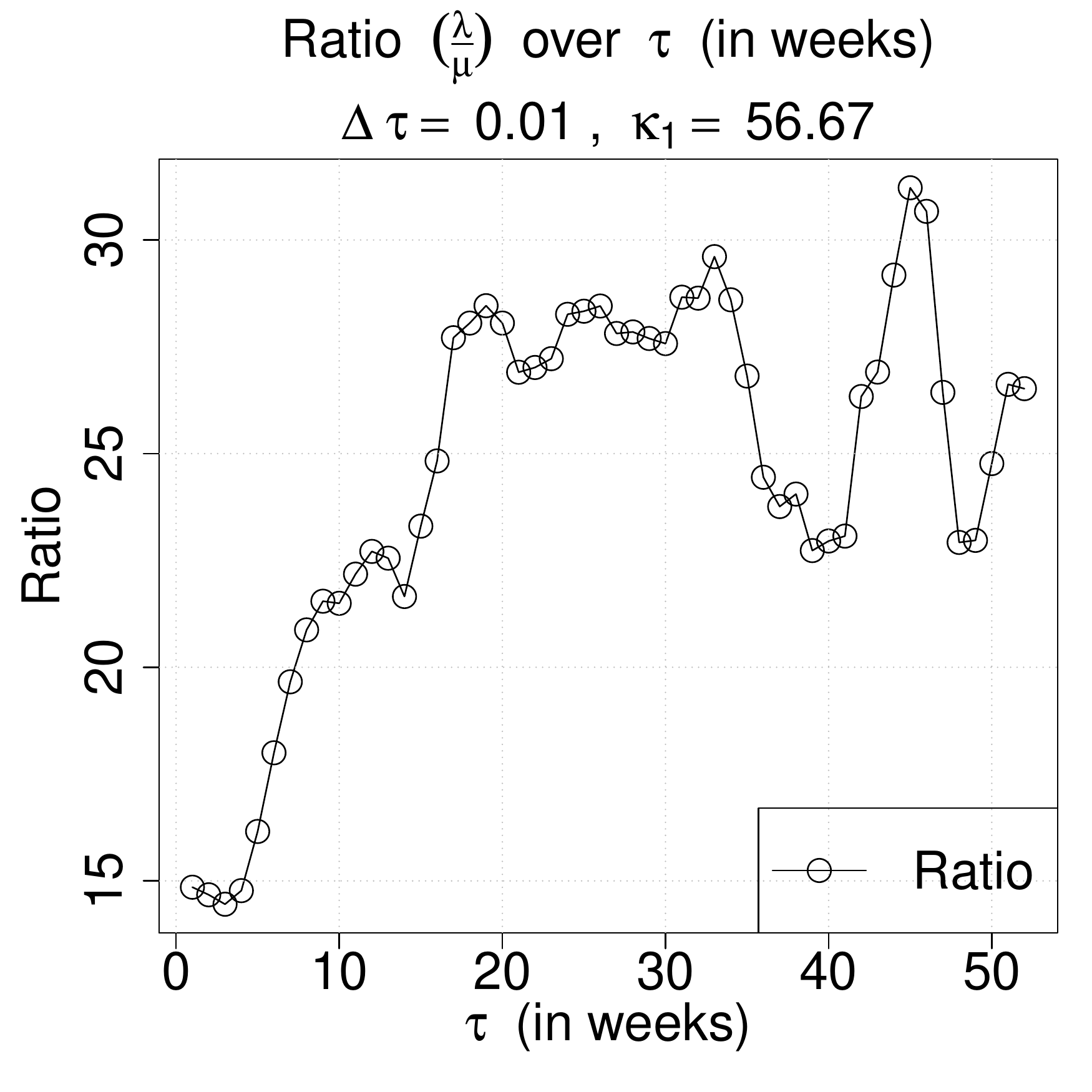}}
\subfigure[English StackExchange Ratios]{\label{fig:ese_ratio}\includegraphics[width=0.24\linewidth]{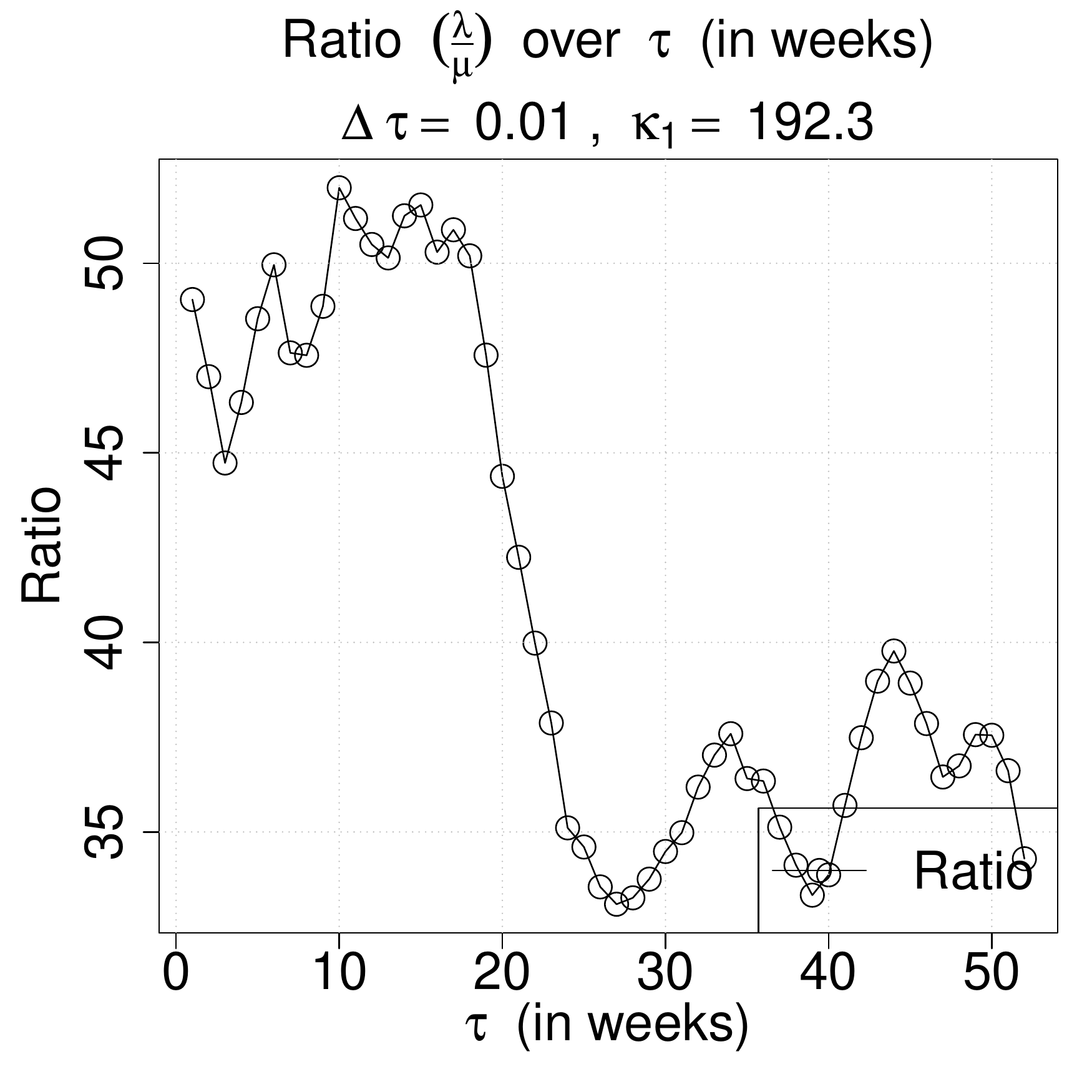}}
\subfigure[Mathematics StackExchange Ratios]{\label{fig:mse_ratio}\includegraphics[width=0.24\linewidth]{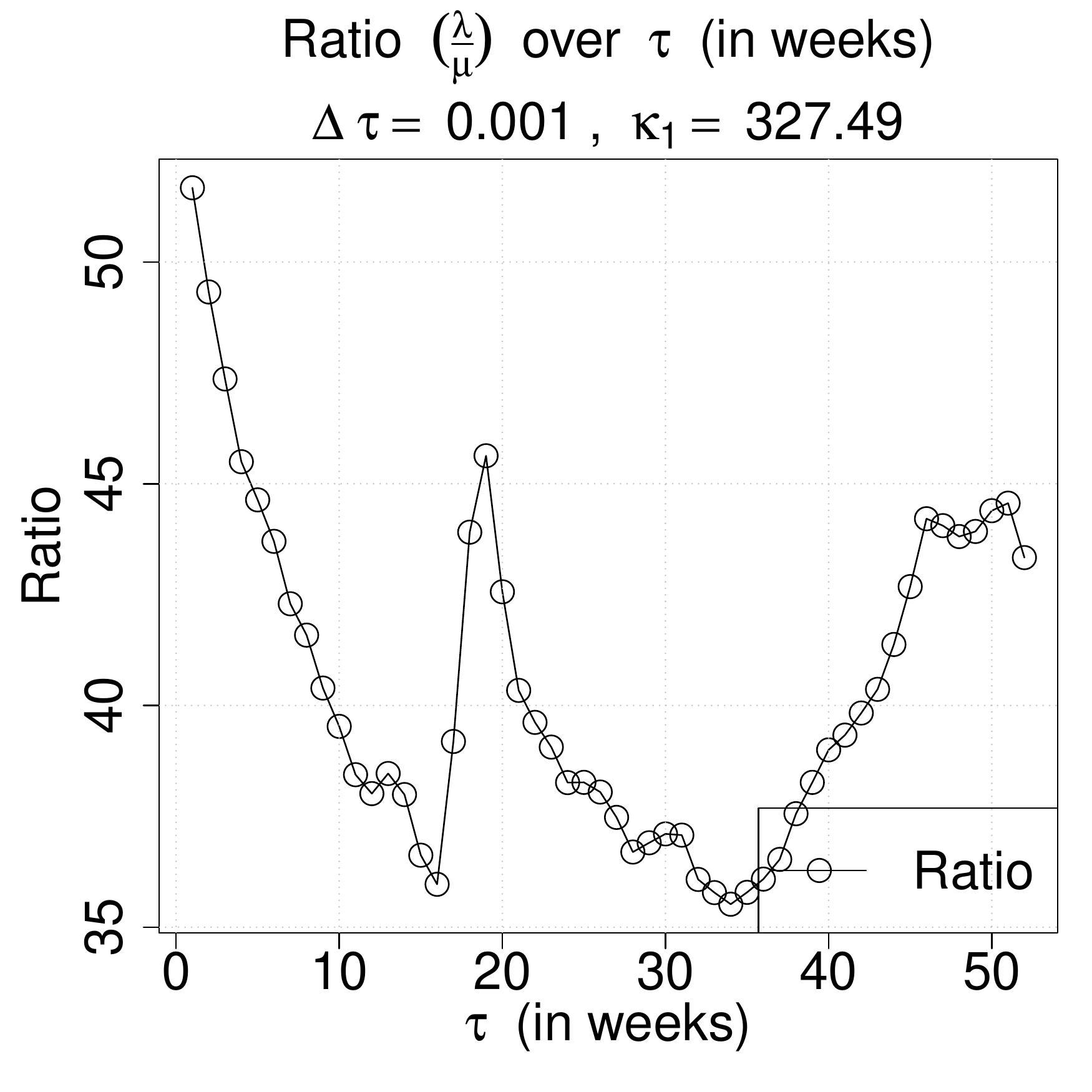}}
\subfigure[Beachapedia Ratios]{\label{fig:bp_ratio}\includegraphics[width=0.24\linewidth]{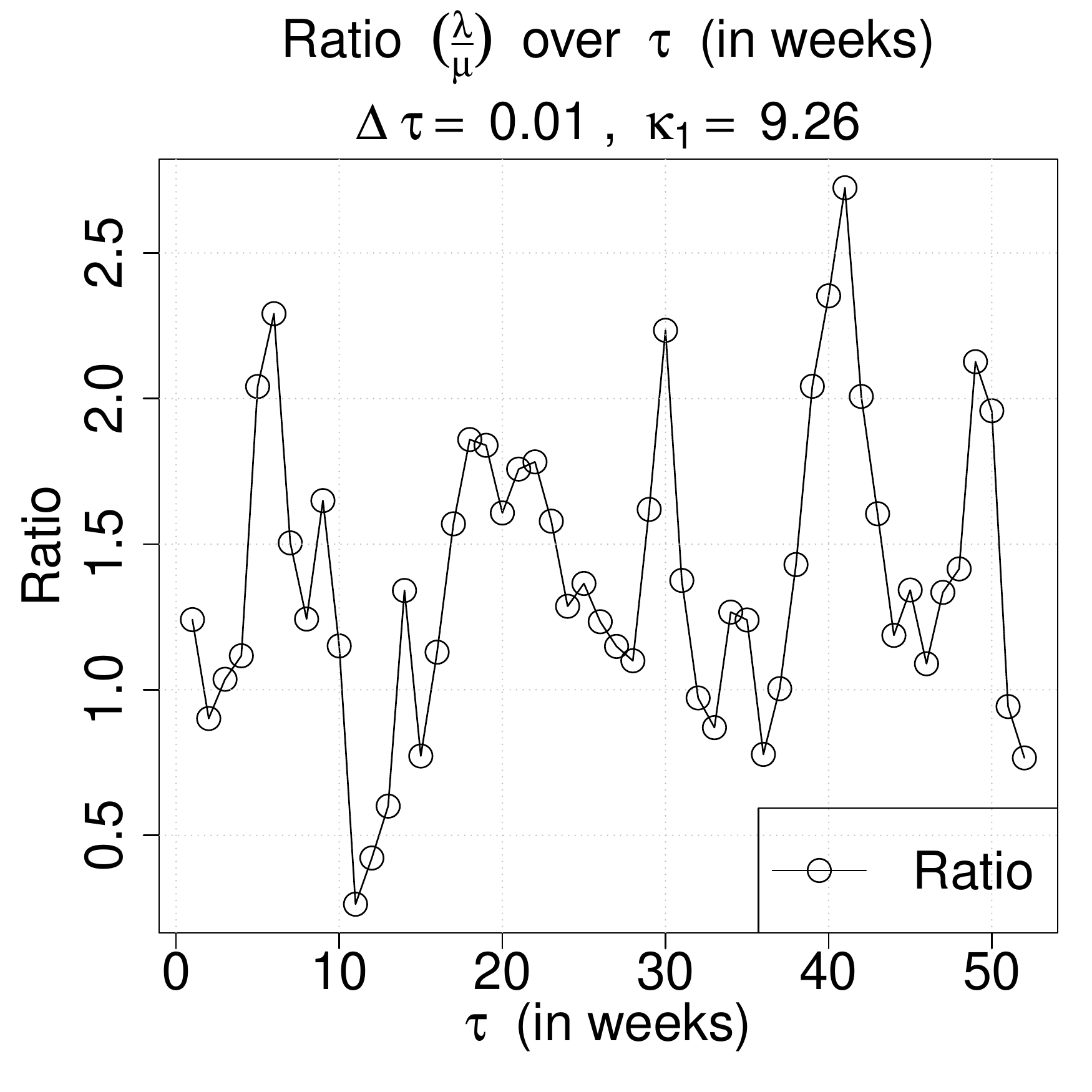}}
\subfigure[NOBBZ Ratios]{\label{fig:nbz_ratio}\includegraphics[width=0.24\linewidth]{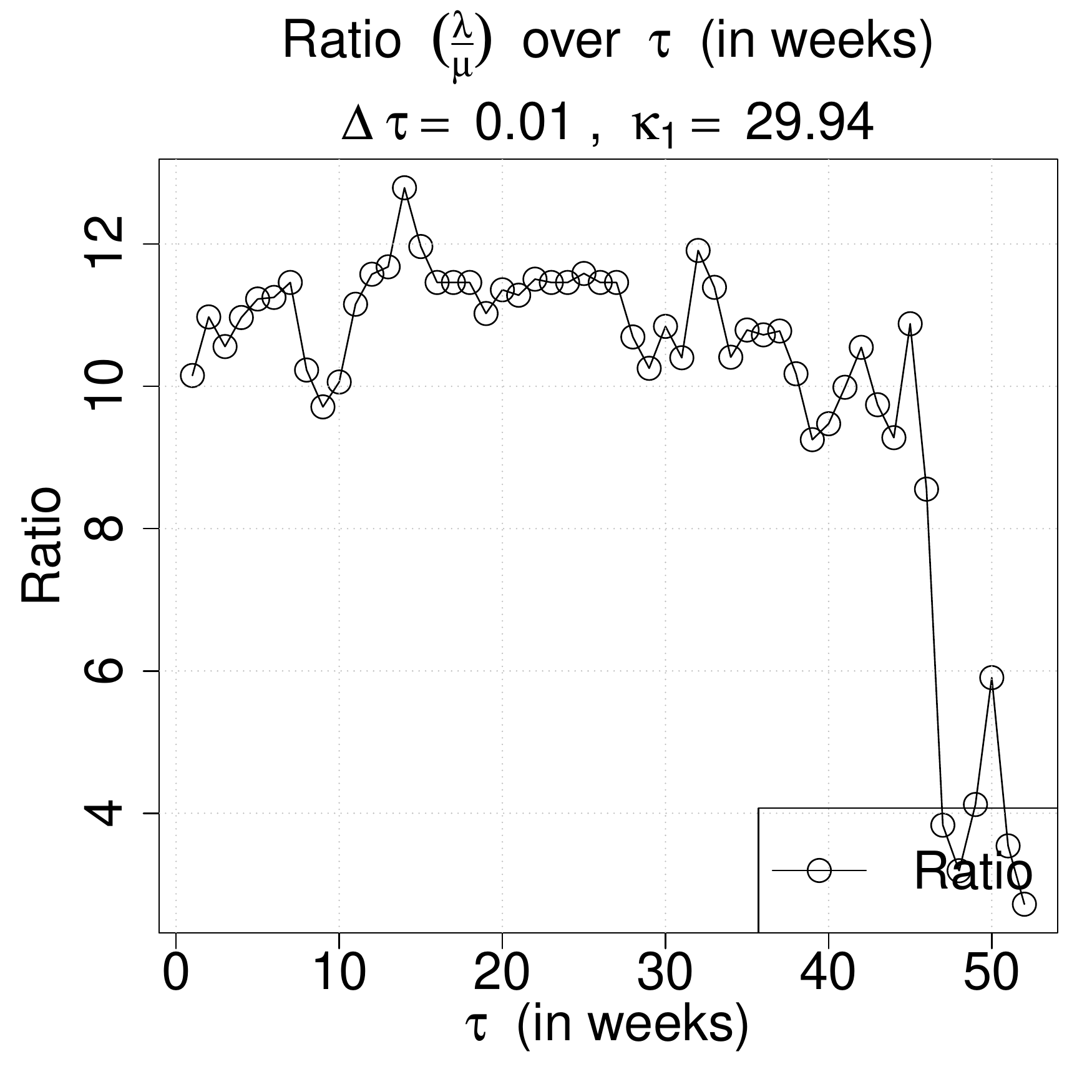}}
\subfigure[NeuroLex Ratios]{\label{fig:nlx_ratio}\includegraphics[width=0.24\linewidth]{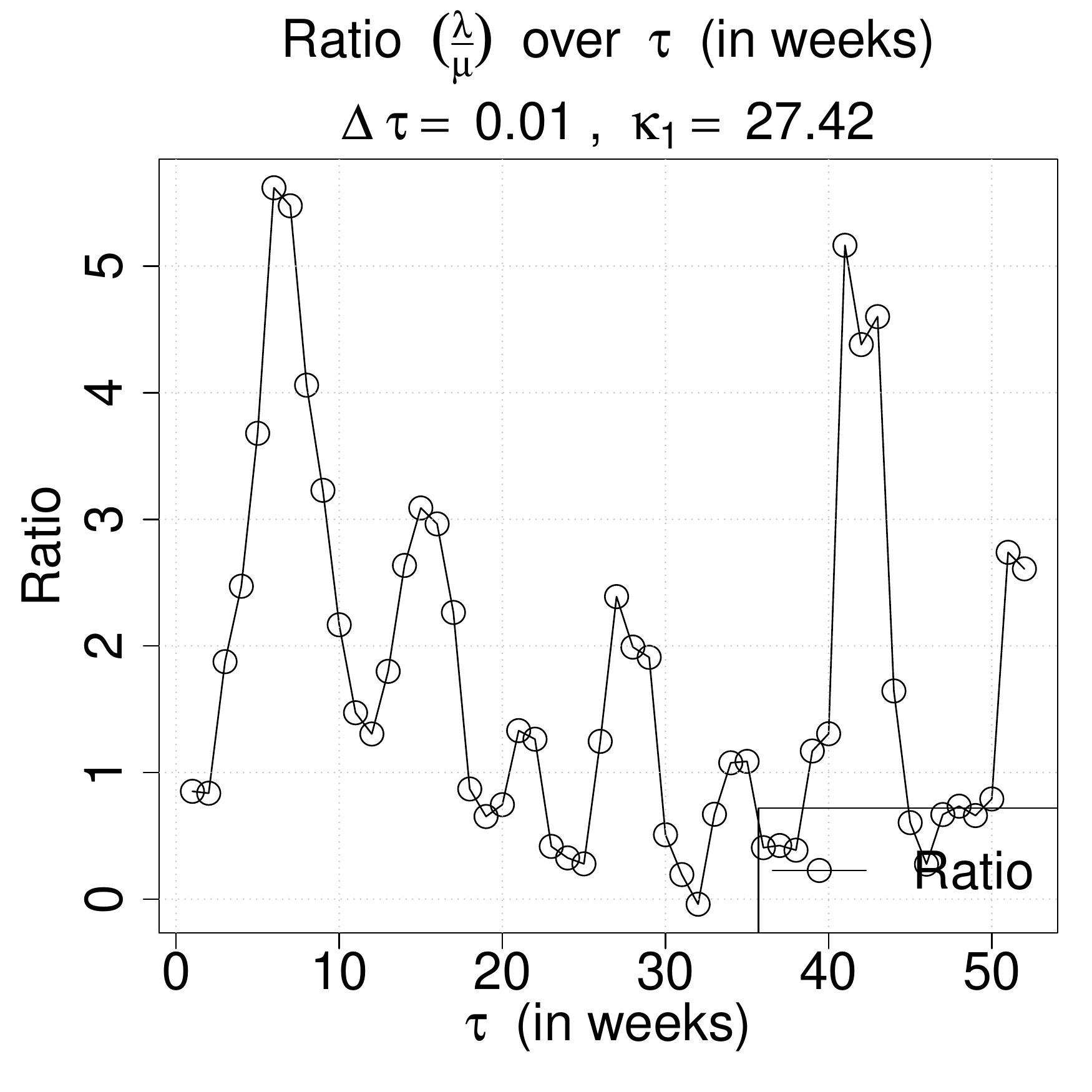}}
\subfigure[15MW Ratios]{\label{fig:w15_ratio}\includegraphics[width=0.24\linewidth]{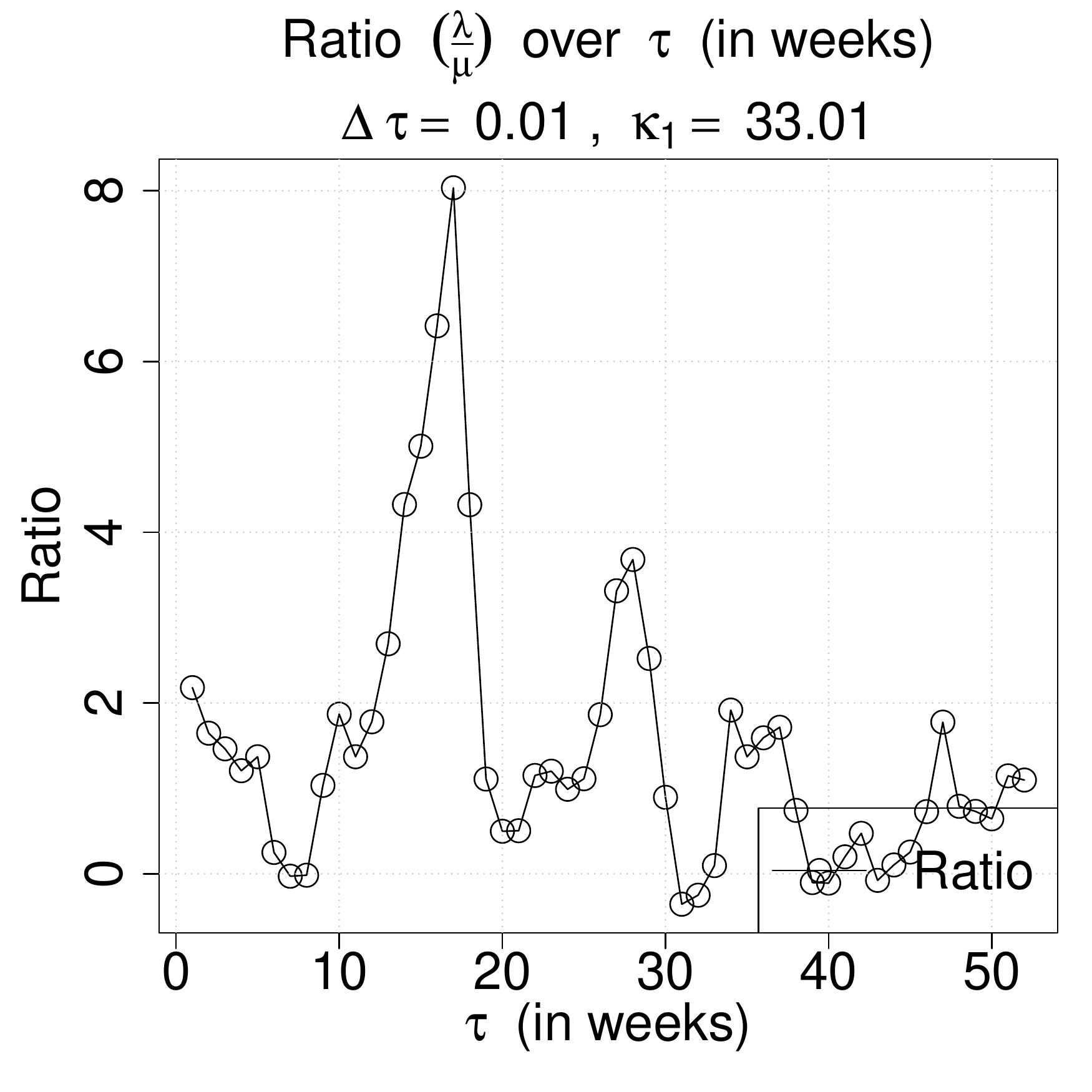}}
\caption{ \textbf{Evolution of ratios $\lambda/\mu$.} The evolution of the ratios $\lambda/\mu$ ($y$-axes) over $\tau$ (in weeks; $x$-axes)  for the StackExchange datasets (\textbf{top row}) and for the Semantic MediaWiki instances (\textbf{bottom row}). \revone{The smaller the ratio, the higher the levels of activity in Figure~\ref{fig:empirical_illustration}. Small variances in \ratio over time indicate that activities of the systems are less influenced by the activity of single individuals than they are by peer influence.}}
\label{fig:empirical_ratios}
\end{figure*}

\section{System Mass and Activity Momentum}
\label{sec:momentum}

We can further analyze the obtained ratios and parameters of our activity dynamics simulation to broaden our understanding of the collaboration networks under investigation. Figure~\ref{fig:empirical_ratios} depicts the value of the calculated ratios \ratio ($y$-axis) for each week ($x$-axis). If the ratio is higher than $\kappa_1$ (denoted in the title of each Figure), our master stability equation holds and the system converges towards zero activity (over time). The amount of activity that is lost per iteration---and hence the speed of activity loss---is proportional to the value of the ratio and the activity already present in the network. In general, a higher ratio results in a higher and faster loss of activity.

If the ratio is smaller than $\kappa_1$, the master stability equation has been invalidated and the system will converge towards a new fixed point of immanent activity (cf. Section~\ref{subsec:linear stability analysis}). If this is the case, we can observe one of three potential behaviors, which are triggered depending on the amount of activity already present in the network and the current ratio:
\begin{enumerate}[label=(\roman*)]
\item \textbf{An increase in activity} if the new fixed point, corresponding to the new ratio, is of higher overall activity than the activity already present in the collaboration network (see $\tau=20-30$ in Figures~\ref{fig:mse_act} and \ref{fig:mse_ratio}). This situation emerges whenever we invalidate the master stability equation from a previously stable fixed point or if the system is already stable in a situation when the new ratio is smaller than the last estimated ratio.
\item \textbf{A decrease in activity} if the new fixed point is of lower overall activity than the activity already present in the collaboration network (see $\tau$ $3-7$ in Figures~\ref{fig:bse_act} and \ref{fig:bse_ratio}). Again this may occur in two specific situations. First, if the ratio increases, so that the master stability equation is now satisfied and the system has been previously in an unstable state. Second, if the system is in an unstable state but the ratio increases slightly without satisfying the stability equation.
\item \textbf{No change in activity} if the new fixed point corresponding to the new ratio is of the same overall activity than the activity already present in the collaboration network (see $\tau$ $20-30$ in Figures~\ref{fig:bse_act} and \ref{fig:bse_ratio}).
\end{enumerate}

\smallskip\noindent\textbf{System Mass.} We can now use the obtained ratios to characterize the collaboration networks and quantify their robustness in terms of their activity dynamics. Robust systems are systems with lively and high levels of activity, which are able to keep that activity even in the cases of small unfavorable changes in the dynamical parameters. Less robust systems are systems that lose their activity very quickly as a consequence of even small changes in the ratio. Thus, we calculate the standard deviation over all ratios $\sigma_{\lambda / \mu}$ over time and normalize it over $\kappa_1$---to account for the size of the collaboration networks---and refer to it as  $\rho$---the normalized standard deviation of the ratio \ratio (see Equation~\ref{eq:sd_ratio}).

\begin{equation}
\label{eq:sd_ratio}
	\rho = \frac{\sigma_{\lambda/\mu}}{\kappa_1}
\end{equation}

The normalized standard deviation is a measure of system sensitivity and its inverse ($1/\rho$) represents a measure of system stability or \textit{inertia} to changes in activity. Analogously to mass in classical mechanics---which defines the inertia or resistance of being accelerated or decelerated for an object by a given force---we call the quantity $1/\rho$ the \emph{System Mass}. We denote this quantity with $m_s$ with the subscript $s$ to distinguish it from the number of links $m$ in a collaboration network (see Table~\ref{tab:momentum}). In systems with a large \textit{System Mass} it is more difficult to induce changes in activity. In particular, this means that it is more difficult to reduce activity in a consistently active system (due to the small standard deviations of \ratio), as well as it is difficult to jump-start the same system if activity levels were consistently low in the past (again, due to small standard deviations of \ratio). 

\begin{table*}[b!]
\center
\tiny
\caption{\textbf{System Mass and Activity Momentum.} The table depicts the results for the activity momentum analysis. $\rho$ is the standard deviation of the calculated ratios normalized over $\kappa_1$. System Mass is represented by $1/\rho$ and Activity Momentum represents System Mass multiplied with Activity. Activity depicts the average activity per week as well as the value for the last observed months in brackets. Activity Momentum follows analogously. MATHSE and ESE exhibit the largest average and current Activity Momenti, followed by 15MW and NLX. Even though 15MW exhibits a System Mass similar to HSE and NZ, its Activity Momentum is much larger.}
\begin{tabular}{ l c c c c }\toprule
Dataset & Activity (last month) & $\rho$ & System Mass & Activity Momentum (last month) \\\midrule
MATHSE & $19,255$ ($70,130$) & $0.0115$ & $86.65$ & $1,674,415$ ($6,076,765$)\\
ESE & $2,952$ ($13,751$) & $0.0344$ & $29.07$ & $85,815$ ($399,742$)\\
BSE & $246$ ($782$) & $0.0762$ & $13.12$ & $3,228$ ($10,260$)\\
HSE & $248$ ($1,110$) & $0.0554$ & $18.10$ & $4,489$ ($20,091$)\\\midrule
15MW & $1,999$ ($4,702$) & $0.0506$ & $19.76$ & $39,500$ ($92,912$)\\
NLX & $668$ ($1,131$) & $0.0532$ & $18.80$ & $12.558$ ($21,263$)\\
NZ & $12$ ($270$) & $0.0802$ & $12.67$ & $152$ ($3,421$)\\
BP & $54$ ($228$) & $0.0547$ & $18.28$ & $987$ ($4,168$)\\
\bottomrule
\end{tabular}
\label{tab:momentum}
\end{table*}

\smallskip\noindent\textbf{Activity Momentum.} After calculating the \emph{System Mass} $m_s$, we are now interested (again analogously to classical mechanics) in calculating the \emph{Activity Momentum} $p$ for our collaboration networks (see Equation~\ref{eq:act_momentum}).

\begin{equation}
\label{eq:act_momentum}
	p = m_sa
\end{equation}
For activity we take (i) the average activity (posts and replies) per week and (ii) the activity in the last month of our observation periods (cf. Table~\ref{tab:momentum}) and calculate (i) the average and (ii) the current momentum.

\revone{The higher the \emph{Activity Momentum} of a collaboration network, the more force is needed to ``stop'' (make it inactive) the system. Hence, the higher the momentum, the more robust a given network.} In particular, if a (sufficiently) small number of users would suddenly stop contributing to a collaboration network that exhibits a very large \emph{Activity Momentum} $p$, activity in the overall network would be minimally influenced. On the other hand, if the same number of users would stop contributing to a collaboration network with a (significantly) smaller \emph{Activity Momentum} $p$, chances are that their actions (or lack thereof) will have a notable influence on the overall trends in activity dynamics of the system. In particular, there are three factors that influence the \emph{Activity Momentum} of collaboration networks:

\begin{enumerate}[label=(\roman*)]
\item \emph{The standard deviation of \ratio.} If the ratio is very stable and does not frequently oscillate, the standard deviation and hence the normalized standard deviation will be very small. This also means that activity, as well as increases and decreases thereof, is equally distributed across $\tau$ and is not (frequently) exercised in bursts.
\item \emph{The largest eigenvalue $\kappa_1$.} Larger and denser collaboration networks exhibit a larger highest eigenvalue $\kappa_1$. As $\rho$ is the normalized variance of the ratios over $\kappa_1$, the largest eigenvalue will directly influence $\rho$. The notion of normalizing $\rho$ over $\kappa_1$ follows the intuition that that large collaboration networks are less likely to exhibit sudden changes in activity than smaller ones.
\item \emph{The activity.} The larger the average activity (posts and replies) per month, the higher the \emph{Activity Momentum} of a collaboration network, and hence the higher the force that is needed to render the collaboration network inactive. Analogously, networks with a small \emph{Activity Momentum} require less force to be influenced (i.e., to either speed up/increase or slow down/decrease activity).
\end{enumerate}

Hence, we can use the calculated \emph{Activity Momentum} $p$ as an indicator of the activity level as well as the tendency of a system to stay at that activity level in the future. For example, MATHSE exhibits the most robust collaboration network of our datasets regarding changes in activity, with an \emph{Activity Momentum} of order $10^6$ (average per week and last month). ESE and 15MW both exhibit similar average \emph{Activity Momenti} of orders $10^4$. However, when looking at the \emph{Activity Momenti} of the last months, ESE is roughly four times as hard to stop as 15MW.

In contrast, HSE and BSE exhibits very similar activity levels for last month, however the corresponding \textit{Activity Momentum} of HSE is twice the one of BSE, indicating that half the force is needed to render BSE inactive than it would be needed to render HSE inactive. The other datasets follow analogously.

On the other hand, BP exhibits a high value for \emph{System Mass} and a very low corresponding \emph{Activity Momentum}, indicating that it will be very difficult to to accelerate or jump-start the system with regards to activity.

\section{Related Work}
\label{sec:related work}

The work presented in this paper was inspired by and builds upon work presented in the areas of \emph{critical mass theory} and \emph{dynamical systems on networks}.

\subsection{Critical Mass Theory}
In 1985 and 1988, \citet{oliver1985theory, oliver1988paradox, marwell1988social} have discussed and analyzed the concept of critical mass theory by introducing so called production functions to characterize decisions made by groups or small collectives. Fundamentally, these production functions represent the link between individual benefits and benefits for the group.

They argue that one very important aspect of critical mass is the natural limitation of collective goods for groups such as housing, food, fuel or oil. Hence, the capacity of users (and thus critical mass) for such a group or system is naturally limited by the corresponding resource. However, collective (digital) goods are not (or only artificially) limited for online communities; theoretically allowing for an infinite increase in users and interest. Without users motivated to contribute, interest will decrease and critical mass will lose momentum and ultimately decelerate until all interest vanishes. In their work they identified multiple different types of production functions, with the most important ones being: \emph{Accelerating}, \emph{decelerating} and \emph{linear} functions. The idea behind accelerating production functions is that each contribution is worth more than its preceding one. In a decelerating production function the opposite would be the case, resulting in each succeeding contribution to be worth less than the preceding one, while contributions to linearly growing functions are always worth the same. Until today it is still mostly unclear what these production functions look like for online communities (e.g., StackOverflow) and online production systems (e.g., Semantic MediaWikis).

Depending on the investigated or desired point of view, different characteristics of these communities and online production systems can be used as basis for calculating production functions. The analysis of \citet{oliver1985theory} also highlights that different production functions can lead to very different outcomes in similar situations. For example, given an accelerating production function, users who contribute to a system are likely to find their potential contribution ``profitable'', as each subsequent contribution increases the value of their own contribution. Naturally, this increases the incentive to make larger contributions to begin with. Given a deceleration production function, users would not immediately see the benefit of large contributions, given that each subsequent contribution is increasing the overall value less, while more effort, in the form of larger contributions, is needed to turn a decelerating production function into an accelerating one.

One approximation for critical mass by \citet{solomon2014critical} involved the investigation of the number of changes -- as activity -- and number of users -- as growth of a community -- for calculating production functions for WikiProjects. The authors argue that activity in online production systems, after certain amounts of time, is the best indicator of a self-sustaining system. In this work, we have extended the analysis presented by Solomon and Wash and specifically define the point of when an online system has reached critical mass and has become self-sustaining in terms of its activity dynamics. \citet{walk2014characterizing} recently conducted a similar analysis to characterize critical mass for Semantic MediaWikis.

\citet{Raban:2010:ESC:1718918.1718932} investigated factors that allow for a prediction of survival rates for IRC channels and identified the production function of these chat channels regarding the number of unique users versus the number of messages posted at certain times, as the best predictor.

\citet{chengcatalyst} have analyzed concepts of activation thresholds, which resemble features that, when achieved, can help to reach and sustain self-sustainability. They created an online platform that allow groups to pitch ideas, which only will be activated if enough people commit to it.

\revone{With regards to activity, \citet{suh2009singularity} have shown that contributions to Wikipedia are slowing down, which is likely a direct consequence of the increase in required coordination activities, as well as comprehensive contribution guidelines which discourage posts by users. Kittur and Kraut \cite{kittur2008harnessing} have demonstrated that when reducing the overhead for editors---effectively minimizing the efforts necessary to contribute to Wikipedia---can help to increase the number of contributions and article quality.  Similarly, \citet{anderson2012discovering} investigated the value and development of contributions to the question answering portal StackOverflow. In contrast, \citet{yang2014sparrows} have investigated the evolution of two different types of users in StackOverflow, namely \textit{sparrows} (very active users) and \textit{owls} (experts) in the discussed topics, and could identify various differences between the two user-groups. }

We use the notion of critical mass to define the barrier, that has to be overcome, for collaboration networks to become self-sustaining in terms of activity.

\subsection{Dynamical Systems on Networks}

Dynamical systems in a non-network context are a well-studied scientific and engineering field. Generally, a dynamical system is any system that changes in time, whose behavior is determined by some specific rules or (differential) equations over a set of quantifiable variables. We distinguish between continuous and discrete as well as deterministic and stochastic systems. \citet{strogatz} and \citet{barrat2008dynamical} provide excellent introductions and analyses of dynamical systems.

Different social and economic processes, which take place both offline and online, have been modeled with the use of dynamical systems. In the context of the Web, the primary focus of dynamical systems was set on analyzing and understanding the diffusion of information in online social networks \cite{leskovec2007dynamics, leskovec2009meme, myers2012information, vespignani2012modelling}, including the analysis of online memes and viral marketing.

On the other hand, the Bass Model \cite{bass1976new} describes how novel products are accepted and adopted in a network and has seen a wide variety of applications in different fields of research and also for practical use. The model consists of two parameters, the propensity for innovation and the propensity for imitation. A product will be successfully accepted and adopted by the community, depending in the ratio between these two parameters.

\citet{acerbi2012logic} investigated factors that determine how social traits propagate within a specific popularity. \citet{iribarren2009impact} conducted a viral email experiment, allowing them to track the diffusion of information in a social network. They showed that due to heterogeneity in human activity, the most common and simple growth equation from epidemic models is not suitable to model information diffusion in social networks.

Recently, in the context of activity dynamics, \citet{Ribeiro:2014:MPG:2566486.2567984} conducted an analysis of the daily number of active users that visit specific websites, fitting a model that allows to predict if a website has reached self-sustainability, defined by the shape of the curve of the daily number of active users over time. He uses two constants $\alpha$ and $\beta$, where $\alpha$ represents the constant rate of active members influencing inactive members to become active. $\beta$ describes the rate of an active member spontaneously becoming inactive. Whenever $\beta/\alpha \geq 1$ a website is unsustainable and without intervention the daily number of active users will converge to zero. If $\beta/\alpha < 1$ and the number of daily active users is initially higher than the asymptotic one, a website is categorized as self-sustaining.

The model presented in this paper to simulate activity dynamics heavily relies on the concept of dynamical systems on networks. We strongly believe that by modeling and understanding activity dynamics, we will gain a better understanding of the processes involved in and around the concept of peer influence in collaboration networks. Other areas of application for dynamical systems on networks are the modeling and simulation of diseases in the form of \emph{epidemic models}, and opinions or traits of a person, also known as \emph{opinion dynamics}.

\subsubsection{Epidemic Models}
Modeling the outbreak of diseases can be seen as a special case of dynamical systems. At first, epidemic models dealt with the spreading of diseases in social (real life) networks \cite{may1984spatial,hethcote1978immunization,anderson1991infectious, bolker1993chaos, bolker1995space, lloyd1996spatial, keeling2002estimating, ferguson2003planning}, ignoring the underlying network aspect, simulating contractions and outbreaks via random encounters of the whole population under investigation. For an exhaustive survey of epidemic models refer to \citet{pastor2014epidemic}.

Henceforth, these models have been extended to include the structure and other aspects of the underlying networks \cite{rvachev1985mathematical, ferguson2003planning, hufnagel2004forecast, longini2005containing, ferguson2005strategies, colizza2006role}, limiting the spread and outbreaks according to different factors. Further, epidemic models were also utilized to simulate the spread for a plethora of properties in different kinds of networks, such as viruses spreading in computer networks \cite{kephart1993computers, kephart1997fighting, pastor2001epidemica, Aron:2002:BNP:2622029.2622535, pastor2007evolution} and information propagation (e.g., memes) \cite{leskovec2007dynamics} among others.

In general, epidemic models are based on the intuition that a disease propagates through a social network with a given infection rate, defining the probability that a neighbor of an already infected node contracts the disease. Different models have been developed and analyzed to simulate epidemic outbreaks in a population or network \cite{bailey1975mathematical, anderson1991infectious, hethcote2000mathematics, newman2010networks}, which can only transfer on contact. Typically, such an outbreak is modeled using a small number of possible states for each node and a fixed probability of contraction (e.g., $\beta$, $\gamma$), which defines the probability or ``threshold'' that has to be reached for a node to change to a different state. For example, the SI model consists of only two states -- \emph{susceptible} and \emph{infected} -- and one probability parameter $\beta$, that determines when the transition from susceptible to infected is initiated. Note that transitions in the SI model can only occur from susceptible to infected while already infected nodes remain infected indefinitely. As the infection rate is relative to the population under investigation, epidemic simulations with a small number of originally infected hosts usually start-off by slowly contracting the disease until exponential growth is reached. Once the majority of the population carries the disease, the infection process slows down again until the whole population is infected.

A more sophisticated extension to the SI model is the SIR model \cite{anderson1991infectious, mathbio}, which additionally introduces the \emph{recovered} (or \emph{removed}) state as well as an additional parameter $\gamma$ to model the transition from infected to recovered. Again, transitions only occur from susceptible to infected to recovered. As the name suggests, this newly introduced state allows nodes to become immune to the disease and will not be infected in the future, nor be able to infect other nodes. Other models for simulating epidemic outbreaks are the SIS and SIRS models, where the population can recover but does not become immune (SIS) or stays immune but still has a chance to become susceptible for infection again (SIRS) \cite{britton2010stochastic, dietz1967epidemics}.

Since their introduction, epidemic models have seen a wide array of application. For example, to analyze how computer viruses spread \cite{kephart1991directed, kephart1993measuring, newman2002email} or the study of epidemics in complex (scale-free, power-law) networks \cite{pastor2001epidemica, pastor2001epidemicb, pastor2002epidemic, moreno2002epidemic}.

Among others \citet{wang2003epidemic} as well as \citet{ganesh2005effect} demonstrated the importance of the networks spectra (eigenvalues and eigenvectors of the network adjacency matrix) for epidemic and dynamical network models \cite{chung2003eigenvalues, chung2003spectra}. We show a similar dependency of activity dynamics on eigenvalues in this paper in Section~\ref{sec:modeling activity dynamics}.

\subsubsection{Collective Behavior \& Opinion Dynamics}
Another important field of application of dynamical systems on networks are opinion dynamics. They are used to model collective behavior and influence, usually in the form of a consensus-reaching task, at every point in time. The main idea behind the concept of social influence is that interacting agents strive to become more alike \cite{festinger1950social}.

For example, agents in the Ising model for ferromagnets \cite{binney1992theory, barthelemy2011spatial} are influenced by the state/opinions of the majority of their peers. This influence naturally drives the system towards an ordered state where all agents are either positive or negative (ferromagnets). Hence, the model can be interpreted as a very simple model for simulating (binary) opinion dynamics. However, the transition probabilities of the Ising model are influenced by temperature, representing the modeling of external or influential factors. In particular, if the temperature is above a certain threshold, consensus-finding, in terms of magnetization, becomes an unstable process that never converges. The Potts model \cite{wu1982potts, dorogovtsev2008critical} further extends the Ising model by increasing the number of potential states an agent can assume from two (positive or negative) to an arbitrary number greater than two. Other factors that might influence the process of reaching consensus is the size of the system under investigation \cite{tessone2009diversity}. In particular, this means that differently sized (or connected) systems potentially need different strategies to reach consensus.

Opinions are usually represented as a set of words or numbers for each agent individually. \citet{weidlich1971statistical} introduced such a model, based on sociodynamics, in 1971.  \citet{galam1982sociophysics,galam1991towards} analyzed the potential applications of the Ising model for simulating opinion dynamics starting in 1982.

The most wide-spread and adapted models to simulate (among others) opinion dynamics are the voter model \cite{clifford1973model,holley1975ergodic}, the Axelrod model \cite{axelrod1997advancing} as well as The Naming Game \cite{baronchelli2006sharp}.

\emph{The voter model} constitutes that each agent is equipped with a binary variable. At each step in time, the binary variable of one (randomly chosen) agent is synchronized with one of its neighbors variable. Introducing the concept of social influence for opinion dynamics. The voter model has since been adapted and extended by many researchers to fit an array of different purposes (e.g., \cite{mobilia2003does, mobilia2005voting, mobilia2007role, vazquez2003constrained, vazquez2004ultimate, castello2006ordering}).

\emph{The Axelrod model} \cite{axelrod1997advancing} combines the notion of social influence -- individuals becoming more similar upon frequent interactions -- and the tendency that similar individuals will have a higher tendency (and frequency) to interact with each other. Each agent is endowed with a set of characterizing variables. The more variables are shared among two agents, the more similar they are. Given this description, one would assume that the described notions are self-reinforcing dynamics and hence, will inevitably produce stable networks with only identical agents. However,  \citet{castellano2000nonequilibrium} have shown that the resulting number of different states is dependent on the number of characterizing variables. Large numbers are likely to result in very few similar individuals (high agent diversity). Analogously to the voter model, the Axelrod model has been extensively adapted, analyzed and expanded by researchers to broaden our understanding of the spread of (cultural) traits across agents (e.g., \citet{klemm2003nonequilibrium,klemm2003global, flache2007local}).

\emph{The Naming Game} originates from idea to analyze and explore the evolution of language \cite{steels1995self}. \citet{baronchelli2006sharp} introduced the most basic version of The Naming Game in 2006, where a group of agents that communicate via a complete network, try to reach consensus when naming an entity. Each agent holds a list of synonyms or words associated with the entity, also referred to as vocabulary, under investigation. Every iteration (or step in time), two agents are chosen. One agent is assigned the role of the speaker, who randomly choses a word of a given/pre-defined vocabulary. If the other agent -- the listener -- knows (i.e., also has the word in the vocabulary) the chosen word, both agents discard all other words in their vocabulary and ``agree'' on the common word. However, if the listeners do not know the word of the speaker, the word is appended to their vocabulary and no words are discarded. In the next step another pair of nodes is chosen and process is repeated until either consensus is found or a predetermined number of steps (time) have passed. The Naming Game has spurred a complete line of dynamical models with a variety of different parameters, that each address different problems and tasks (e.g., \citet{abrams2003linguistics, minett2008modelling, wang2005invasion, castello2006ordering}). For an excellent and comprehensive introduction to opinion dynamics (among others) we refer the interested reader to \citet{castellano2009statistical}.

\section{Discussion, Limitations \& Future Work}
\label{sec:discussion}

We have developed a model\footnote{We have released a Python implementation of our model, to estimate empirical parameters and run activity dynamics simulations, as Open Source Software at \url{https://github.com/simonwalk/ActivityDynamics}.} to simulate and characterize the intricate dynamics of activity in collaboration networks, consisting of an \ldesc and \mdesc. First, we applied it on Zachary's Karate Club (see Figure~\ref{fig:Karate}) dataset to illustrate its core mechanics. Subsequently, we continued with a linear stability analysis (cf. Section~\ref{subsec:linear stability analysis}) and depicted the behavior that can occur when the master stability equation is invalidated (see Figure~\ref{fig:Karate}). Using our proposed model to simulate activity dynamics, we have shown that the overall activity in collaboration networks appears to be a composite of the \ldesc and the \mdesc, as described in Section~\ref{sec:modeling activity dynamics}. In Section~\ref{sec:illustrative examples}, we have fitted our model on \revone{synthetic and} empirical datasets to simulate activity dynamics trends.

The presented results are destined to be interpreted only and solely as an indicator for trends in activity dynamics, rather than absolute values that can be used for accurately predicting the activity for a given system. This is a direct result of the different approximations and simplifications (cf. Section~\ref{sec:illustrative examples}) that we have made when estimating the parameters for our activity dynamics simulation. 

\revone{Note that one advantage of our model over other existing approaches, such as autoregression, is the interpretability of the ratio \ratio. For example, a ratio of $4$ means that users intrinsically lose activity $4$ times faster than they can get back from one of their peers, while the coefficients of the autoregression lack such interpretable characteristics. Further, using the concept of dynamical systems we can represent the underlying mechanisms in a closed form, allowing for detailed analytical analyses (i.e., the linear stability analysis), which is much harder (if not impossible) to conduct for other models, such as agent-based models, autoregression or more complex models based on dynamical systems.}

For future work we plan on extending the ability of our model to not only reflect on changes in activity dynamics but also properly cope with structural changes in the underlying collaboration networks. One additional limitation of the presented approach is the fact that nodes with a very small degree, which are not connected to the largest connected component, inevitably will lose activity until they reach the point of total inactivity. Including the structural evolution of a collaboration network in our analyses will allow us to mitigate this effect, as users will only be added to the collaboration network and considered in our calculations, once they have actually become active. One potential approach involves the investigation of snapshots of the collaboration networks at every $\tau$, providing additional insights into the evolution of the parameters of our model and the investigated systems. Additionally, we assume that peer influence is a symmetric property. This means that posts and replies exercise the same amount of influence on peers as we do not differentiate between different types of activity and influence will always traverse along both directions of the edges in our collaboration networks. \revone{Further, tasks that do not trigger entries in the change-logs (i.e., reading articles, posts or replies) are not considered in our experiments due to a lack of available data.}   

\revone{The fact that the \adm only requires a single parameter to be configured represents not only an advantage, but also a limitation. Given that there is only one parameter that determines the evolution of activity in a system, we are not be able to model periodic fluctuations with only one ratio. Instead, we have to calculate ratios for multiple points in time. For future work we plan on extending the \adm by adding parameters, for example, to model different external influences. With this extended model, we will be able to simulate such periodic patterns with a single configuration. On the other hand, we are only able to model additional (social) mechanisms with the use of additional parameters. For example, one reason for the decreasing levels of activity in Wikipedia might also be related to a very high barrier for newly registered users to add content due to comprehensive guidelines for contributions and a very concentrated and active community of power users. Over time, these power users leave Wikipedia for various reasons while new contributors are lacking to fill in the gaps.}

Furthermore, all of our estimated parameters are calculated for the collaboration networks as a whole. Future work will also include extending the activity dynamics model to calculate the ratio \ratio on a user level, rather than on a network level. This modification not only potentially increases the accuracy of our model but would also allow us to gather additional information for each user of the corresponding networks. Further, with an increased accuracy in our simulations it will be possible to conduct activity prediction experiments and emulate network attacks as well as optimize (arbitrary) cost-strategies for increasing activity in these systems.

\revone{In this context it is also worth mentioning that decreasing levels of activity for collaboration networks can also signal that the community has completed their work and no further actions are required as the intended goal has been achieved. Further analyses are required to determine if completeness and quality of content affect activity in collaboration networks. One could even argue that, once we are able to calculate \ratio for each user, we could potentially observe the evolution of users and categorize different types of users in collaboration networks (e.g., early adopters or experienced users versus new and inexperienced users).}

The ratio \ratio---describing how fast users lose activity (\ldesc $\lambda$) over how fast they regains activity over their neighbors (\mdesc $\mu$)---fluctuates below the corresponding highest eigenvalue $\kappa_1$ for all investigated empirical datasets. Negative peaks in this ratio represent periods of time ($\tau$; in our case weeks) where activity grew faster than could be compensated by the \mdesc. It naturally follows that a decrease of $\lambda$---resulting in less activity-loss per contribution for each user---is necessary to accomplish such drastic increases of activity. If the network itself is of a smaller scale and/or these negative peaks occur on a frequent basis, the activity dynamics of the corresponding networks are depending on the contributions (and thus influence) of single (individual) users. To compare the stability of the activity dynamics across multiple networks we calculated the \emph{System Mass} and \emph{Activity Momentum} $p$---indicating the required force to accelerate or render the corresponding collaboration networks inactive.

When comparing $p$ and the results of our empirical illustration (cf. Figures~\ref{fig:empirical_illustration} and \ref{fig:empirical_ratios}) between the different datasets, we can see that the \emph{Activity Momentum} is very small for datasets that either (i) exhibit only a very small number of changes and are close to inactivity or (ii) exhibit a small $\kappa_1$ (see Figure~\ref{fig:empirical_illustration} and \ref{fig:empirical_ratios}). This suggests that we can use \emph{Activity Momentum} as an indicator for the robustness of a collaboration network with regards to its activity dynamics.

Further, we can characterize the potential of a collaboration network to become self-sustaining by comparing the calculated ratios of \ratio with the corresponding $\kappa_1$ and \emph{Activity Momentum}. If the ratio is below $\kappa_1$, our master stability equation is invalidated, pushing the system towards a new fixed point where the forces of the \ldesc and the \mdesc reach an equilibrium so that the network converges towards a state of immanent and lasting activity (see Figure~\ref{fig:Karate}). If such a state is reached and combined with a high \emph{Activity Momentum}, the corresponding collaboration network has reached critical mass of activity and has become self-sustaining; no external impulses are required to keep the network active. Of course, in real world scenarios, activity will not last forever without providing additional incentives as interest (and thus activity) in a system potentially decays over time. As a consequence, this would first result in an increase of $\mu$ and inevitably, with a sufficiently large $\mu$, the collaboration network would return to its stable fixed point, once our master stability equation holds again, and activity would once more converge towards zero. Once we extend our model to allow for user-based calculations, we will be able to not only calculate \emph{Activity Momentum} for collaboration networks, but also for single and individual users.

\section*{Acknowledgements}
This research was in part funded by the FWF Austrian Science Fund research projects P24866.

\bibliographystyle{numbers, open={[}, close={]}}

\end{document}